\documentclass[preprint]{aastex}
\usepackage{natbib}
\usepackage{emulateapj5}
\newcommand{\Halpha}{H$\alpha$}
\newcommand{\HI}{\ion{H}{1}}
\newcommand{\HII}{\ion{H}{2}}
\newcommand{\NII}{[\ion{N}{2}]}
\newcommand{\kms}{\mbox{km~s$^{-1}$}}
\newcommand{\Msol}{\mbox{M$_\odot$}}

\newcommand{\tgas}{\ensuremath{\tau_{\rm gas}}}

\shorttitle{Gas Kinematics and Star Formation in NGC 4736}
\shortauthors{Wong \& Blitz}

\slugcomment{accepted 17 April 2000 by ApJ}

\begin{document}

\title{Non-circular Gas Kinematics and Star Formation \\
	in the Ringed Galaxy NGC 4736}
\author{Tony Wong and Leo Blitz}
\affil{Astronomy Department and Radio Astronomy Laboratory, 
University of California, Berkeley, CA 94720}
\email{twong, blitz@astro.berkeley.edu}

\begin{abstract}

We analyze the gas kinematics and star formation properties of the
nearby RSab galaxy NGC 4736 using interferometric and single-dish
CO(1-0) data and previously published \Halpha\ and \HI\ data.  The CO
morphology is dominated by a central molecular bar and tightly wound
spiral arms associated with a bright ring of star formation.  Strong
\HI\ emission is also found in the ring, but \HI\ is absent from the
central regions.  Comparison of the \HI\ and \Halpha\ distributions
suggests that \HI\ in the ring is primarily dissociated H$_2$.
Modeling of the CO kinematics reveals gas motion in elliptical orbits
around the central bar, and we argue that the ring represents both the
OLR of the bar and the ILR of a larger oval distortion.  The
\HI\ kinematics show evidence for axisymmetric inflow towards the ring
and are inconsistent with streaming in aligned elliptical orbits, but
the highly supersonic ($\sim$40 \kms) inflow velocities required,
corresponding to mass inflow rates of $\sim$2 \Msol\ yr$^{-1}$,
suggest that more sophisticated models (e.g., gas orbiting in precessed
elliptical orbits) should be considered.  The radial CO and
\Halpha\ profiles are poorly correlated in the vicinity of the nuclear
bar, but show a better correlation (in rough agreement with the Schmidt
law) at the ring.  Even along the ring, however, the azimuthal
correspondence between CO and \Halpha\ is poor, suggesting that massive
stars form more efficiently at some (perhaps resonant) locations than
at others.  These results indicate that the star formation rate per
unit gas mass exhibits strong spatial variations and is not solely a
function of the available gas supply.  The localization of star
formation to the ring is broadly consistent with gravitational
instability theory, although the instability parameter $Q \sim 3$ on
average in the ring, only falling below 1 in localized regions.
Large-scale dynamical effects, by concentrating gas at resonances and
influencing the star formation rate, appear to play a key role in this
galaxy's evolution.

\end{abstract}


%% INTRODUCTION

\section{Introduction}\label{intro}

The process of star formation in present-day galaxies, although still
poorly understood, appears to require the accumulation of significant
amounts of cold neutral gas.  Thus, in galaxies where star formation is
concentrated in distinct structures such as rings, the cold gas is also
found concentrated in such structures, as revealed by high-resolution
radio interferometry \citep[see review by][]{Ken97}.  Exceptional
examples of such rings include those in NGC 4314 \citep*{Ben96} and NGC
4321 \citep{Sak95}.  Two key questions that are raised by the
study of these systems are the following: what are the physical
processes that lead to the formation of gaseous ring structures, and
how does gas accumulation ultimately lead to star formation?

In the usual interpretation of dynamical rings (where no collision
between galaxies is indicated), a rotating bar or oval distortion
induces a radial flow of gas toward orbital resonances, leading to
ring formation.  While this explanation has been lent considerable
support by numerical simulations \citep{Sch81,Com85,Byrd94} as well as
observations that show that rings are preferentially found in barred
galaxies \citep[see review by][]{But96}, direct kinematic evidence that
bars drive {\it net} radial flows has been lacking.  Indeed this is hardly
surprising: \citet{Ken94} has pointed out a number of problems with
measuring gas inflow rates directly, among them the tendency for gas in
a barred potential to follow elliptical orbits that can mimic net inflow
when only the line-of-sight velocity component is observed.  As a
result, inflow rates have been calculated only indirectly, by
comparison with hydrodynamic models \citep*{Reg97} or by estimating the
gravitational torque exerted by the stellar potential on the gas
\citep{Qui95}.  These studies have yielded net inflow velocities of
$\sim$10--20 \kms.

Once gas is available, are there general ``laws'' that govern the rate
at which it is converted into stars?  \citet{KC89} investigated this
question from a primarily global perspective, using disk-averaged
quantities to derive a Schmidt (\citeyear{Schmidt59}) type law, in
which the star formation rate (SFR) per unit area is roughly
proportional to a power $N \sim 1.5$ of the gas surface density
($\Sigma_{\rm SFR} \propto \Sigma_{\rm gas}^N$).  Yet he found that a
simple power law tended to overpredict the SFR in the outer parts of
galaxies, and consequently revived a suggestion by \citet{Quirk72} that
star formation could only occur in regions where the gas density
exceeded a critical threshold for gravitational instability.  In this
scenario, a controlling factor in star formation lies in the
large-scale properties of the galaxy disk, in contrast to theories in
which star formation is regulated by conditions in the local
environment \citep[e.g.][]{Dop85}.  Further observations, at higher
spatial and velocity resolution than those available to \citet{KC89},
should provide a clear test of the generality of the Schmidt law and
the star formation threshold, distinguishing them from alternative star
formation laws that depend on the orbital frequency \citep{Wyse89} or
tidal shear \citep*{Ken93}.

A puzzling result also highlighted by \citet{KC89} is the lack of
correlation between disk-averaged \Halpha\ and CO line intensities;
indeed, he found a substantially better correlation between \Halpha\
and \HI\ emission.  Two explanations were suggested: a variation in
the CO-to-H$_2$ conversion factor \citep[hereafter the ``X-factor'',
following][]{Bloem86}, or a bimodal star formation process that leads
to low-mass star formation being more strongly coupled to the
molecular gas than high-mass star formation.  More recently,
\citet{Elm93} has argued that both diffuse and self-gravitating
molecular clouds can exist, and as a result the H$_2$ mass may be a
poor indicator of star formation.  Although these ideas may be
difficult to test observationally, a simple comparison of how \HI\ and
CO are distributed {\it within} galaxies---relative to recent star
formation---may provide clues as to why, on large scales, CO appears
to be only weakly correlated with star formation.

Motivated by these questions, we have undertaken a detailed study of
the gas kinematics and star formation in NGC 4736, a nearby ($d$ = 4.2
Mpc using $H_0$ = 75 \kms\ Mpc$^{-1}$) Sab galaxy with a bright ring of
star formation about 45\arcsec\ (1 kpc) from its center.  A faint
outer ring, about 5\arcmin\ (6 kpc) in radius, is also seen in deep optical
images \citep{San61}.  Previous \HI\ synthesis observations
\citep*{Bos77} and optical and near-infrared imaging \citep*[][hereafter
MMG95]{Mol95} have demonstrated that the disk of the galaxy is
non-axisymmetric, and hence the rings may occur at the inner and outer
Lindblad resonances (ILR and OLR) of an oval potential \citep*{Ger91}.
In addition, MMG95 have suggested that the inner ring may coincide
with the OLR of a central stellar bar, $\sim$30\arcsec\ in extent,
seen in optical and near-infrared isophotes.  We summarize the basic
observational parameters of the galaxy in Table \ref{proptable}.

\begin{center}
Table 1. Global Properties of NGC 4736\label{proptable}
\vskip 0.1in
\begin{tabular*}{3in}{llr}
Parameter	& Value & Ref.\\ \hline\hline
Right ascension (J2000) & $12^h50^m53.\!\!^s06$ 	& 1 \\
Declination (J2000)	& $41^\circ 07^\prime 13$\farcs65 & 1 \\
Morphological type	& (R)SA(r)ab			& 2 \\
Systemic LSR velocity	& 315 \kms 			& 3 \\
Adopted distance, $d$	& 4.2 Mpc 			& 3 \\
Angular scale		& 1\arcsec\ = 20 pc		& 3 \\
Inclination, $i$	& 35\arcdeg 			& 4 \\
Position angle, $\phi_{maj}$	& $\sim$295\arcdeg 	& 3 \\
Optical diameter, $D_{25}$	& 11\farcm2 (13.5 kpc)	& 2 \\
\hline\hline
\end{tabular*}
\vskip 0.1in
{\footnotesize
(1) \citealt*{Bec95};
(2) \citealt{RC3};
(3) This paper;
(4) \citealt{Mol95}.}
\vskip 0.1in
\end{center}

The nucleus of NGC 4736 has also been studied intensively.  Classified
as a LINER based on optical emission lines \citep{Hek80}, several
indications suggest that it harbors an active galactic nucleus (AGN): a
compact nuclear source has been detected at 5 and 15 GHz
\citep[1\arcsec\ resolution,][]{Tur94}, with a spectral index
indicative of synchrotron emission, as well as in the ultraviolet with
{\sl HST} \citep[0\farcs1 resolution,][]{Maoz95}, and the X-ray with
the {\sl ROSAT}\/ PSPC \citep*[$\sim$30\arcsec\ resolution,][]{Cui97}.
The lack of emission lines and dust features in the 8--13 $\mu$m
spectrum led \citet{Roche85} to conclude that there was relatively
little star formation occurring in the nucleus.  However, the presence
of strong Balmer absorption lines indicative of A-type stars
\citep{Prit77} and strong CO absorption indicative of young red giants
\citep*{Walk88} suggests that a nuclear starburst occurred roughly 1 Gyr
ago.  If NGC 4736 indeed harbors an AGN and recent starburst, it would
be an important testbed for theories relating nuclear bars to the
fueling of nuclear activity \citep*[e.g.,][]{Sim80, Shl89}.

The results of our investigation are presented as follows.
\S\ref{obs} describes our new CO observations and the \HI\ and
\Halpha\ datasets used for comparison.  \S\ref{morph} compares and
contrasts the distribution of atomic and molecular gas in this galaxy,
revealing some striking differences in both the radial and azimuthal
profiles.  In \S\ref{kin} we discuss the gas kinematics as traced by
CO and \HI\ and compare them with simple models of radial inflow, a
warp in the disk, and oval orbits.  In \S\ref{sfr} we analyze the star
formation rate and its relation to the observed gas density.  The
results and implications of this work are discussed in \S\ref{disc},
and our conclusions are summarized in \S\ref{conc}.
Preliminary results from this study have appeared in \citet{Wong99}.

%% OBSERVATIONS

\section{Observations and Data Reduction}\label{obs}

\subsection{BIMA CO Observations\label{obs_bima}}

Observations with the Berkeley-Illinois-Maryland Association\footnote{The
BIMA Array is funded in part by the National Science Foundation.} (BIMA)
interferometer were conducted in 1996 October and 1997 April in the C
array configuration (projected baselines 2.2--36 k$\lambda$) with 9
antennas.  Additional observations in the D configuration (2.3--11
k$\lambda$) with 10 antennas were made in 1999 July and August as part
of the BIMA Survey of Nearby Galaxies (BIMA SONG).  To ensure that the
entire \Halpha\ ring was imaged, a 7-pointing hexagonal mosaic was
observed, with the pointings separated by 50\arcsec\ (half of the
primary beam FWHM).  This provides fairly uniform sensitivity out to a
radius of $\sim$50\arcsec.  The total on-source observing time was 25 hours.

The receiver was tuned to the CO ($J=1\rightarrow 0$) transition at
115.2712 GHz ($\lambda$=2.6 mm), and the correlator was configured to
have 4 independently positioned spectral windows, each with 100 MHz
bandwidth and 64 channels.  The LSR velocity range covered was
150--650 \kms\ at a resolution of 4.06 \kms.  The typical SSB system
temperature corrected to above the atmosphere ($T_{sys}^*$) was
400--600 K on all tracks.

The data were calibrated and reduced using the MIRIAD package
\citep*{Sau95}.  An online linelength calibration was applied to
correct for variations in the length of the signal path from the
antenna to the correlator; remaining drifts in the antenna phase gains
were then corrected by observing the nearby quasar 1310+323 every half
hour and fitting a low-order polynomial to the antenna-based
self-calibration solutions.  Flux calibration was performed using
observations of Mars taken on each track.  A bandpass measurement was
made on 3C273 for each track, to verify that the online bandpass
calibration was working properly.  Following the initial calibration,
the data were subjected to an additional round of phase-only
self-calibration using a deconvolution model.

The uncertainty in the flux scale is likely to be dominated by the
effects of atmospheric phase decorrelation.  Although the Mars
observations, when phase-corrected over very short ($\sim$1 s)
timescales, provide a measurement of the antenna gains to within a few
percent (apart from uncertainties in the Martian flux model), rapid
phase correction cannot likewise be applied to the source data.  Hence
the source flux will be attenuated when averaged over time.  To make a
partial correction for decorrelation, we have instead derived the
antenna gains by a bootstrap method: a comparison with Mars was used to
derive an estimate for the flux of 1310+323, and this value was then
compared to the visibility amplitudes of the 1310+323 data, when
averaged over 6-minute timescales, to derive the antenna gains.  The
resulting gains should scale up the amplitudes for antennas that share
many long baselines and thus suffer greater decorrelation.  Comparison
of source spectra from the four best tracks suggests an uncertainty of
$\lesssim$20\% in the flux scale; the uncertainty is somewhat larger
for the poorer tracks, but data from those tracks is weighted less due
to their high noise variance.

After calibration of the visibilities, two datacubes were produced
with 10 \kms\ channels, one with a robust weighting \citep{Briggs95},
which provides a good compromise between the sensitivity of natural
weighting and the resolution and sidelobe suppression of uniform
weighting, and the second with a Gaussian taper applied to weight down
the longer baselines.  These will be referred to as the ``robust'' and
``tapered'' datacubes respectively.  The cubes were deconvolved using
a variant of the CLEAN algorithm developed by \citet*{Steer84} to
improve the deconvolution of spatially extended emission, and
implemented in the MIRIAD task MOSSDI.\@ The robust datacube had a
resolution of 6\farcs86 $\times$ 5\farcs02 (FWHM of synthesized beam).
The tapered datacube, although having poorer resolution (15\arcsec),
was more sensitive to extended structure and less affected by
atmospheric decorrelation and resulting calibration errors because it
emphasizes shorter baselines.  (For these observations the
``short-baseline'' data come primarily from antennas that were
physically close together, not from projection effects due to low source
elevation.) The rms noise in the robust and tapered datacubes were
roughly 70 and 80 mJy beam$^{-1}$ (190 and 33 mK) respectively for
each 10 \kms\ channel.

\subsection{Kitt Peak CO Observations\label{obs_kp}}

Single-dish CO observations were performed in 1997 June at the 
NRAO\footnote{The National Radio Astronomy Observatory is a facility
of the National Science Foundation, operated under cooperative
agreement by Associated Universities, Inc.}
12~m telescope on Kitt Peak, Arizona.  With the 3 mm SIS receivers,
typical system temperatures were 300--400 K.  Spectra from orthogonal
polarizations were recorded by two 256-channel filterbanks (one for
each polarization) with a channel width of 2 MHz (5.2 \kms).
Calibration was performed using the chopper wheel method
\citep[e.g.,][]{Kut81} approximately every 20 minutes.  Pointing was
checked approximately every 2--2.5 hours with observations of Mars;
the typical change in pointing offsets was 8\arcsec\ (rms) and ranged
from 3\arcsec\ to 12\arcsec.

A total of 43 grid points were observed, comprising a 7 $\times$ 7
square grid centered on NGC 4736 with 3 points each at the NE and SW corners
omitted.  The spacing of the grid was 27\arcsec, approximately half of
the FWHM of the primary beam.  Nyquist sampling of all spatial
frequencies (``baselines'') from 0 to 12 m requires a grid spacing of
$\lambda/2D$ = 22\arcsec; nonetheless, because the dish illumination
falls to zero towards the edges, the amount of signal aliased by our
undersampling should be small.  On-source integration time was
typically 10 minutes for the inner 5 $\times$ 5 grid points (1\farcm8
$\times$ 1\farcm8) and 4 minutes for the outer points.

Spectra were output on the $T_R^*$ scale, which represents the source
antenna temperature corrected for ohmic losses, atmospheric absorption,
spillover, and scattering \citep{Kut81}.  Linear baselines were fit to
line-free channels of the spectra and subtracted; higher-order
variations in the baseline were generally not seen.  The resulting
datacube (hereafter the ``KP cube'') had an rms noise of 11 mK in each
5.2 \kms\ channel.  The $T_R^*$ temperatures were converted to Janskys
using a nominal telescope gain of 33 Jy K$^{-1}$.  This gain is
strictly appropriate for point sources only; it was found that the
fluxes had to be scaled up by 10\% to produce a more satisfactory
combination with the BIMA data.

\subsection{Combining the BIMA and KP Data\label{obs_kpbima}}

The total CO flux measured in the KP cube was 1620 Jy \kms, which is
comparable to the flux of 1600 Jy \kms\ detected by \citet{Ger91} using
the IRAM 30~m telescope.  For comparison, the tapered BIMA cube
recovered a flux of 1000 Jy \kms, $\sim$60\% of the total flux, while the
robust BIMA cube recovered a flux of 620 Jy \kms.  These results
illustrate that deconvolution errors, which become more severe at higher
resolution because of greater decorrelation, can
affect the CLEAN algorithm's ability to recover the actual flux.

Two techniques were employed to combine the interferometer and
single-dish data.  In the ``linear combination'' technique, the BIMA
and KP data are combined in the image plane before deconvolution, and
then deconvolved with a composite beam, as described by
\citet{Stan99}.  In the ``Fourier'' technique, implemented in the
MIRIAD task IMMERGE, the KP and deconvolved BIMA maps are transformed
to the Fourier plane, and low spatial frequencies in the BIMA maps are
partially replaced by single-dish data in the inner part of the Fourier
plane.  The results of the two methods agreed well (cf.\ discussion
below), especially considering the inherent limitations of the data
(uncertain relative flux calibration, single-dish pointing errors,
etc.).  Further discussion of techniques for combining single-dish and
interferometer data and their application to the BIMA SONG will follow
in a future paper (Regan et al.\ 2000, in preparation).

Following the data combination, an integrated intensity (hereafter
``moment-0'') map for each cube was made by first smoothing each
datacube to 20\arcsec\ (robust) or 30\arcsec\ (tapered) resolution and
masking out regions in the original cube where the signal fell below
the 3$\sigma$ level in the smoothed cube.  The channel maps of the
original cube were then summed, including only unmasked pixels that
rose above the 1$\sigma$ level.  Although this masking technique
introduces some bias against finding isolated compact emission in the
datacube, it is valuable in practice because the emission at any given
point in the galaxy is generally confined to a small number of spectral
channels, and will be strongly attenuated if the noise from all of the
channels is added in.

The line-of-sight velocity field was derived by making Gaussian fits to
the spectrum at each pixel of the datacube.  This provides greater
robustness than a first moment map, which can be strongly affected by
noise in outlying channels \citep[for a full discussion
see][]{Mythesis}, without having to impose an arbitrary mask or
clipping level.  The assumption that the profiles are Gaussian was
found to be a reasonable one: inspection of the profiles for the
robustly weighted cube showed no clear indication of double-peaked
profiles, although asymmetries in profiles near the center do occur as
a result of beam smearing of the galaxy's rotation.  To reduce noise in
the velocity field, only Gaussian fits with integrated intensities
of $>$6 Jy beam$^{-1}$ \kms\ and uncertainties in mean velocity of 
$<$20 \kms\ were included.

%%%%%%%%%%%%%%%%%%%%%%%%%%%%%%%%%%%%%%%%%%%%%%%%%%%%%%%%%%%%%
%%%%%%%%%%%%%%%%%%%%%%%%   FIG. 1   %%%%%%%%%%%%%%%%%%%%%%%%%
%%%%%%%%%%%%%%%%%%%%%%%%%%%%%%%%%%%%%%%%%%%%%%%%%%%%%%%%%%%%%

\vskip 0.25truein
\includegraphics[width=3.25in]{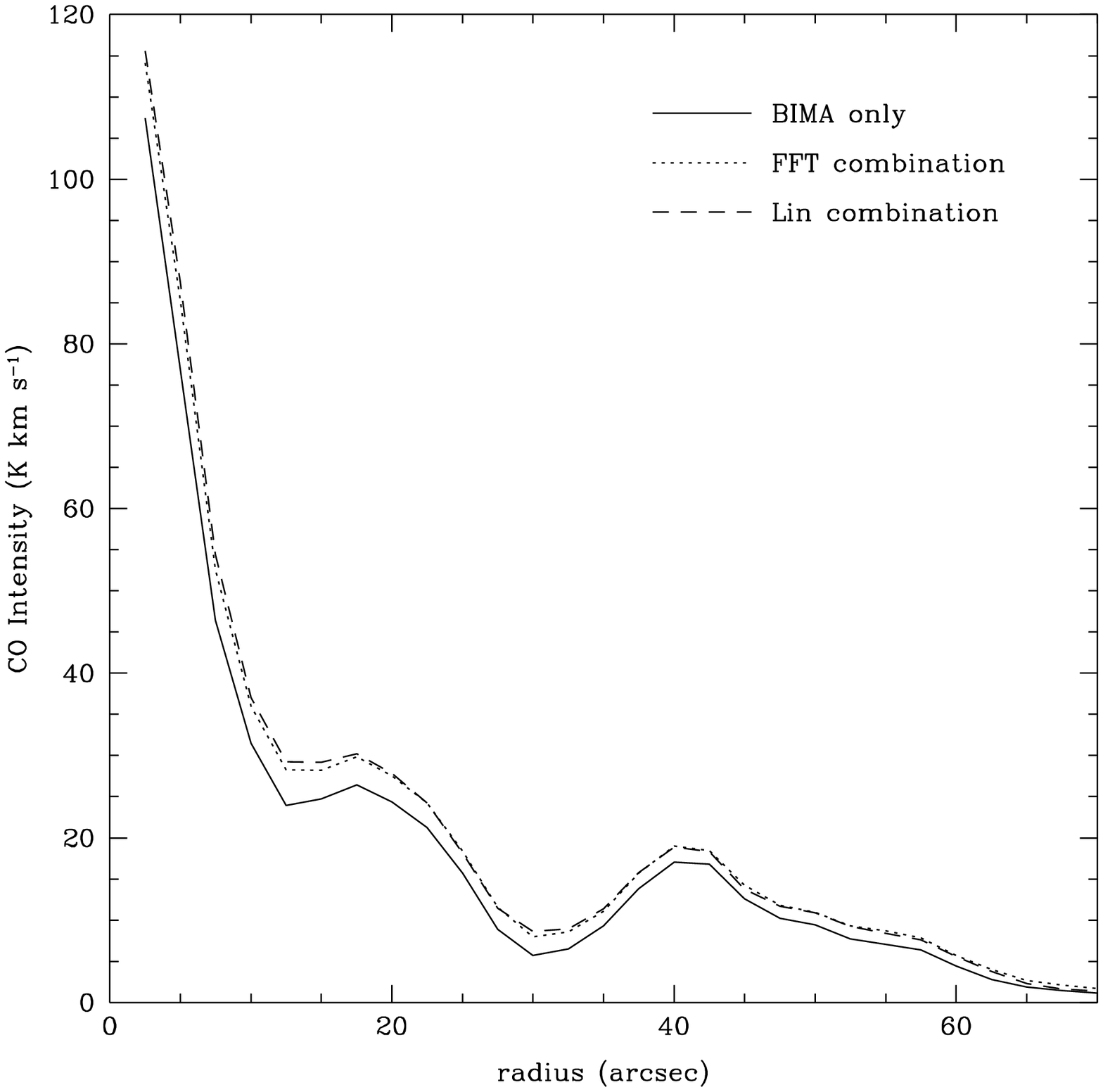}
\figcaption[datacomb.eps]{
Radial CO profiles, averaged in 2\farcs5-wide elliptical rings, for
the BIMA-only robustly weighted data, as well as the combined
(KP+BIMA) data using two data combination techniques (the ``Fourier''
method and the ``linear combination'' method, see text).  No
inclination correction has been made.
\label{datacomb}}
\vskip 0.25truein

%%%%%%%%%%%%%%%%%%%%%%%%%%%%%%%%%%%%%%%%%%%%%%%%%%%%%%%%%%%%%

In Figure~\ref{datacomb} we compare the radial CO profiles derived from
the BIMA-only moment-0 maps to those derived from each of the two data
combination techniques.  Not only do the two combination methods show
good agreement, but the inclusion of the single-dish data has
surprisingly little impact on the radial profile.  Two factors appear
to be responsible for this result.  First of all, due to the short
baselines provided by BIMA, only very extended ($>$30\arcsec), low
surface brightness structure is resolved out, resulting in little
change to the {\it shape} of the profile over the region observed.
This also implies that the derived velocity field is relatively
insensitive to the inclusion of the single-dish data, since it is
weighted toward high-brightness regions.  Secondly, the low measured
fluxes in the BIMA cube are mainly due to large-scale negative regions
in the channel maps that are anticorrelated with the actual emission,
and have therefore been excluded in the process of forming the moment-0
map.  (These negative regions result from imperfect deconvolution,
probably due to residual calibration errors and phase decorrelation.)
As a result, while the inclusion of the single-dish data has a
noticeable impact on the flux of each channel map, it has much less
effect ($\lesssim$20\%) on the flux of the moment-0 map, even for the
robust datacube.  Constructing the moment-0 maps using Gaussian fits
instead yielded similar results.

%%%%%%%%%%%%%%%%%%%%%%%%%%%%%%%%%%%%%%%%%%%%%%%%%%%%%%%%%%%%%
%%%%%%%%%%%%%%%%%%%%%%%%   FIG. 2   %%%%%%%%%%%%%%%%%%%%%%%%%
%%%%%%%%%%%%%%%%%%%%%%%%%%%%%%%%%%%%%%%%%%%%%%%%%%%%%%%%%%%%%

\setlength{\textfloatsep}{6pt}
\begin{figure*}
\begin{center}
\includegraphics[width=4in,angle=-90]{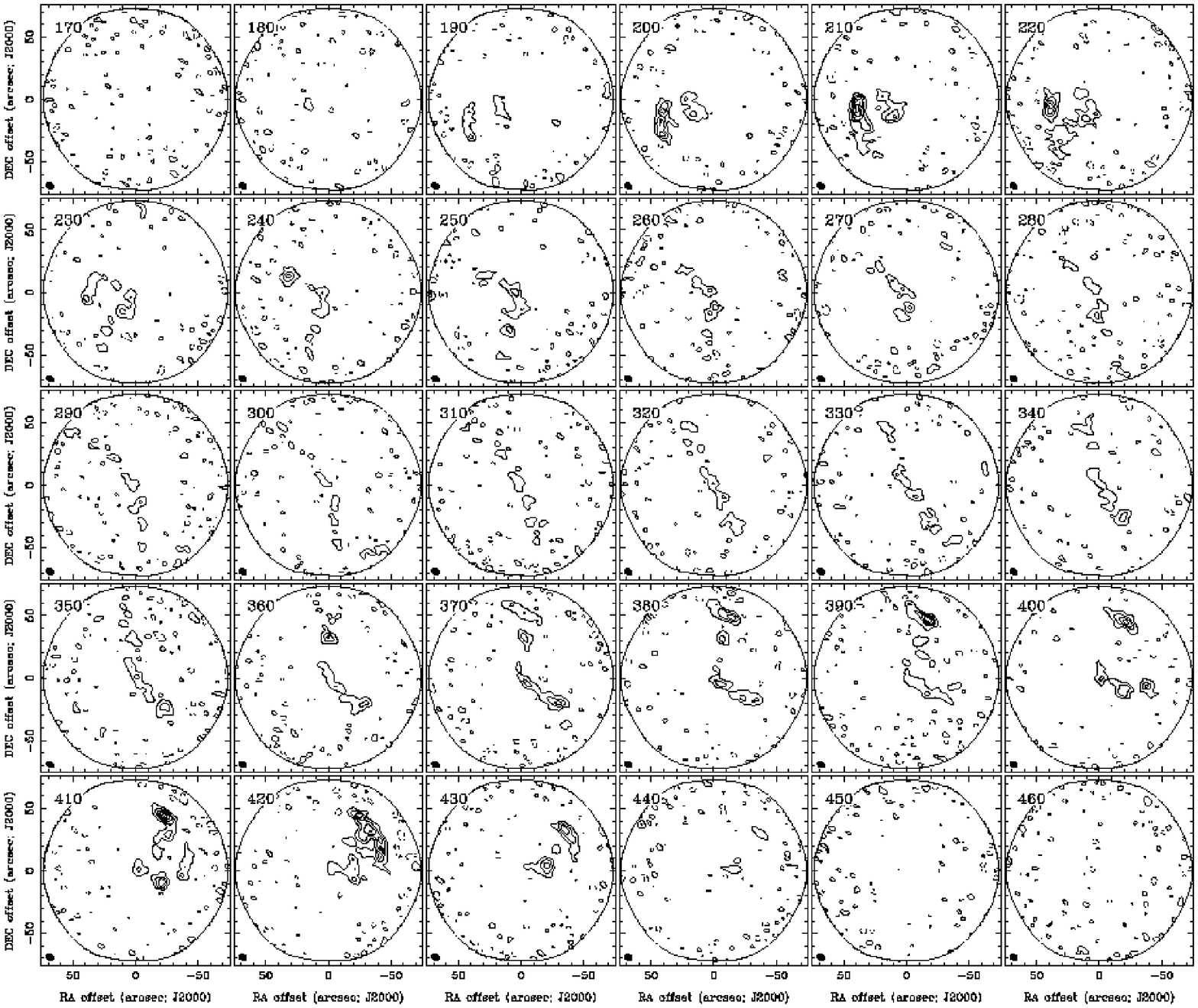}
\figcaption[robmaps.bit.ps]{
Robustly weighted channel maps of CO emission in NGC 4736.
Contours are spaced by 3$\sigma$ starting at 2$\sigma$, where
$\sigma$=0.07 Jy bm$^{-1}$.  Maps have been
primary beam corrected within the region shown, so contours represent
absolute fluxes.  The synthesized beam, shown in the lower left
corner of each panel, has a FWHM of 6\farcs9 $\times$ 5\farcs0.
The LSR velocity of each plane is shown in the upper left.
\label{robmaps}}
\end{center}
\end{figure*}

%%%%%%%%%%%%%%%%%%%%%%%%%%%%%%%%%%%%%%%%%%%%%%%%%%%%%%%%%%%%%
%%%%%%%%%%%%%%%%%%%%%%%%   FIG. 3   %%%%%%%%%%%%%%%%%%%%%%%%%
%%%%%%%%%%%%%%%%%%%%%%%%%%%%%%%%%%%%%%%%%%%%%%%%%%%%%%%%%%%%%

\begin{figure*}
\begin{center}
\includegraphics[width=4in,angle=-90]{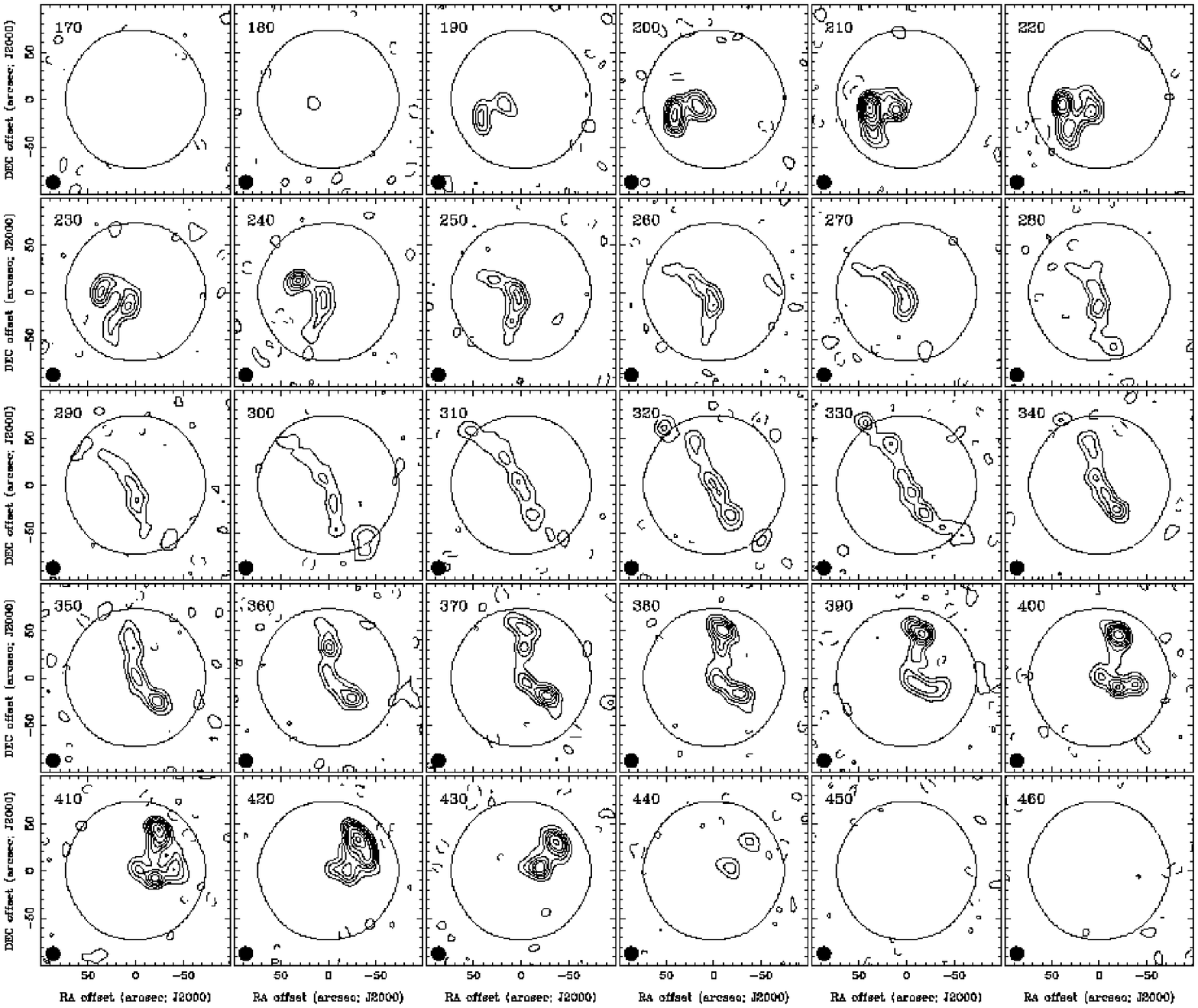}
\figcaption[tprmaps.bit.ps]{
Channel maps of CO emission in NGC 4736 at 15\arcsec\ resolution.
Contours are spaced by 3$\sigma$, where $\sigma$=0.08 Jy bm$^{-1}$.
Maps have been primary beam corrected within the region enclosed by
the circular contour; absolute fluxes are unreliable outside of this 
contour.
\label{tprmaps}}
\end{center}
\end{figure*}

%%%%%%%%%%%%%%%%%%%%%%%%%%%%%%%%%%%%%%%%%%%%%%%%%%%%%%%%%%%%%

In this paper we perform our analysis on the moment-0 maps and velocity
fields generated with the Fourier combination method.  The
corresponding channel maps from the robust and tapered cubes are shown
in Figures \ref{robmaps} and \ref{tprmaps} respectively.  As discussed
above, it would make little difference if we used the linear
combination method or even the BIMA-only maps for our analysis, since
their fluxes agree within the $\sim$20\% calibration uncertainties, not
to mention the potentially much larger uncertainties in the X-factor.
The differences in the velocity fields are similarly quite small ($<$4
\kms) except near the nucleus where the profiles are broad and the mean
velocity is more uncertain.

\subsection{\Halpha\ Data\label{obs_ha}}

We were kindly provided with \Halpha+\NII\ images from a number of
sources, including \citet{Pog89} and \citet{Gon97}.  For this study we
have used the image of \citet{Pog89}, taken with the Lick 1~m telescope
in 1987, which provides a field of view (4\farcm5 $\times$ 4\farcm5)
comparable to our BIMA observations.  Absolute fluxes for this image
are given by \citet{Smi91}, who derived \Halpha\ fluxes of $8.3 \times
10^{-12}$ erg s$^{-1}$ cm$^{-2}$ for the ring
($r$=30\arcsec--60\arcsec) and $1.2 \times 10^{-11}$ erg s$^{-1}$
cm$^{-2}$ for the entire image.  Astrometric coordinates were derived
by comparison with a VLA 1.4 GHz image provided by N. Duric
\citep{Dur88} and are accurate to 2\arcsec.  Of some concern is the
much stronger emission in the nuclear region found by \citet{Gon97}
relative to \citet{Pog89}: the former finds that 15\% of the flux
within $r$=60\arcsec\ lies within $r$=15\arcsec, while the latter finds
only 3.3\%.  The continuum subtraction may be more reliable in Pogge's
image, since the line-free continuum was measured at 6435 \AA\ rather
than 5960 \AA; furthermore, the integrated Fabry-P\'{e}rot image of
\citet{Mul95}, which excludes \NII\ from the LINER, also shows 
very little \Halpha\
near the nucleus.  Nonetheless, given that the effects of \NII\
emission, stellar \Halpha\ absorption, and dust extinction will all be
most severe at the nucleus, caution must be exercised in drawing
conclusions about the SFR in the nuclear region from an \Halpha\ image
(see \S\ref{sfr}).

\subsection{H\thinspace I Data\label{obs_hi}}

A VLA \HI\ datacube for NGC 4736 was provided courtesy of R. Braun
\citep{Brn95}.  The observations were taken in 1989--90 in the B, C,
and D configurations; with uniform weighting, a resolution of
$\sim$6\arcsec\ was achieved.  The cube used here was smoothed to a
resolution of 15\arcsec, resulting in an rms noise in each 5.2
\kms\ channel of about 1.7 mJy bm$^{-1}$ or 4.6 K.  The integrated
intensity and line-of-sight velocity maps were constructed from the
deconvolved datacube in the same way as for the CO data
(\S\ref{obs_bima}).  The total fluxes in the cube and moment-0 maps
were 90 and 57 Jy \kms\ respectively, indicating that the masking
technique missed some of the flux in the cube.  However, in the inner
5\arcmin\ $\times$ 5\arcmin, the region of interest for this study, the
agreement is substantially better (31 Jy \kms\ in the cube and 28 in
the moment-0 map).  Although single-dish data were not included \citep[a
single-dish flux of 94$\pm$5 Jy \kms\ was measured by][]{Hucht85}, the
inclusion of a D-array mosaic at 65\arcsec\ resolution provides
sensitivity to most of the extended flux, especially in the inner
regions of the galaxy where the emission in any one velocity channel is
spatially confined due to the rotation of the galaxy.

%% MORPHOLOGY

\section{Distribution of the Neutral Gas}\label{morph}

%%%%%%%%%%%%%%%%%%%%%%%%%%%%%%%%%%%%%%%%%%%%%%%%%%%%%%%%%%%%%
%%%%%%%%%%%%%%%%%%%%%%%%   FIG. 4   %%%%%%%%%%%%%%%%%%%%%%%%%
%%%%%%%%%%%%%%%%%%%%%%%%%%%%%%%%%%%%%%%%%%%%%%%%%%%%%%%%%%%%%

\vskip 0.25truein
\includegraphics[width=3.25in]{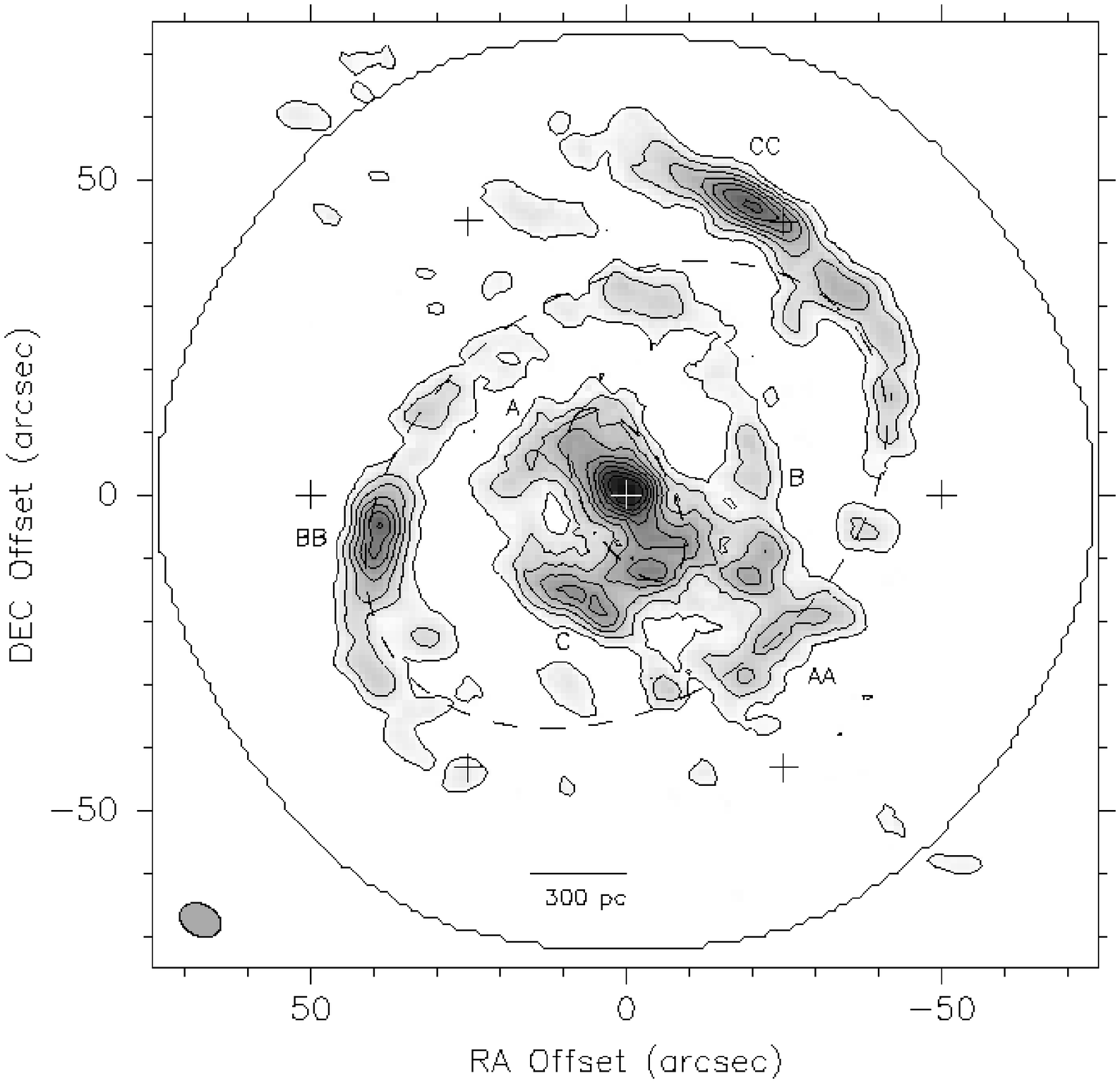}
\figcaption[bimom0.bit.ps]{
Integrated intensity map derived from the robustly weighted CO
datacube.  The contours are spaced by 4.5 Jy bm$^{-1}$ \kms.  The
crosses represent the pointing centers of the mosaic, and the outer
contour is the boundary of the region that has been primary-beam
corrected (corresponding to $\ge 56$\% of the central sensitivity).
Several of the major gas complexes are labeled to highlight the
threefold symmetry of the CO distribution (see text).  Prominent
optical features (the nuclear bar and the \Halpha\ ring) are drawn
as ellipses.
\label{bimom0}}
\vskip 0.25truein

%%%%%%%%%%%%%%%%%%%%%%%%%%%%%%%%%%%%%%%%%%%%%%%%%%%%%%%%%%%%%

\subsection{Integrated Intensity Images\label{morph_mom}}

The moment-0 image for the robustly weighted CO cube is shown in
Figure~\ref{bimom0}.  Our map is generally consistent with previous CO
observations by \citet{Ger91} using the IRAM 30~m and \citet{Sak99}
using the Nobeyama Millimeter Array.  We note that the combination of
high angular resolution and a large field of view in the BIMA map allow
for a much clearer view of CO emission in the ring than has been
previously available.  Moving outward from the center, the CO
morphology is characterized by a strong central peak, an elongated
feature (the ``molecular bar''), arc-like extensions from the ends of
the bar, and tightly wound arms that are associated with the
\Halpha\ ring (outer dashed ellipse).  

{\it The bar.}---That
the apparent ``bar'' is a true distortion in the potential of the
galaxy is not obvious from the CO image, but is confirmed by
near-infrared images \citep[MMG95;][]{Block94a} and the kinematic
analysis presented in \S\ref{kin}.  There is no clear indication of an
angular offset between the stellar bar (inner dashed ellipse) and its
molecular counterpart, although the gas bar may be trailing slightly.
This contrasts with the general expectation that the gas bar will {\it
lead} the stellar bar, based on numerical simulations
\citep{Com85,San80} and a number of previous observations (e.g., NGC
7479, \citealt{Qui95}; M101, \citealt*{Ken91a}).

{\it Arc-like extensions from the bar.}---These
features extend azimuthally from the bar ends and exhibit a
complex morphology.  One interpretation is that a single spiral arm
begins at the northern end of the bar (labeled A in
Figure~\ref{bimom0}), wraps around by 270\arcdeg\ to point B, and
merges into the ring CO component.  Alternatively, the arm beginning at
A may merge with the ring at point AA, and the emission at B could be
associated with a distinct spiral arm beginning at the southern end of
the bar.  Optical/near-infrared color index maps (MMG95;
\citealt{Block94b}), which trace the distribution of cold dust, appear
to show a continuous spiral arm that does not approach the ring until it
has wrapped around to the northern side, thus favoring the first
interpretation.  Note the gap in CO brightness between the bar+spiral
arm region and the ring.

{\it CO in the ring.}---While
the \Halpha\ ring morphology might be described as a pointy oval,
the clumpy CO emission in this region traces a pair of tightly wound
spiral arms, which on the northern side extends well outside the ring.
These arms are continuous with the \HI\ arms seen in the maps presented
by \citet{Brn95}.  Apparent ring features formed by tightly wound
spiral arms, dubbed ``pseudorings,'' are commonly found in galaxies
\citep{dV80}, and support the interpretation that galactic rings are
manifestations of spiral density waves \citep{But96}.

Also notable in Figure~\ref{bimom0} is the lack of twofold symmetry in
the CO brightness.  Whereas the locations at which the CO arms depart
from the ring are approximately symmetric, reflecting the twofold
symmetry of the \HI\ arms which they are continuous with, the locations
of peak emission along these arms do not show this symmetry.  Rather,
there is a noticeable {\it threefold} symmetry in the CO intensity
distribution, with emission peaks labeled A, B, and C in the inner disk
(Figure~\ref{bimom0}) lying opposite to the peaks labeled AA, BB, and
CC in the ring.  We discuss the azimuthal symmetries in the ring
further in \S\ref{sfr_az} and \S\ref{disc_molaz}.

%%%%%%%%%%%%%%%%%%%%%%%%%%%%%%%%%%%%%%%%%%%%%%%%%%%%%%%%%%%%%
%%%%%%%%%%%%%%%%%%%%%%%%   FIG. 5   %%%%%%%%%%%%%%%%%%%%%%%%%
%%%%%%%%%%%%%%%%%%%%%%%%%%%%%%%%%%%%%%%%%%%%%%%%%%%%%%%%%%%%%

\begin{figure*}
\begin{center}
\includegraphics[width=6in]{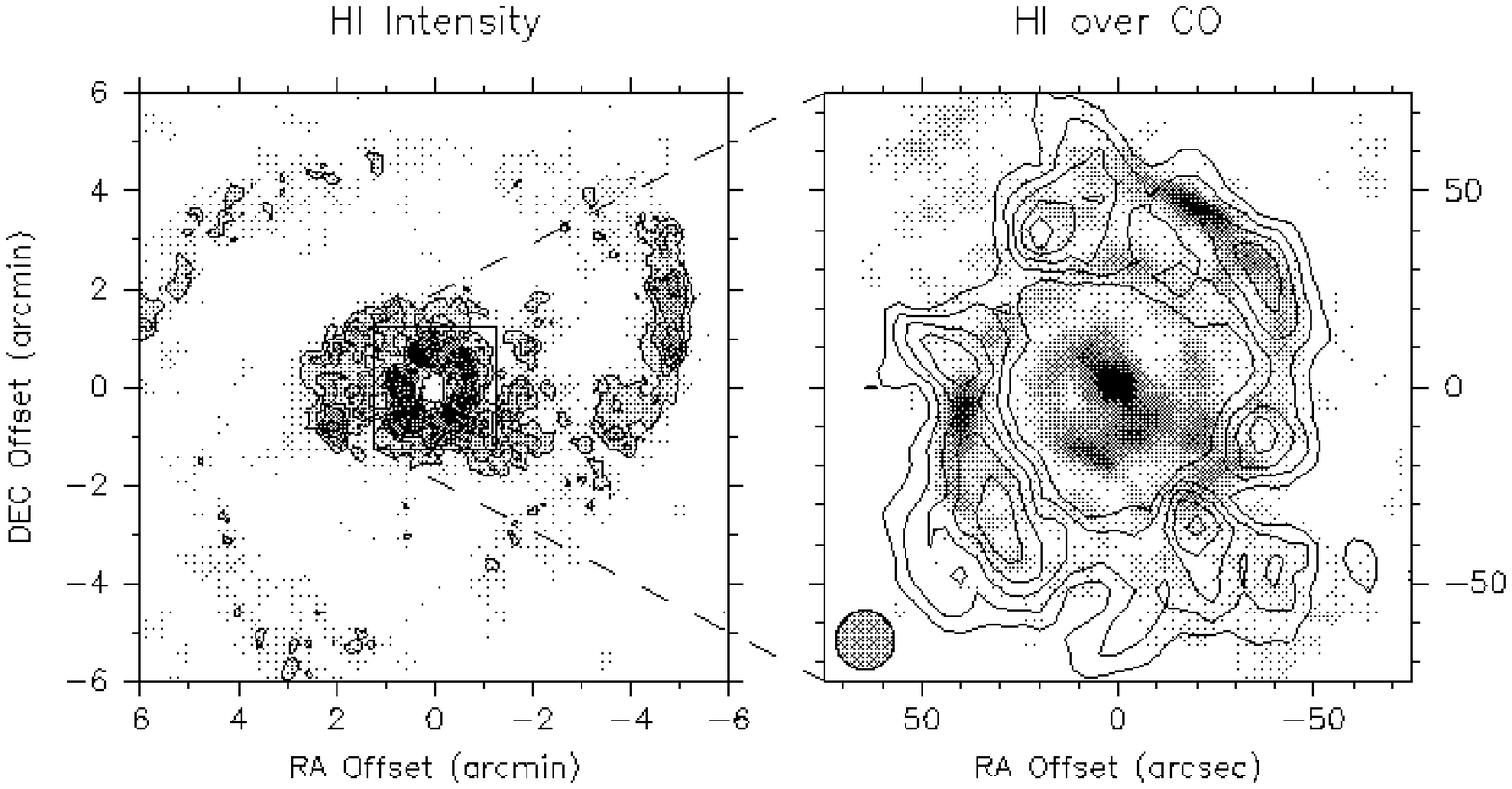}
\figcaption[coh1.bit.ps]{
\HI\ integrated intensity image (contours spaced by 70 mJy bm$^{-1}$ \kms\
starting at 140) overlaid on CO intensity image (greyscale).
The 15\arcsec\ \HI\ beam is shown on the lower left.
\label{coh1}}
\end{center}
\end{figure*}

%%%%%%%%%%%%%%%%%%%%%%%%%%%%%%%%%%%%%%%%%%%%%%%%%%%%%%%%%%%%%

Contours of integrated \HI\ emission are overlaid on a greyscale image
of the CO emission in Figure~\ref{coh1}.  In contrast to the CO, the
\HI\ distribution in the ring is strongly bisymmetric.  Atomic gas is
largely absent interior to the ring, where most of the gas is
molecular, and even within the ring the CO and \HI\ peaks avoid each
other, with the possible exception of the \HI\ peak in the northwest
(upper right of Fig.~\ref{coh1}).  This detailed anticorrelation was
already hinted at by \citet{Ger91} based on lower resolution data and
is unexpected if one assumes that molecular clouds are associated with
regions of excess \HI\ column density.  Rather, it suggests that
the \HI\ peaks may represent regions where molecular gas has been
dissociated by UV radiation and has not yet recombined to form H$_2$,
a possibility that is discussed in \S\ref{disc_at}.

\subsection{Radial Gas Profiles\label{profile}}

The radial profiles of \HI\ and H$_2$ surface density at 15\arcsec\
resolution are plotted in Figure~\ref{radprof} on a log-linear plot.
Also shown is the radial profile of the $K_s$-band (2.2 $\mu$m) light,
wth arbitrary scaling, derived from a Two Micron All Sky Survey (2MASS)
Atlas image smoothed to 15\arcsec\ resolution.  For $r\le60$\arcsec,
the unsmoothed $K_s$ profile from the 2MASS image is consistent with 
the profile derived by MMG95 from a more sensitive 2.2 $\mu$m image.

%%%%%%%%%%%%%%%%%%%%%%%%%%%%%%%%%%%%%%%%%%%%%%%%%%%%%%%%%%%%%
%%%%%%%%%%%%%%%%%%%%%%%%   FIG. 6   %%%%%%%%%%%%%%%%%%%%%%%%%
%%%%%%%%%%%%%%%%%%%%%%%%%%%%%%%%%%%%%%%%%%%%%%%%%%%%%%%%%%%%%

\vskip 0.25truein
\includegraphics[width=3.25in]{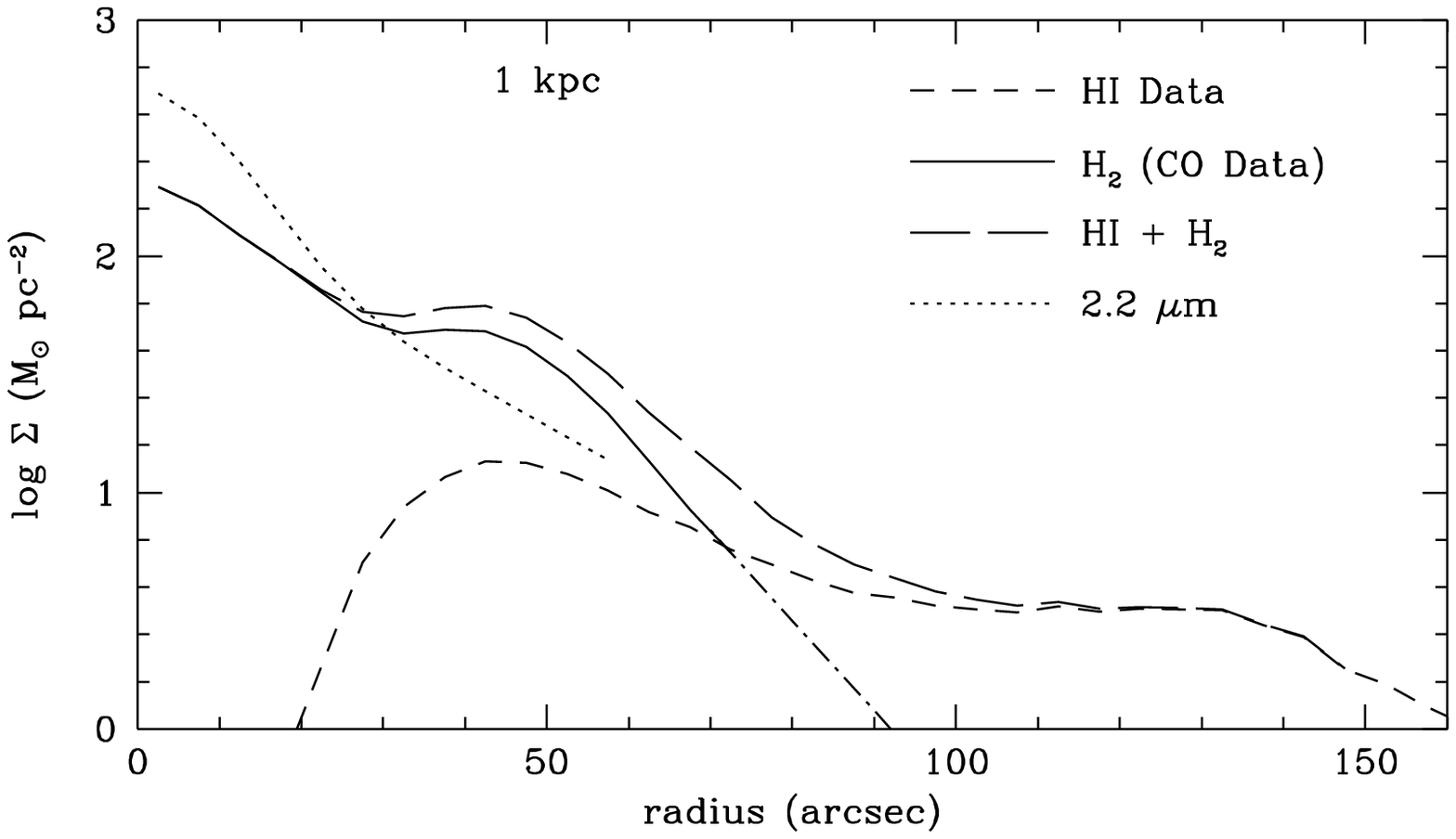}
\figcaption[radprof.eps]{
Azimuthally averaged radial gas profiles, corrected for inclination
and including helium, on a logarithmic scale.  The stellar light
profile at 2.2 $\mu$m is shown as a dotted line for comparison, with
an arbitrary vertical offset.  The CO profile has been extrapolated
beyond 70\arcsec\ (dash-dot line).  All images are at 15\arcsec\
resolution, hence the flattening near $r$=0\arcsec.
\label{radprof}}
\vskip 0.25truein

%%%%%%%%%%%%%%%%%%%%%%%%%%%%%%%%%%%%%%%%%%%%%%%%%%%%%%%%%%%%%
%%%%%%%%%%%%%%%%%%%%%%%%   FIG. 7   %%%%%%%%%%%%%%%%%%%%%%%%%
%%%%%%%%%%%%%%%%%%%%%%%%%%%%%%%%%%%%%%%%%%%%%%%%%%%%%%%%%%%%%

\begin{figure*}[t]
\begin{center}
\includegraphics[width=6in]{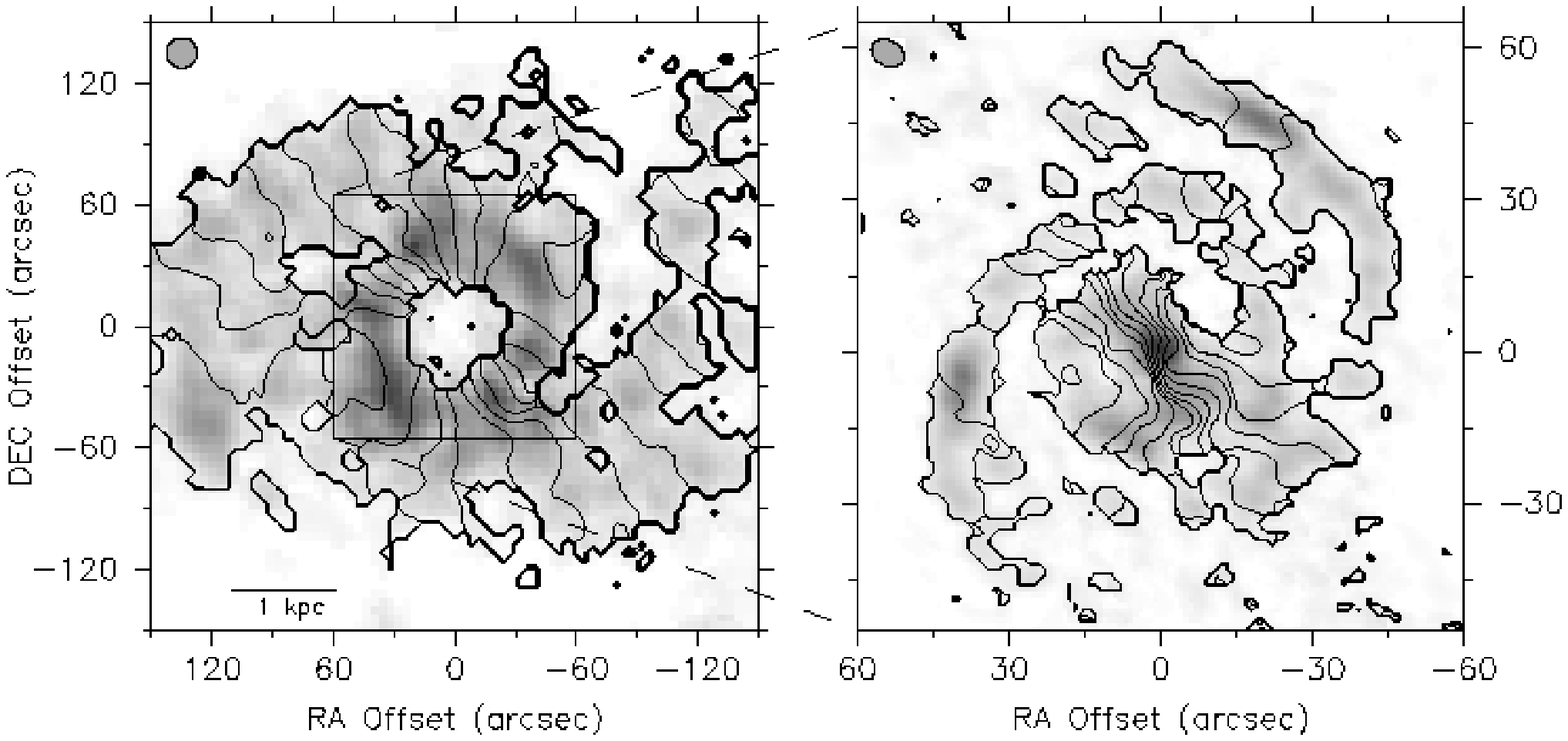}
\figcaption[mom1.bit.ps]{
\HI\ ({\it left}) and CO ({\it right}) line-of-sight velocity maps.
The contours are spaced by
20 \kms\ and range from 170 \kms\ (east side) to 450 \kms\ (west side).
\label{mom1}}
\end{center}
\end{figure*}

%%%%%%%%%%%%%%%%%%%%%%%%%%%%%%%%%%%%%%%%%%%%%%%%%%%%%%%%%%%%%

In converting from CO brightness to H$_2$ column density we have adopted
an X-factor of:
\[N_{\rm H_2} = 2 \times 10^{20} 
	\left(\frac{I_{\rm CO}}{\rm K\;\kms}\right) \rm cm^{-2}\;.\]
This is consistent with the value of $(1.9\pm 0.2) \times 10^{20}$ 
cm$^{-2}$ (K \kms)$^{-1}$ estimated by \citet{Str96} from diffuse 
Galactic $\gamma$-ray emission.  The corresponding mass surface density is
\begin{eqnarray*}
\Sigma_{\rm H_2} & = & 2\, N_{\rm H_2}\,m_H\,\cos i \\ & = & 330
	\left(\frac{I_{\rm CO}}{\rm Jy\;\kms\,arcsec^{-2}}\right) \cos
	i \;\;\; \Msol\; \rm pc^{-2},
\end{eqnarray*}
where the conversion of 1~K=$9.6 \times 10^{-3}$ Jy~arcsec$^{-2}$ has
been applied for $\lambda$=2.6 mm.  The analogous expressions for \HI\
in the optically thin approximation are:
\begin{eqnarray*}
N_{\rm HI} & = & 1.82 \times 10^{18} \left(\frac{I_{21}}{\rm
	K\;\kms}\right) \rm cm^{-2}\;, \\ 
\Sigma_{\rm HI} & = & N_{\rm
	HI}\,m_H\,\cos i \\ & = & 1.0 \times 10^4 \left(\frac{I_{21}}{\rm
	Jy\;\kms\,arcsec^{-2}}\right) \cos i \;\;\; \Msol\; \rm
	pc^{-2},
\end{eqnarray*}
using the conversion 1~K=$1.5 \times 10^{-6}$ Jy~arcsec$^{-2}$.  For
both \HI\ and CO profiles, the galaxy has been assumed to have an
inclination $i$=35\arcdeg\ (MMG95) and a position angle $\phi$=295\arcdeg\
(\S\ref{kin_rot}).

The total gas surface density (corrected for helium) is then given by:
\[\Sigma_{\rm gas} = 1.36 (\Sigma_{\rm HI} + \Sigma_{\rm H_2})\,,\] 
a quantity which is independent of the adopted distance.  Note that the
\HI~+~H$_2$ profile in Figure~\ref{radprof} includes an extrapolation
of the CO profile by an exponential with scalelength 11\farcs5
(dot-dashed line), in order to avoid a spurious kink in the \HI~+~H$_2$
profile resulting from the fact that the CO data are less sensitive
than the \HI\ data to the same gas surface density.  (The CO
observations at 15\arcsec\ resolution detect, at the 3$\sigma$ level in
each of two consecutive channels, surface densities down to $\sim$5
\Msol\ pc$^{-2}$ when corrected to face-on, a factor of 3 higher than
the \HI\ limit.)  The extrapolated H$_2$ mass is $1.1 \times 10^7$
\Msol, less than 10\% of the $1.7 \times 10^8$ \Msol\ of H$_2$
(excluding helium) measured within $r$=70\arcsec.  When convolved to
the appropriate resolution, our radial profiles agree well with
previous observations of \HI\ by \citet{Mul93} using the WSRT, and of
CO by \citet{Ger91} using the IRAM 30~m.

The assumption that \HI\ is optically thin is widely made in
extragalactic astronomy, although \citet{Brn97} has argued that
emission from dense, cool clouds in the inner disks of galaxies are
likely to be optically thick, leading to an underestimate of the column
density by factors of $\sim$2 \citep{Walt96}.  In his high-resolution
datacube for NGC 4736, \citet{Brn95} finds maximum emission brightness
temperatures of $\sim$100 K, comparable to the thermal temperatures of
cool \HI\ clouds.  Our adopted radial profile should therefore be
considered a {\it lower limit} to the true \HI\ surface density,
although the correction is unlikely to be large since at 15\arcsec\
($\sim$300 pc) resolution
we are averaging over both cold and warm clouds.  The adoption of a
``standard'' Galactic value for the X-factor is considerably more
uncertain, and has been shown to underestimate the H$_2$ mass in the
Magellanic Clouds and other low-metallicity galaxies \citep{Wils95}
while possibly overestimating the mass near the Galactic Center by up to an
order of magnitude \citep{Sod95,Dah98}.  Since our observations of NGC
4736 cover the central 1--2 kpc where the metallicity is at or above
the solar value \citep{Oey93} and physical conditions may be similar to
those in the Galactic Center, we expect that the H$_2$ surface density
we derive to be an {\it upper limit} to the true value.

With these assumptions in mind, a comparison of the \HI\ and CO
profiles indicates that most ($\sim$75\%) of the gas at the location of
the ring ($r$=45\arcsec) is molecular.  An exponential fit to the total
gas profile gives a scalelength of $\sim$30\arcsec; however, the fit is
rather poor due to the presence of the ring at 45\arcsec\ and a
``plateau'' from 100\arcsec--130\arcsec.  These features suggest that
gas may have been redistributed across the disk, as discussed in the
following section.  Comparison of the gas profiles to the $K_s$-band
profile, which traces the stellar surface density, shows that for
$r<50$\arcsec\ the stellar profile is considerably steeper than the gas
profile.  This is partly due to a secondary enhancement at the location
of the ring which is much more prominent in the gas than in the stars.
Even interior to the ring ($r<30$\arcsec), however, the stellar profile
is steeper, presumably due to a bulge component.

%% KINEMATICS

\section{Gas Kinematics}\label{kin}

\subsection{Rotation Curve\label{kin_rot}}

The CO and \HI\ velocity fields derived from the Gaussian fits are
shown in Figure~\ref{mom1}.  The dominant signature of circular
rotation is apparent, although there are clearly non-circular motions
along the kinematic minor axis (systemic velocity contour), which runs
almost north-south.  Rotation curves were fit to the velocity fields
using a tilted-ring model \citep{Beg89} as implemented in the program
ROTCUR within the GIPSY software package \citep{vdH92}.  In this model,
the line-of-sight\footnote{Throughout this paper we refer to measured
Doppler velocities as {\it line-of-sight} velocities, since {\it
radial} velocities could be interpreted as being in the plane of the
galaxy.  We also use $V$ to refer to Doppler velocities and $v$ to
refer to velocities in the plane of the galaxy.} velocity at each point
is decomposed as
\begin{equation}
V(x,y) = V_{\rm sys} + v_{\rm rot}(r) \sin i \cos \theta
\end{equation}
where $V_{\rm sys}$ is the systemic velocity of the galaxy, $v_{\rm
rot}(r)$ is the rotation velocity of a ring at radius $r$ from the
center, $i$ is the inclination of the ring from face-on, and $\theta$
is the azimuthal angle in the plane of the galaxy, which depends on
the location of the point in relation to the rotation center as well
as the position angle in the sky of the receding major axis ($\phi$)
and the inclination ($i$).  We refer to $V_{\rm sys}$, $\phi$, $i$, and
the rotation center $(x_0,y_0)$ as the {\it disk parameters} because
for a circularly rotating disk they are fixed at all radii.

To determine the disk parameters, the velocity field of the tapered CO
cube was fit in annuli from 20\arcsec--60\arcsec\ in intervals of
5\arcsec.  Initial estimates for the rotation velocity, inclination,
and position angle were taken from MMG95 and used to determine the
rotation center and systemic velocity.  Then the rotation velocity,
center, and systemic velocity were fixed to determine the inclination
and position angle, and the process was repeated to improve the
estimates.  Due to the low inclination of the galaxy, $v_{\rm rot}$ and
$i$ are tightly coupled and could not be determined independently; we
therefore fixed the inclination to a value of 35\arcdeg\ as determined
by MMG95 from surface photometry.  The systemic velocity was found to
be 315$\pm$4 \kms\ and the rotation center was consistent with a
location at the position of the radio continuum nucleus (Table
\ref{proptable}).  The remaining disk parameter, $\phi_{kin}$ (also
referred to as the kinematic position angle or PA), assumed an average
value of 295\arcdeg, but showed a noticeable variation with radius,
rising from 290\arcdeg\ to 300\arcdeg\ between $r=25\arcsec$ and
60\arcsec.

Fixing the disk parameters to their average values, the remaining free
parameter, $v_{\rm rot}$, was determined by fitting to the velocity
field of the robust CO cube for $r\le 25$\arcsec\ (where beam smearing
in the tapered cube is significant) and the tapered CO cube from
$r$=30\arcsec--55\arcsec.  A $\cos\theta$ weighting was applied to emphasize
points close to the major axis, and points within 30\arcdeg\ of the
minor axis were excluded from the fit.  Fitting was performed for the
approaching and receding halves of the galaxy separately as well as
for the entire galaxy.  No statistically significant differences were
found between the rotation curves of the two halves.  

Using the same disk parameters, $v_{\rm rot}(r)$ was then determined
from the \HI\ velocity field.  Beyond $r$=80\arcsec\ the
rotation curve was derived from the \HI\ data after smoothing to
45\arcsec\ resolution, in order to improve the signal-to-noise ratio.
Figure~\ref{rotcur} shows the adopted rotation curve, which is a
composite of the CO and \HI\ data at various resolutions assuming the
same disk inclination and PA.  Outside the CO disk ($r>60\arcsec$), the
derived rotation curves for the approaching and receding sides differ
by as much as 25 \kms, yet these differences are still within the
errors at these radii.  The large error bars in the outer disk probably
reflect a combination of decreasing azimuthal coverage and azimuthal
streaming motions near the major axis.

%%%%%%%%%%%%%%%%%%%%%%%%%%%%%%%%%%%%%%%%%%%%%%%%%%%%%%%%%%%%%
%%%%%%%%%%%%%%%%%%%%%%%%   FIG. 8   %%%%%%%%%%%%%%%%%%%%%%%%%
%%%%%%%%%%%%%%%%%%%%%%%%%%%%%%%%%%%%%%%%%%%%%%%%%%%%%%%%%%%%%

\vskip 0.25truein
\includegraphics[width=3.25in]{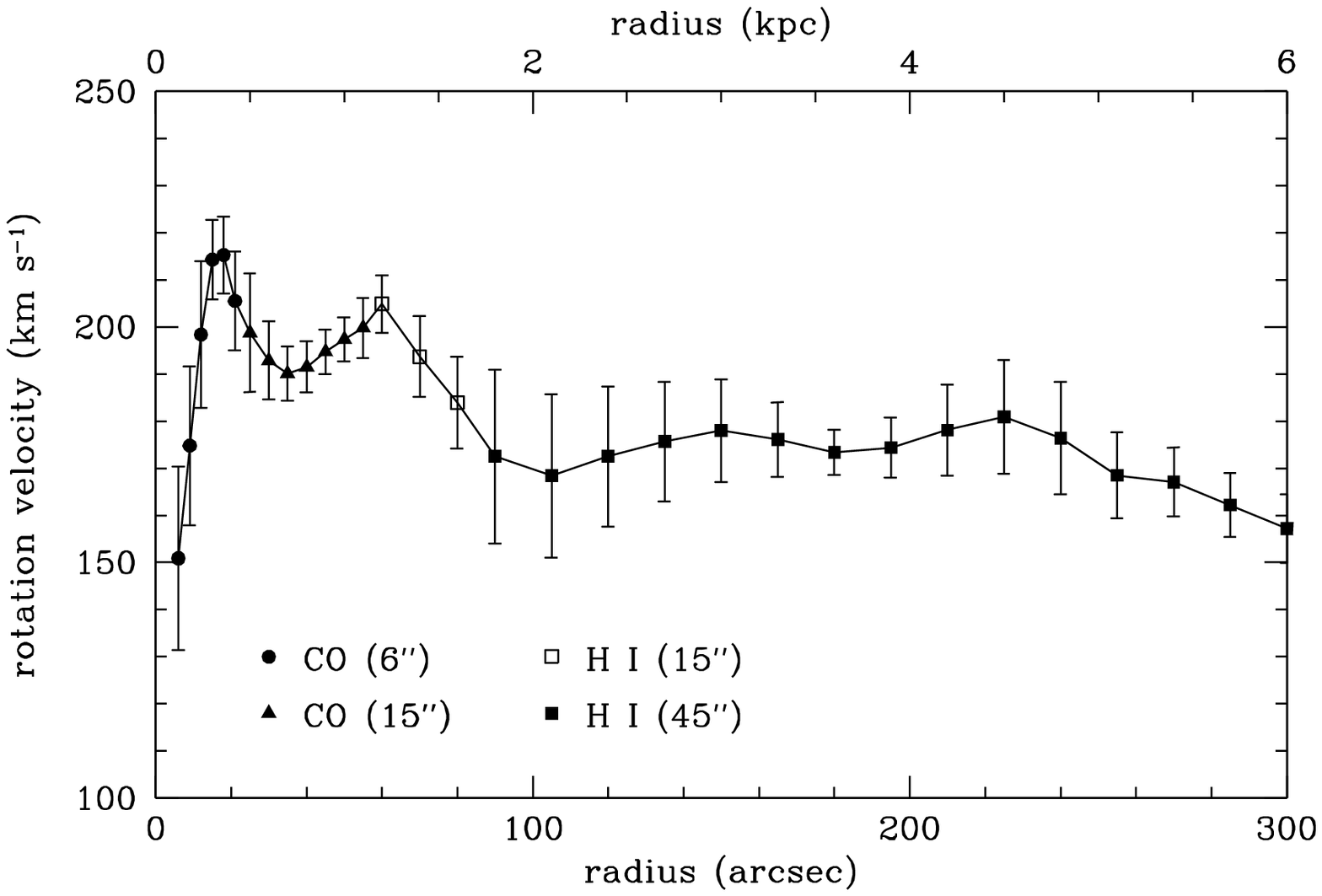}
\figcaption[rotcur.eps]{
Adopted rotation curve for NGC 4736, derived from the 6\arcsec\ CO datacube
for $r<30\arcsec$, the 15\arcsec\ CO datacube for $30\arcsec\le r < 60\arcsec$,
the 15\arcsec\ \HI\ datacube for $60\arcsec\le r \le 80\arcsec$, and
the 45\arcsec\ \HI\ datacube for $r>80\arcsec$.  An inclination of
35\arcdeg\ has been taken into account.
\label{rotcur}}
\vskip 0.25truein

%%%%%%%%%%%%%%%%%%%%%%%%%%%%%%%%%%%%%%%%%%%%%%%%%%%%%%%%%%%%%

The adopted rotation curve is characterized by a number of distinct
humps and dips.  The curve begins with a steep rise in the inner
15\arcsec, with no correction having been applied for the effects of
beam smearing or the influence of the nuclear bar.  While the former
will tend to soften the rise, the latter will tend to increase it,
since the kinematic major axis is nearly orthogonal to the bar major
axis.  It is therefore likely that the first peak in $v_{\rm rot}$ at
$r \approx 20\arcsec$ is due to gas moving in elliptical orbits aligned
with the bar (\S\ref{kin_ell}).  A secondary peak in $v_{\rm rot}$ at
$r \approx 60\arcsec$, apparent in both the CO and \HI\ datasets, also
occurs in a region of strong streaming motions, especially on the
southern side of the galaxy (\S\ref{kin_pv}).  On the other hand, the
rotation curve is already rising across the ring
($r$=40\arcsec--50\arcsec), where the gas motions are predominantly
circular (see \S\ref{kin_res}), suggesting that this second peak
reflects a genuine mass concentration in the ring.  This additional
mass must be provided by stars, since the gas mass alone is only
$\sim$2\% of the dynamical mass at the radius of the ring.  Consistent
with this picture, MMG95 noted a ``shelf'' feature in their $K$ and
$I$-band light profiles at $r$=40\arcsec, although it is hard to
identify such a feature in the overall $K$ profile (Figure~\ref{radprof}).

%%%%%%%%%%%%%%%%%%%%%%%%%%%%%%%%%%%%%%%%%%%%%%%%%%%%%%%%%%%%%
%%%%%%%%%%%%%%%%%%%%%%%%   FIG. 9   %%%%%%%%%%%%%%%%%%%%%%%%%
%%%%%%%%%%%%%%%%%%%%%%%%%%%%%%%%%%%%%%%%%%%%%%%%%%%%%%%%%%%%%

\vskip 0.25truein
\includegraphics[width=3.25in]{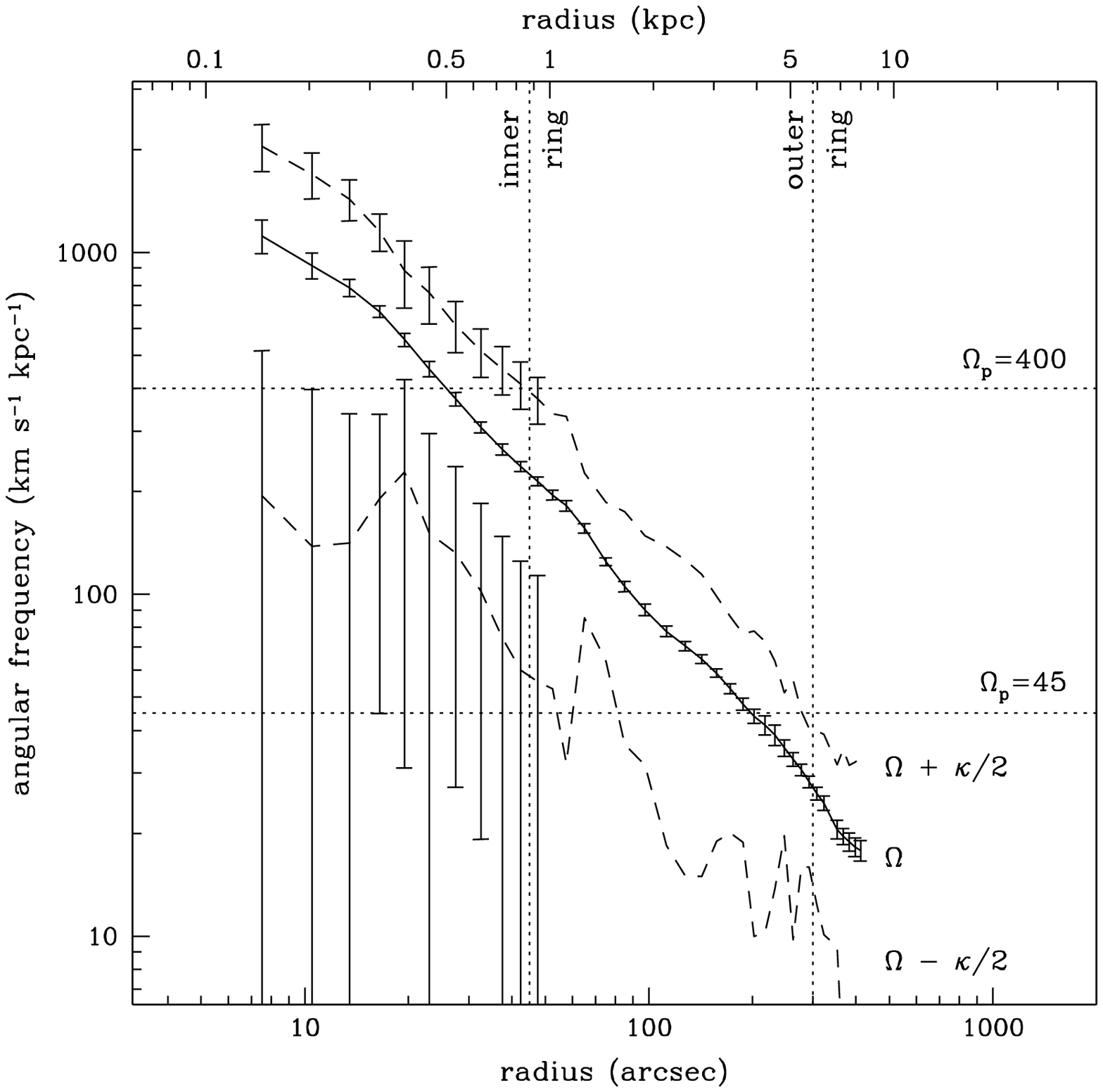}
\figcaption[reso.eps]{
Resonance diagram showing radial variation of $\Omega$ and $\Omega \pm
\kappa/2$ derived from the adopted rotation curve.  The locations of
the inner and outer rings are shown, as well as the assumed pattern
speeds for the nuclear bar (400) and outer oval (45).  The formal
errors on $\Omega \pm \kappa/2$ become too large beyond 1 kpc to plot.
\label{reso}}
\vskip 0.25in

%%%%%%%%%%%%%%%%%%%%%%%%%%%%%%%%%%%%%%%%%%%%%%%%%%%%%%%%%%%%%

%%%%%%%%%%%%%%%%%%%%%%%%%%%%%%%%%%%%%%%%%%%%%%%%%%%%%%%%%%%%%
%%%%%%%%%%%%%%%%%%%%%%%%   FIG. 10  %%%%%%%%%%%%%%%%%%%%%%%%%
%%%%%%%%%%%%%%%%%%%%%%%%%%%%%%%%%%%%%%%%%%%%%%%%%%%%%%%%%%%%%

\begin{figure*}
\begin{center}
\includegraphics[width=6in]{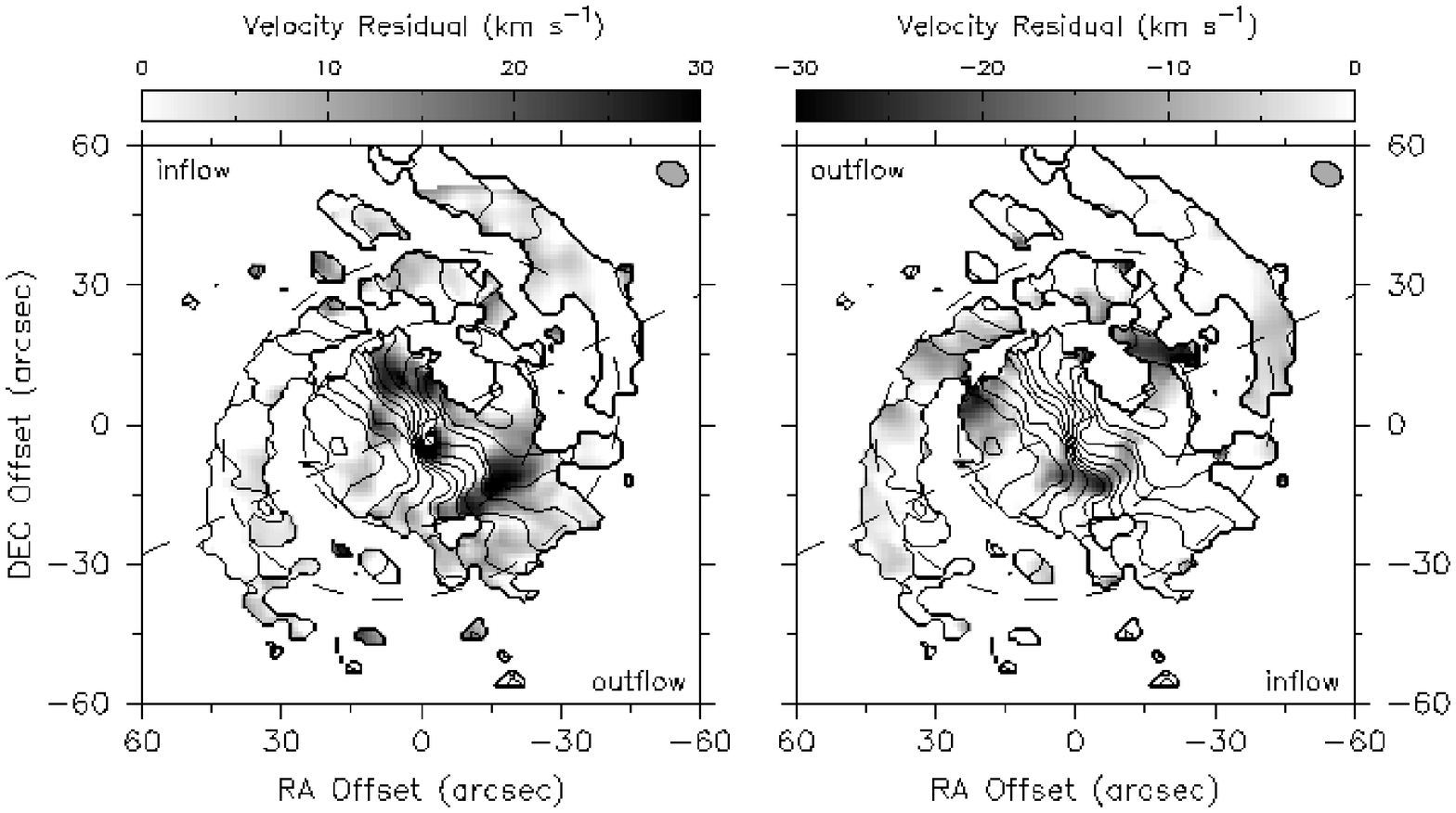}
\figcaption[cores.bit.ps]{
Velocity residuals $V_r(obs)-V_r(model)$, for the CO velocity field (shown in
contours).  Positive residuals are shown on the left and negative on the
right; their interpretation as inflow or outflow are shown in the plot
corners.  The assumed line of nodes (galaxy major axis) is shown as
a dashed line, and the dashed ellipses at 26\arcsec\ and 44\arcsec\ are
the CR and OLR of the kinematic model discussed in the text.
\label{cores}}

%%%%%%%%%%%%%%%%%%%%%%%%%%%%%%%%%%%%%%%%%%%%%%%%%%%%%%%%%%%%%
%%%%%%%%%%%%%%%%%%%%%%%%   FIG. 11  %%%%%%%%%%%%%%%%%%%%%%%%%
%%%%%%%%%%%%%%%%%%%%%%%%%%%%%%%%%%%%%%%%%%%%%%%%%%%%%%%%%%%%%

\vskip 1truein
\includegraphics[width=6in]{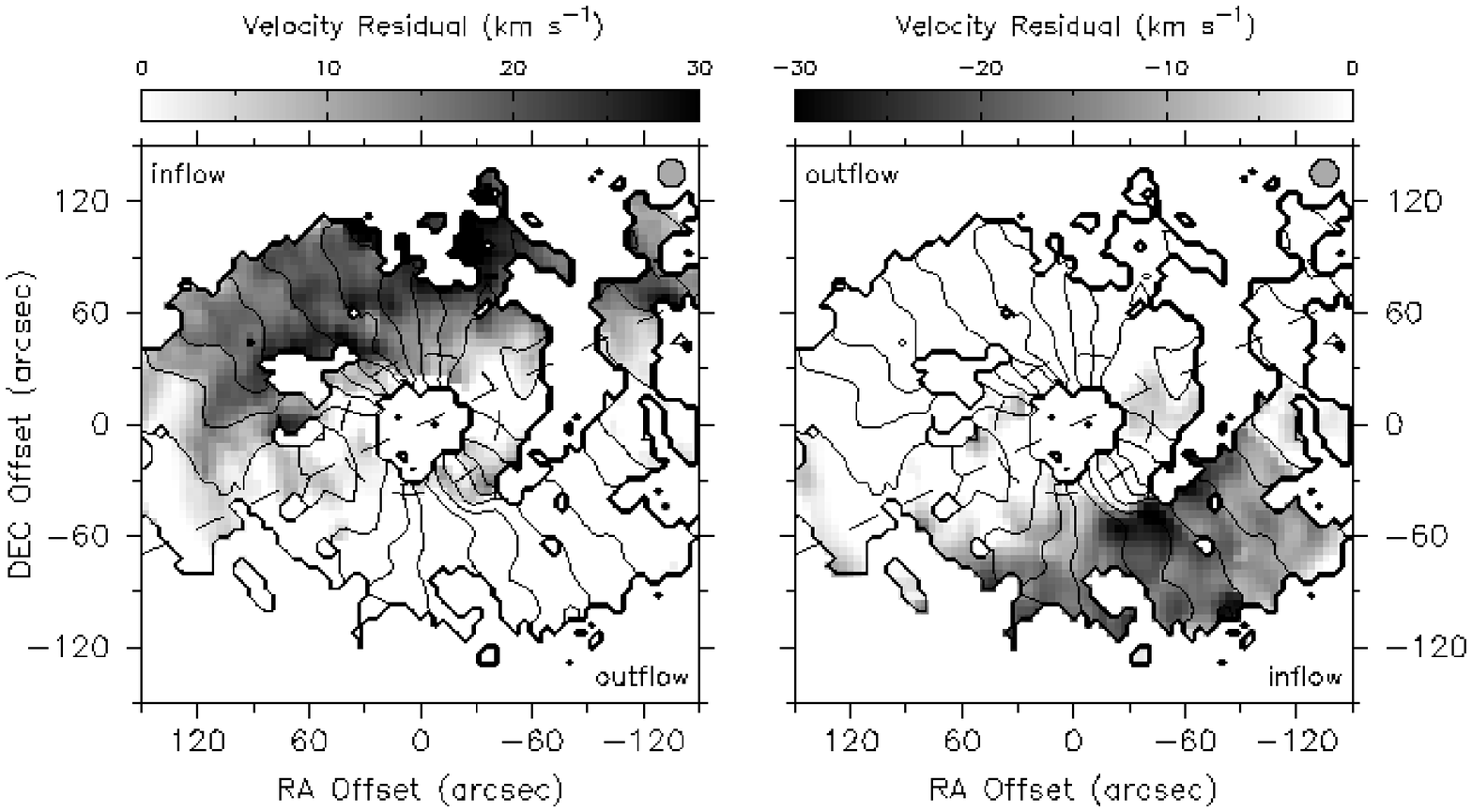}
\figcaption[h1res.bit.ps]{
Velocity residuals $V_r(obs)-V_r(model)$, for the \HI\ velocity field (shown in
contours).  Positive residuals are shown on the left and negative on the
right; their interpretation as inflow or outflow are shown in the plot
corners.  The assumed line of nodes is shown as
a dashed line, and a dashed ellipse at $r$=44\arcsec\ is shown for
comparison with Figure~\ref{cores}.
\label{h1res}}
\end{center}
\end{figure*}

%%%%%%%%%%%%%%%%%%%%%%%%%%%%%%%%%%%%%%%%%%%%%%%%%%%%%%%%%%%%%

Figure~\ref{reso} is a ``resonance diagram'' showing $\Omega(r) =
v_{\rm rot}/r$, $\Omega-\kappa/2$, and $\Omega+\kappa/2$ as a function
of radius.  Here $\kappa$ is the epicyclic frequency, defined as
\[\kappa = \sqrt{\frac{2v}{r}\left(\frac{dv}{dr}+\frac{v}{r}\right)}\;.\]
The larger error bars in the \HI\ rotation curve resulted in a poorer
determination of $\Omega \pm \kappa/2$ at large radii, both because
$\Omega$ is smaller and because $\kappa$ depends on the slope of the
rotation curve.  The data are clearly consistent with the proposal by
\citet{Ger91} that the inner and outer rings (vertical dotted lines)
correspond to the ILR ($\Omega_p=\Omega-\kappa/2$) and OLR
($\Omega_p=\Omega+\kappa/2$) of an oval distortion with pattern speed
$\Omega_p=45$ \kms\ kpc$^{-1}$, but the uncertainties are large.
Further kinematic evidence in support of this interpretation is given
in \S\ref{kin_inflow}.  At smaller radii, the CO data confirm that if
the inner ring corresponds to the OLR of the nuclear bar, as suggested
by MMG95, then the CR is located at $r\approx 26$\arcsec, just beyond
the end of the nuclear bar (the bar ends at $r\approx 20$\arcsec\ if
deprojected).  A corresponding pattern speed of 400 \kms\ kpc$^{-1}$
is plotted, equivalent to the pattern speed of 290 \kms\ kpc$^{-1}$
derived by MMG95 if scaled to their adopted distance (6.6 Mpc).  We
discuss the justification for these pattern speeds in
\S\ref{disc_bar}.

\subsection{Velocity Residuals\label{kin_res}}

The adopted rotation curve was used to generate a model velocity field,
in which the position angle of the line of nodes (defined as the line
formed by the intersection of the galaxy plane with the ``plane'' of
the sky) was fixed at 295\arcdeg, the average value of $\phi_{kin}$ for
the inner disk.  The model field was then subtracted from the observed
velocity fields to generate maps of residual velocities.  Assuming the
spiral arms are trailing, we infer that the near side of the galaxy is
the northern side.  Thus, if interpreted as motions in the galaxy
plane, positive velocity residuals on the northern half of the galaxy's
minor axis indicate a {\it radially inward} velocity component (i.e. local
inflow), while they correspond to local outflow on the southern half.
The CO and \HI\ residual velocity maps are shown in Figures \ref{cores}
and \ref{h1res} respectively.  The uncertainty in the mean
line-of-sight velocity from the Gaussian fits is typically less than 5
\kms, increasing to $\sim$10 \kms\ in regions with poor signal-to-noise
or in the circumnuclear region where line profiles are broad.

The CO velocity residuals in the vicinity of the molecular bar
($r<40\arcsec$) reveal strong non-circular motions, regardless of the
choice of PA.  As expected, the residuals are largest near the minor
axis (25--30 \kms), since points near the major axis are weighted more
heavily in determining the rotation curve.  The sign of the residuals
corresponds to local inflow near the ends of the nuclear bar and
outflow between the southern end of the bar and the ring.  However,
these large flow velocities (40--50 \kms\ in the plane of the galaxy)
need not correspond to {\it net} radial flows if we are viewing
elliptical orbits in projection.  This indeed appears to be the case,
as discussed in \S\ref{kin_ell}.

The residuals are generally $<$10 \kms\ in the vicinity of the ring
($r$=40\arcsec--50\arcsec), where both the \HI\ and CO velocity fields
are well-described by pure circular rotation.  This is somewhat
surprising given that \citet{vdK76} and \citet{But88} found clear
evidence for an asymmetry in the \Halpha\ velocity field at the ring,
with ``a steep linear rise in the velocity across the ring on the
southeast side'' \citep{But88}.  The ionized gas may possess
small-scale velocity structure due to recent star formation to
which the neutral gas is less sensitive.  Furthermore, the interpretation by
\citet{vdK76} of the ring as an expanding feature was based on an
assumed line-of-nodes position angle of 302\arcdeg.  In this paper we
have adopted a PA of 295\arcdeg, which eliminates the outflow
signature at the ring at the expense of producing an inflow signature
outside the ring.

Indeed, as shown in Figure~\ref{h1res}, the \HI\ velocity residuals
become large ($\gtrsim$20 \kms) outside the ring, showing a
clear dipole pattern indicative of inward motion on both sides
of the major axis.  While this pattern is characteristic of radial
inflow \citep{vdK78}, it can also be interpreted as an error in
$\phi_{kin}$.  In Figure~\ref{pavar} we show the results of allowing
the PA to vary with radius in the tilted-ring fit.  An increase in
$\phi_{kin}$ by about 15\arcdeg\ between $r$=50\arcsec\ and
100\arcsec\ is sufficient to remove most of the dipole signature and
leave only isolated residuals, typically $<$10 \kms.  Note that the
isophotal PA also exhibits a variation with radius, but in the {\it
opposite} direction, a point which we return to in \S\ref{kin_warp} and
\S\ref{kin_ell}.

%%%%%%%%%%%%%%%%%%%%%%%%%%%%%%%%%%%%%%%%%%%%%%%%%%%%%%%%%%%%%
%%%%%%%%%%%%%%%%%%%%%%%%   FIG. 12  %%%%%%%%%%%%%%%%%%%%%%%%%
%%%%%%%%%%%%%%%%%%%%%%%%%%%%%%%%%%%%%%%%%%%%%%%%%%%%%%%%%%%%%

\vskip 0.4truein
\includegraphics[width=3.25in]{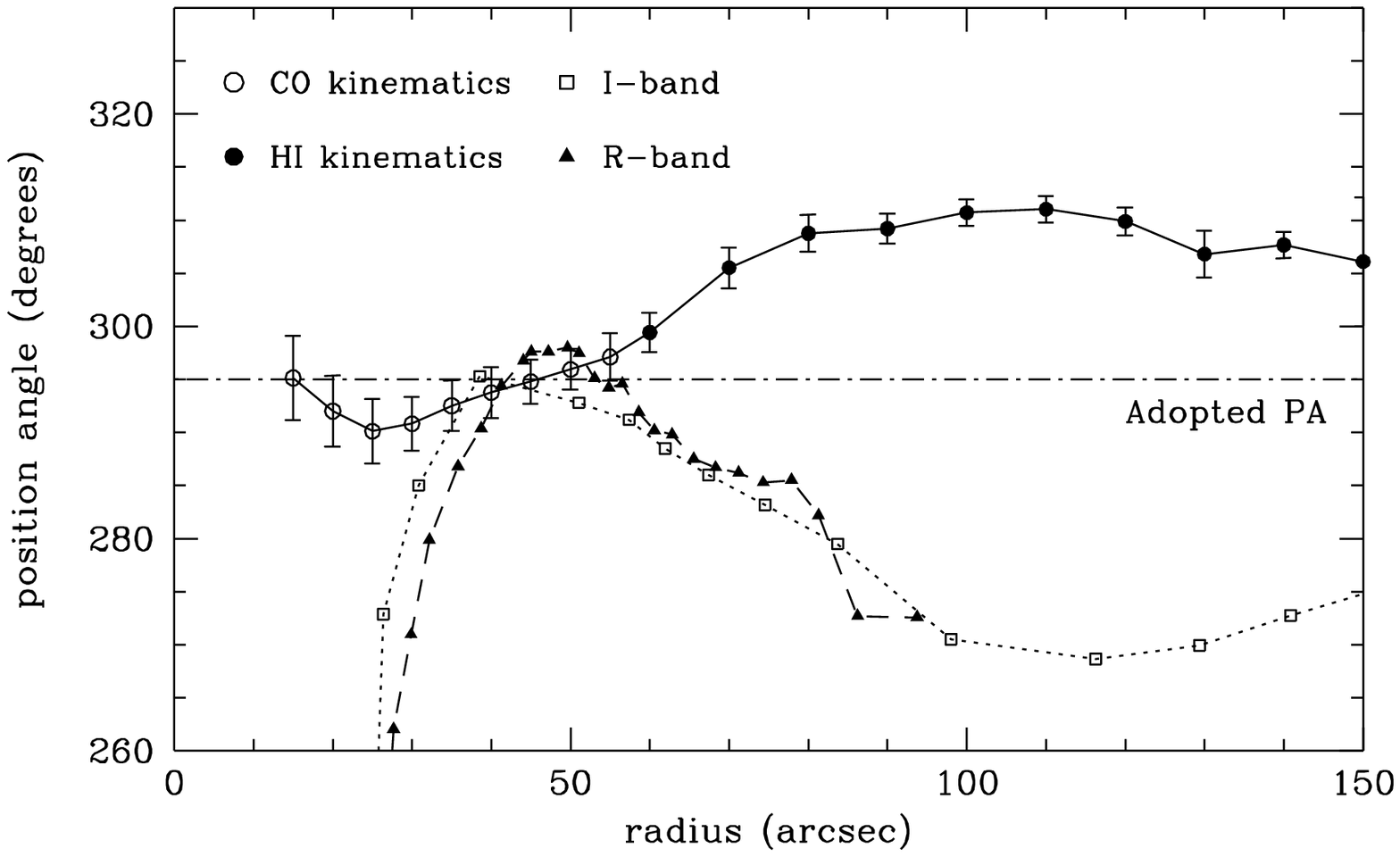}
\figcaption[pavar.eps]{
Major axis position angle (defined from 180\arcdeg--360\arcdeg) as a
function of radius, derived from isophotal data as well as from CO and \HI\
kinematics.  The $I$-band data is from MMG95 and the $R$-band data is from
\citet{Mul95}.
\label{pavar}}
\vskip 0.25truein

%%%%%%%%%%%%%%%%%%%%%%%%%%%%%%%%%%%%%%%%%%%%%%%%%%%%%%%%%%%%%

\subsection{Position-Velocity Diagrams\label{kin_pv}}

Since two-dimensional velocity fields are insensitive to the presence of
multiple velocity components or changes in spectral line profiles, we
constructed position-velocity (p-v) maps for slices at various angles
through the CO and \HI\ datacubes.  Figure~\ref{barpv} shows a slice
through the CO cube along the nuclear bar (PA=28\arcdeg).  This slice
is close to the kinematic minor axis (PA=25\arcdeg), so the predicted
velocities from the fitted circular rotation model are close to $V_{\rm
sys}$=315 \kms, as shown by the heavy solid line.  The actual CO
emission, however, departs strongly from this model.  Part of the
discrepancy is due to beam smearing of the galaxy's rotation, which is
especially severe in the central regions of the galaxy where the
velocity gradient is large.  This will increase the velocity extent of
the emission in the p-v diagram, and contributes to the large
linewidths in the region from $-5$\arcsec\ to $+5$\arcsec\ (delimited
by the dashed lines).  In addition, however, there are regions where
the CO emission is offset in velocity by $\gtrsim$20 \kms\ from the
circular rotation model.  These include the regions labeled X in
Figure~\ref{barpv}, which occur at the ends of the bar where the local
gas velocities are {\it inward}, as indicated by the negative velocity
offset on the southwest side and the positive offset on the
northeast side (cf.\ Figure~\ref{cores}).  Similarly, the region
labeled Y represents local {\it outward} velocities between the bar and
the ring.  The presence of these features in the p-v diagram confirms
our earlier deductions from the velocity residual maps (\S\ref{kin_res}).

%%%%%%%%%%%%%%%%%%%%%%%%%%%%%%%%%%%%%%%%%%%%%%%%%%%%%%%%%%%%%
%%%%%%%%%%%%%%%%%%%%%%%%   FIG. 13  %%%%%%%%%%%%%%%%%%%%%%%%%
%%%%%%%%%%%%%%%%%%%%%%%%%%%%%%%%%%%%%%%%%%%%%%%%%%%%%%%%%%%%%

\vskip 0.25truein
\includegraphics[width=3.25in]{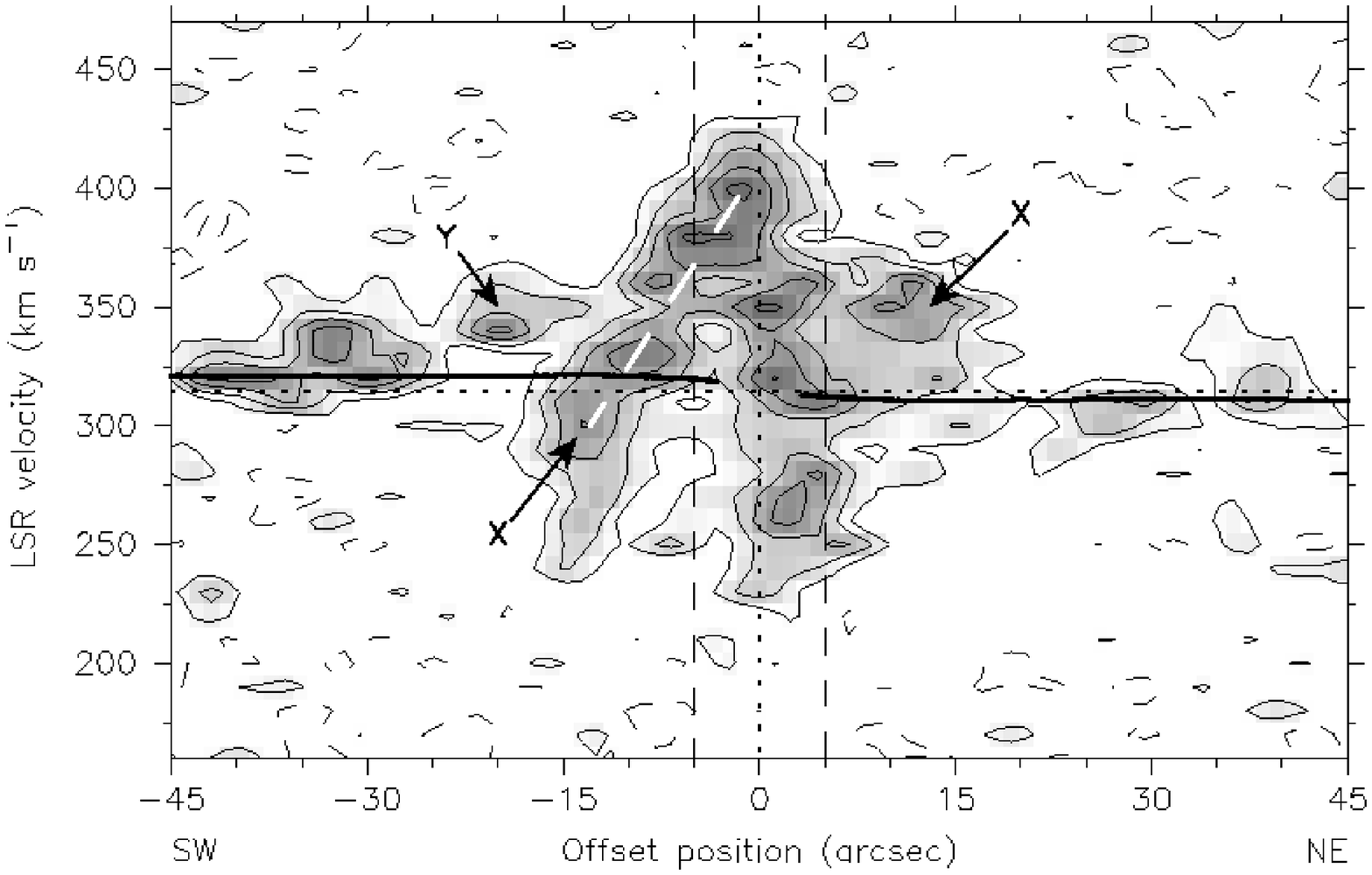}
\figcaption[barpv.bit.ps]{
Position-velocity map of CO emission along a cut having a position
angle of 28\arcdeg\ (parallel to the nuclear bar) and passing through
the galaxy center.  Contour levels are $\pm \sigma \ldots \pm6\sigma$,
where $\sigma$=70 mJy bm$^{-1}$.  Dotted horizontal and vertical lines
correspond to the systemic velocity and galaxy center.  Dashed vertical
lines at $\pm5$\arcsec\ delineate the circumnuclear region where beam
smearing of the galaxy's rotation may dominate.  Regions of local
inflow and outflow are labeled X and Y respectively.  The thick solid
line is the velocity curve predicted by the fitted rotation model.
\label{barpv}}
\vskip 0.25truein

%%%%%%%%%%%%%%%%%%%%%%%%%%%%%%%%%%%%%%%%%%%%%%%%%%%%%%%%%%%%%

The large velocity difference between regions X and Y on the southern
side of the bar may indicate that the streaming motions associated with
the bar (region X) are different from those in the spiral arm close to
the bar end (region Y).  Such a change in the direction of radial
streaming motions is expected when crossing a bar's corotation radius
\citep{Com88}.  Using our assumed pattern speed (\S\ref{kin_rot}), the
CR occurs at an offset position of 21\arcsec\ along this slice.  Also
apparent in Figure~\ref{barpv} is a transition from locally inward to
outward gas motion in going from region X on the SW side ($-15$\arcsec)
towards the center (along the white dashed line).  Since the inward velocity
component at X is well-modeled by gas moving in elliptical streamlines
(\S\ref{kin_ell}), the transition to radially outward motion may
indicate a change in the orientation of the elliptical gas orbits.
Higher resolution data would be valuable in exploring this
possibility.

%%%%%%%%%%%%%%%%%%%%%%%%%%%%%%%%%%%%%%%%%%%%%%%%%%%%%%%%%%%%%
%%%%%%%%%%%%%%%%%%%%%%%%   FIG. 14  %%%%%%%%%%%%%%%%%%%%%%%%%
%%%%%%%%%%%%%%%%%%%%%%%%%%%%%%%%%%%%%%%%%%%%%%%%%%%%%%%%%%%%%

\vskip 0.25truein
\includegraphics[width=3.25in]{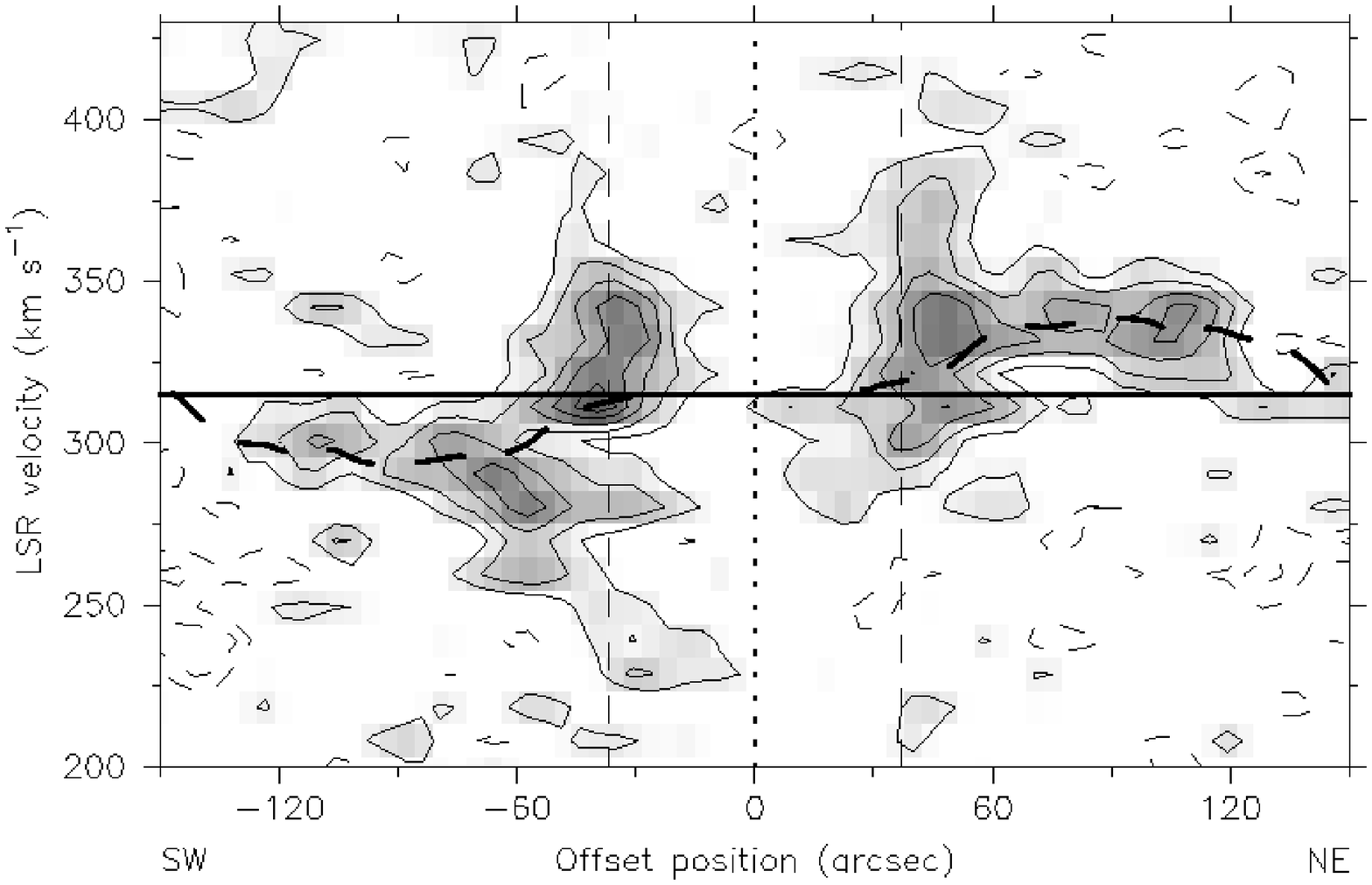}
\figcaption[atpv.bit.ps]{
Position-velocity map of \HI\ emission, binned to a velocity resolution
of 10.4 \kms, along the assumed minor axis (PA=25\arcdeg).  Contour
levels are $\pm\sigma \ldots \pm6\sigma$, where $\sigma$=1.5 mJy
bm$^{-1}$.  The thick solid line is the systemic velocity, which should
be the centroid of the emission according to the disk model with PA
fixed.  The thick dashed line is the velocity curve predicted if the PA
is allowed to vary with radius (cf.\ Figure \ref{pavar}).  The dashed
vertical lines mark the location of the H$\alpha$ ring.
\label{atpv}}
\vskip 0.25truein

%%%%%%%%%%%%%%%%%%%%%%%%%%%%%%%%%%%%%%%%%%%%%%%%%%%%%%%%%%%%%

Figure~\ref{atpv} displays a p-v slice through the \HI\ cube along the
assumed minor axis (PA=25\arcdeg).  While the circular rotation model
predicts that the emission will be centered at $V_{\rm sys}$ (heavy
solid line), the \HI\ emission deviates quite strongly from this model,
displaying the inflow signature that was already noted in the
velocity residuals (\S\ref{kin_res}).  On the southwestern side, there
is a striking transition from positive to negative velocity residuals
as one crosses the ring (vertical dashed line), corresponding to
relative outflow on the inner side of the ring (where the CO also shows
outflow) and inflow on the outer side.  Precisely such a signature is
expected when crossing a Lindblad resonance, as discussed in
\S\ref{disc_bar}.  The heavy dashed line represents a best-fit model
in which the PA is allowed to vary with radius, as in Figure~\ref{pavar}.
Note that even this model does not completely account for the velocity
offsets at around $\pm$60\arcsec.

We conclude that while the p-v diagrams allow for a more detailed view
of the kinematics in certain regions, they are generally consistent
with the non-circular motions inferred from the velocity residual maps.

\subsection{Radial Inflow Model\label{kin_inflow}}

In the following sections we explore three physical models which may
account for the non-circular motions seen in the velocity residuals and
p-v diagrams.  The first model we consider has gas moving in circular
coplanar orbits superposed on a component of uniform radial contraction
or expansion.  In the ROTCUR program it is possible to fix the disk
parameters and fit simultaneously for rotation and expansion
velocities:
\begin{equation}
V(x,y) = V_{\rm sys} + v_{\rm rot}(r) \sin i \cos \theta
	+ v_{\rm exp}(r) \sin i \sin \theta \;.
\end{equation}

The resulting inflow velocities ($v_{\rm in}=-v_{\rm exp}$) are shown in 
Figure~\ref{vexp}a.  The inclusion of this additional
velocity term virtually eliminates the dipole signature seen in
Figure~\ref{h1res}, since the effect of axisymmetric inflow on the
velocity field of a galaxy is almost indistinguishable from an error
in $\phi_{kin}$, at least for a nearly face-on galaxy such as NGC
4736.  In the presence of radial inflow, the kinematic major axis 
($\phi_{kin}$) rotates by an angle defined by
\[\tan\Delta\phi_{kin} = \alpha \cos i\,,\] 
where $i$ is the inclination from face-on and $\alpha$ is the ratio of
the inflow speed to the circular speed, while the kinematic minor axis
(systemic velocity contour) rotates by a somewhat larger angle:
\[\tan\Delta\phi_{min} = \alpha /(\cos i)\,.\] 
Note that for a more highly inclined galaxy, it would be easier to distinguish
inflow from an error in $\phi_{kin}$ because inflow would lead to a 
change in $\phi_{min}$ different from that in $\phi_{kin}$.

The zero point for $v_{\rm in}$ in Figure~\ref{vexp}a is quite
sensitive to the assumed value of $\phi_{kin}$: each 5\arcdeg\ increase
in the adopted $\phi_{kin}$ is equivalent to roughly a 15
\kms\ decrease in $v_{\rm in}$ at all radii.  Since the isophotal
position angle decreases from about 295\arcdeg\ to 270\arcdeg\ between
45\arcsec--100\arcsec, there appears to be a great deal of freedom in
setting this zero point (although adopting a smaller $\phi_{kin}$ would
only {\it increase} the inflow velocity).  Under the assumption that
the galaxy is not warped in these regions, we still consider
295\arcdeg\ the most likely value for the true line-of-nodes PA,
because outside the ring the kinematic and isophotal position angles
disagree, but both are consistent with $\phi$=295\arcdeg\ at the ring
(Figure~\ref{pavar}).  This choice also agrees with theoretical
expectations in the context of an inflow model: (1) it causes the
inflow velocity at the radius of the ring to vanish, with outflow
interior to the ring and inflow exterior to it, conditions which seem
appropriate for the formation of a gaseous ring; (2) the resulting sign
changes in $v_{\rm in}$ occur at the probable resonances of the oval
distortion:  a transition from outflow to inflow at $r \approx
45$\arcsec\ (ILR) and from inflow to outflow at $r \approx
200$\arcsec\ (CR), consistent with the predicted torques exerted on gas
by a bar \citep{Com88,Sch81}; (3) there is no significant $v_{\rm in}$
outside the OLR of the oval ($r \approx 330$\arcsec), as expected from
its reduced gravitational influence.

%%%%%%%%%%%%%%%%%%%%%%%%%%%%%%%%%%%%%%%%%%%%%%%%%%%%%%%%%%%%%
%%%%%%%%%%%%%%%%%%%%%%%%   FIG. 15  %%%%%%%%%%%%%%%%%%%%%%%%%
%%%%%%%%%%%%%%%%%%%%%%%%%%%%%%%%%%%%%%%%%%%%%%%%%%%%%%%%%%%%%

\vskip 0.25truein
\includegraphics[width=3.25in]{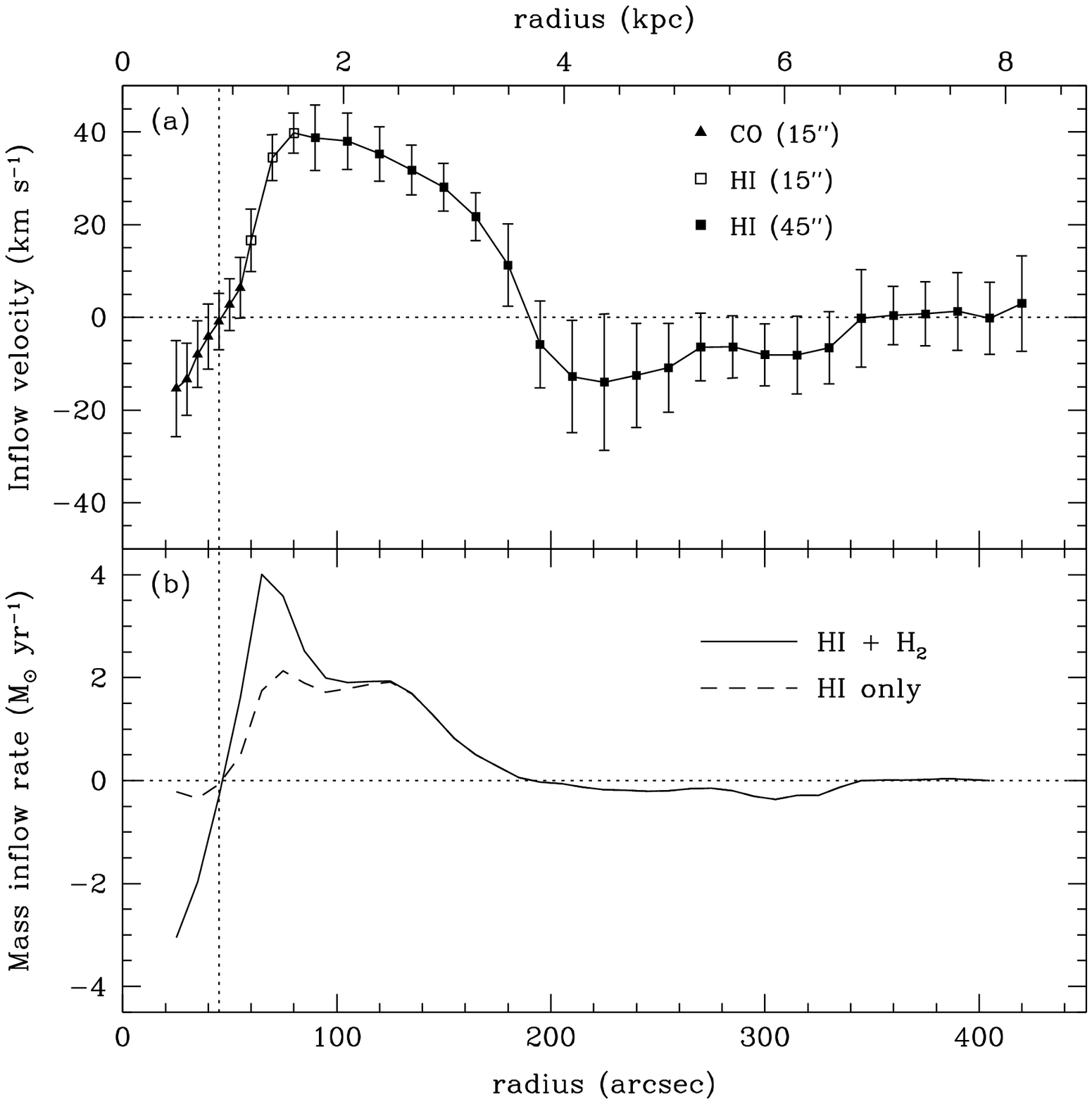}
\figcaption[vexp.eps]{
(a) Fitted inflow velocity ($v_{\rm in}=-v_{\rm exp}$) as a function of
radius, with disk parameters fixed at $\phi=295\arcdeg$,
$i=35\arcdeg$.  The curve is based on the same datasets used for
Figure~\ref{rotcur}.  (b) Inferred mass inflow rate as a function of
radius, based on panel (a) and the 15\arcsec\ resolution gas profiles.  The
vertical dotted line represents the location of the inner ring.
\label{vexp}}
\vskip 0.25truein

%%%%%%%%%%%%%%%%%%%%%%%%%%%%%%%%%%%%%%%%%%%%%%%%%%%%%%%%%%%%%

Figure~\ref{vexp}b shows the inferred mass accretion rate as a 
function of radius,
\[\dot{M}(r) = 2\pi r \Sigma_{\rm gas} v_{\rm in}\,.\]
If the molecular gas is assumed to participate in the same radial
inflow as the atomic gas, the mass inflow rate reaches a maximum of 4
\Msol\ yr$^{-1}$ at $r$=65\arcsec.  However, the validity of this
assumption is unclear due to the lack of CO detected outside the ring.
Considering only the atomic gas, an inflow rate of $\sim$2
\Msol\ yr$^{-1}$ is implied.  This is within a factor of 2 of the
estimates given by \citet{Reg97} for NGC 1530 (1$^{+2}_{-0.5}$
\Msol\ yr$^{-1}$) and \citet{Qui95} for NGC 7479 (4$\pm$2
\Msol\ yr$^{-1}$), both strongly barred galaxies.  In neither of these
cases, however, was the inflow assumed to be axisymmetric.  For
comparison, the rate at which gas is consumed by star formation in the
ring is approximately 0.2 \Msol\ yr$^{-1}$ (\S\ref{disc_evol}).

The most disturbing aspect of a pure inflow model is the high (up to 40
\kms) inflow velocities required to explain the residuals outside the
ring.  These speeds are well above the likely sound speed and would
therefore create strong shocks.  Of course, such shocks are precisely
what is needed to drive rapid inflows in the first place, but one would
expect them to be confined to localized regions, e.g.\ where the gas
first encounters a spiral arm \citep*{Rob69,Shu73}, contrary to what is
seen in Figure~\ref{h1res}.  Another problem is the relatively short
timescale on which the \HI\ outside the ring would flow inward.  The
mass-weighted mean flow velocity in the annulus from
$r$=80\arcsec--190\arcsec---where the radial flow is inward and where
most of the \HI\ outside the ring lies---would be 32 \kms, sufficient
to clear out that region in $\sim$70 Myr.  This timescale is much
shorter than the Hubble time or the gas consumption timescale
(\S\ref{sfr_tau}) and indeed is comparable to the orbital (dynamical)
timescale.

Our analysis highlights the intrinsic difficulty in measuring
{\it subsonic} inflow speeds in galaxy disks.  For a moderately
inclined galaxy, detection of a 5 \kms\ inflow velocity against a
circular speed of $\sim$200 \kms\ requires being able to determine a
shift in the kinematic position angle of $\sim$1\arcdeg.  Large
inclination angles ($i>55\arcdeg$) would be necessary to create a shift
of several degrees in the kinematic minor axis, but then beam-smearing
effects and the finite thickness of the disk complicate the
interpretation.  

\subsection{Warped Disk Model\label{kin_warp}}

The increase in kinematic position angle with radius, in the region
outside the ring, may be an indication of a warp in the disk---in other
words, the gas orbits may not be coplanar.  Warps in the stellar disk
are generally not seen well inside the optical radius of a galaxy
\citep{Korm82}; on the other hand, a strong ($\sim$25\arcdeg) tilt in
the molecular and atomic gas layer has been inferred for the inner 2
kpc of our Galaxy \citep{Burt78,Liszt80}, suggesting that the gas disk
can deviate substantially from the midplane of the galaxy.  The classic
signature of a warp is a progressive variation in the values of both
$i$ and $\phi$ produced by a tilted-ring fit, usually occurring beyond
a certain radius.  Unfortunately, we find that allowing both parameters
to vary with radius results in a poor fit unless the rotation
velocities were also fixed, another indication of the tight coupling
between $v_{\rm rot}$ and $i$.  Hence the data do not allow a clear
determination of how $i$ may change with radius.

The most serious objection to a warp model is that while there does
seem to be a connection between the change in optical morphology and
the change in kinematics outside the ring, they cannot both be
explained by a warp, since the isophotal PA decreases as the kinematic
PA increases (Figure~\ref{pavar}).  For a warped disk one expects the
two angles to change in the same direction.  Furthermore, since the
change in $\phi_{kin}$ reverses itself at larger radii
(cf.\ Figure~\ref{vexp}a), the direction of the warp would also have to
reverse direction.

\subsection{Elliptical Streaming Model\label{kin_ell}}

The divergence of the isophotal and kinematic position angles shown in
Figure~\ref{pavar} is best explained as an effect of elliptical orbits
when seen in projection.  This is because the kinematic major axis is
defined as the locus of maximum line-of-sight velocity, which for
elliptical orbits will generally {\it not} occur when a parcel of gas
is at its largest distance from the galaxy's center
(apogalacticon)---which in turn may not even coincide with the major
axis of the ellipse as viewed in the sky!  To illustrate this property,
we used the NEMO software package \citep{Teu95} to generate a simple
model of oval orbits by stretching circular orbits by $\epsilon^{1/2}$
and $\epsilon^{-1/2}$ along the $x$ and $y$ axes respectively and
scaling the circular velocities by $\epsilon^{-1}$ and $\epsilon$, the
correct scaling for a harmonic potential ($\kappa=2\Omega$).  A value
of $\epsilon$=1.2 was used, based on the ellipticity of the oval
distortion in NGC 4736, and the orbits were rotated and inclined to the
line of sight so that they appeared as ellipses with a position angle
of $-27$\arcdeg\ relative to the true line of nodes---this value should
approximate the isophotal PA if the orbits are aligned with a barred
potential.  The resulting {\it kinematic} position angle as a function
of radius is shown by the solid line in Figure~\ref{majmin}(c).  Due to
projection effects, the true line of nodes ($\phi$=0\arcdeg) has a PA
greater than the isophotal PA ($\phi$=$-27$\arcdeg, not shown) but less
than the kinematic PA ($\phi$=5\arcdeg--10\arcdeg), similar to what is
seen in the NGC 4736 data (Figure~\ref{pavar}).

%%%%%%%%%%%%%%%%%%%%%%%%%%%%%%%%%%%%%%%%%%%%%%%%%%%%%%%%%%%%%
%%%%%%%%%%%%%%%%%%%%%%%%   FIG. 16  %%%%%%%%%%%%%%%%%%%%%%%%%
%%%%%%%%%%%%%%%%%%%%%%%%%%%%%%%%%%%%%%%%%%%%%%%%%%%%%%%%%%%%%

\vskip 0.25truein
\includegraphics[width=3.25in]{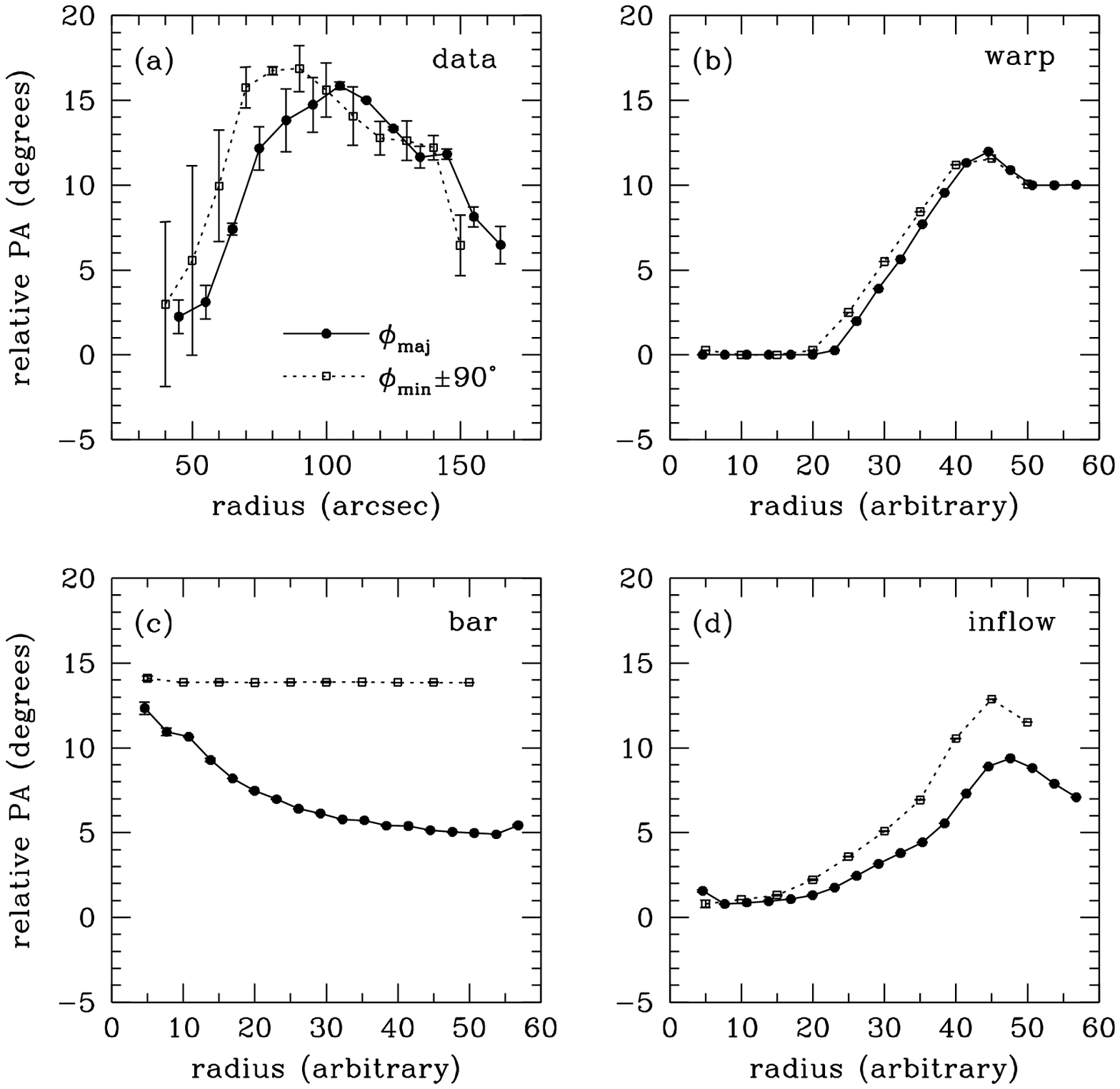}
\figcaption[majmin.eps]{
Position angles of the kinematic major and minor axes as a function of
radius for the \HI\ data and three kinematic models (warped disk,
elliptical orbits aligned with a bar, and radial inflow).  For each axis
the PA has been determined independently on each side of the galaxy center;
the plotted points are the average values and the error bars represent
the difference between the two measurements.  The assumed line of nodes
is taken to be 0\arcdeg.
\label{majmin}}
\vskip 0.25truein

%%%%%%%%%%%%%%%%%%%%%%%%%%%%%%%%%%%%%%%%%%%%%%%%%%%%%%%%%%%%%

Figure~\ref{majmin} also reveals another characteristic of this simple
bar model, namely that the kinematic major and minor axes do not remain
perpendicular---a property that is commonly observed in barred galaxies
\citep{Bos78}.  For aligned elliptical orbits, the minor axis is
expected to remain constant with radius, while the major axis drifts
away from it.  Yet for NGC 4736, the \HI\ data outside the ring
(Figure~\ref{majmin}a) show that the two axes stay approximately
perpendicular even as they drift by 10\arcdeg--15\arcdeg.  In this
respect, the data agree better with the warp and inflow models
discussed above and shown schematically in panels (b) and (d).

Elliptical streaming should also produce distinctive signatures in 
the velocity residuals as a function of $\theta$ (the angle in
galactic plane), as discussed by \citet{Teu91}.  In the simplest
approximation, the requirement of no net inflow suggests that the
expansion velocity in Equation 2 (\S\ref{kin_inflow}) takes the form
\[v_{\rm exp} \propto \sin 2(\theta+\theta_0)\,\]
where $\theta_0$ is a phase shift dependent on the orientation of the
orbits.  In other words, the expansion velocity is positive in two
quadrants and negative in the other two.  In the special case where we
view the orbits end-on or edge-on ($\theta_0$=0\arcdeg\ or 90\arcdeg),
the velocity residuals have a $\cos 3\theta$ form (cf.\ Equations 1 \& 2), and
the oval orbits mimic circular orbits with a different value for the
disk inclination \citep{vdK78}.  In the more general case where the bar
has some arbitrary angle to the line of nodes, the residuals will be
dominated by a linear combination of $\sin \theta$ and $\cos
(3\theta+2\theta_0)$, with the $\cos 3\theta$ wave being the stronger
of the two.  Then the velocity field {\it cannot} be modeled in terms
of circular orbits, since one can remove the $\sin \theta$ term with an
appropriate choice of $\phi_{kin}$, but attributing the remaining $\cos
3\theta$ term to inclination would have required a different $\phi_{kin}$
due to the phase shift $\theta_0$.

%%%%%%%%%%%%%%%%%%%%%%%%%%%%%%%%%%%%%%%%%%%%%%%%%%%%%%%%%%%%%
%%%%%%%%%%%%%%%%%%%%%%%%   FIG. 17  %%%%%%%%%%%%%%%%%%%%%%%%%
%%%%%%%%%%%%%%%%%%%%%%%%%%%%%%%%%%%%%%%%%%%%%%%%%%%%%%%%%%%%%

\vskip 0.25truein
\includegraphics[width=3.25in]{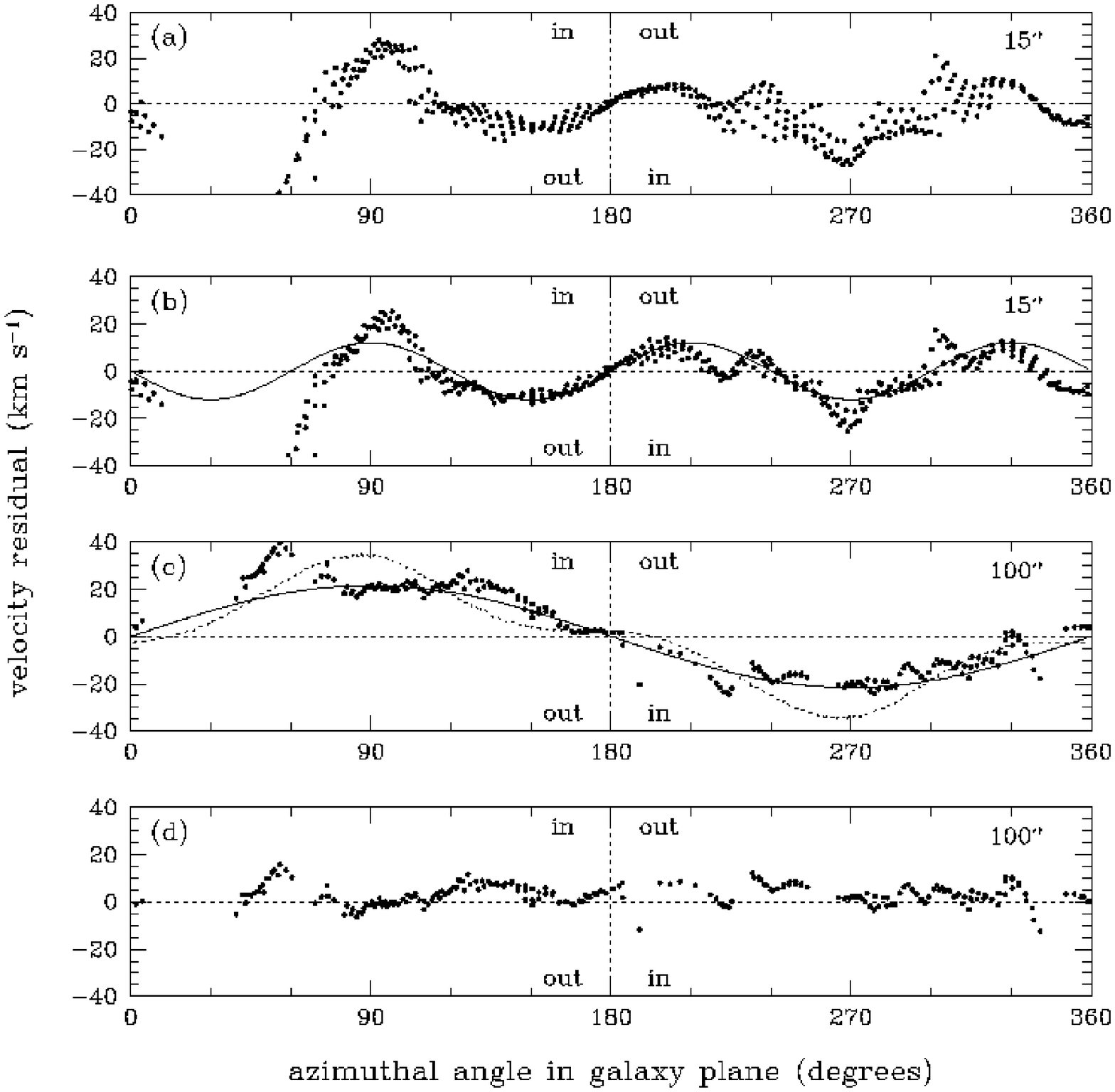}
\figcaption[azres.bit.ps]{
Velocity residuals plotted in galactocentric rings.  Panels (a) and (b)
show the CO residuals at $r$=15\arcsec--20\arcsec\ before and after
fitting a $\sin\theta$ (varying PA) component.  A $\sin 3\theta$
curve is shown in panel (b).  Panels (c) and (d) show the same analysis
for the \HI\ residuals at $r$=100\arcsec--110\arcsec.  The solid line in
panel (c) approximates the best-fit $\sin\theta$ curve, while the dotted
line is a prediction for an elliptical streaming model.
Residuals correspond to the directions of inflow and outflow as marked.
See text for discussion.
\label{azres}}
\vskip 0.25truein

%%%%%%%%%%%%%%%%%%%%%%%%%%%%%%%%%%%%%%%%%%%%%%%%%%%%%%%%%%%%%

These simple predictions for elliptical streaming agree qualitatively
with the kinematics in the inner CO disk.  As Figure~\ref{azres} shows,
in the annulus from $r$=15\arcsec--20\arcsec\ there is a clear $\sin
3\theta$ term in the residuals.  This persists in roughly the same form
even after removal of a $\sin \theta$ term (i.e. freely varying PA) in
panel (b).  The curve drawn has the form $V_{res}=-12\sin 3\theta$
\kms\ and is a reasonably good fit to the data.  We conclude that the
velocity residuals in this region are due primarily to elliptical
streaming in the bar potential.  In light of the preceding discussion,
the presence of a $\sin 3\theta$ rather than $\cos 3\theta$ residual
suggests that the major axis of the orbits would be oriented
$\sim$45\arcdeg\ away from the line of nodes if the galaxy were viewed
face-on.  While this is somewhat puzzling since we appear to be looking
at the bar close to end-on, we do not expect our rough analysis to
yield exact agreement with the data (for instance, we have neglected
the $\theta$ dependence of $v_{\rm rot}$ in Eq.\ 2).

In contrast, the \HI\ residuals outside the ring are dominated by a
$\sin \theta$ term (Figure~\ref{azres}c), which upon removal reveals
little if any $\cos 3\theta$ component (Figure~\ref{azres}d).  This is
inconsistent with gas streaming in aligned elliptical orbits, in which
case one would expect the strong $\cos 3\theta$ term to produce
residuals with a more cusped appearance (dotted line in
Figure~\ref{azres}c).  Rather, the dominance of the $\sin \theta$ (dipole)
term agrees better with the radial inflow model.  In \S\ref{disc_kin} we
discuss possible ways in which an elliptical streaming model could be
made more consistent with the observations.

%% STAR FORMATION

\section{Star Formation}\label{sfr}

\subsection{Determination of the SFR\label{sfr_cal}}

As noted by \citet{KC98b}, \Halpha\ is a sensitive tracer of the Lyman
continuum flux, which is primarily due to stars with masses $>$10
\Msol\ and lifetimes $<$20 Myr. Ignoring extinction and adopting a
Salpeter IMF, we convert the H$\alpha$ flux to an SFR using the
formula \citep{KC98b}:
\[\frac{\rm SFR}{\rm \Msol\; yr^{-1}} = 7.9 \times 10^{-42}\;
	\left(\frac{L_{\rm H\alpha}}{\rm erg\; s^{-1}}\right)\;.\] 

Although uncertainties in the IMF and the escape fraction of ionizing
photons can significantly affect the SFR, probably the largest source
of uncertainty when comparing regions within a galaxy is variable
extinction due to dust in the galaxy's disk.  (For NGC 4736, extinction
due to dust in our Galaxy is likely to be minimal at a Galactic
latitude of 76\arcdeg.)  A rough estimate of the extinction at
\Halpha\ can be derived by assuming an extinction law of $N_H/A_V = 2
\times 10^{21}$ cm$^{-2}$ mag$^{-1}$ and $A_R/A_V = 0.75$
\citep*{Bohl78,Rieke85}.  For NGC 4736, if the \HII\ regions are located
in the midplane, this scaling gives typical values of $A_R \sim 0.75$
at the ring ($r$=45\arcsec), $\sim$2.5 in the molecular bar
($r$=10--20\arcsec), and $\sim$4--9 at the nucleus ($r<5$\arcsec).
However, this assumes a constant gas-to-dust ratio across a huge range
of environments as well as uniform mixing of dust.  A better way to
test the importance of extinction is to measure the thermal
bremsstrahlung emitted in the radio, which is transparent to dust; the
main difficulties are that synchrotron emission from supernova remnants
is likely to dominate and good sensitivities are difficult to achieve.
Still, the VLA observations of \citet{Tur94} and \citet{Dur88} appear
to show very little thermal emission interior to the ring;
\citet{Tur94} estimate a thermal flux of 2 mJy for the central
1\arcmin\ $\times$ 1\arcmin, compared to 13 mJy in the ring.  These
results are consistent with the non-detection of Br$\gamma$ emission
from the nucleus by \citet{Walk88}.  We conclude that while extinction
is likely to be important in the nuclear region, infrared and radio
observations show no indication that high levels of star formation are
occurring there.  On the other hand, there are spectroscopic
indications of a young stellar population in the nucleus, implying that
the SFR in the nucleus was much higher $\sim$1 Gyr ago
\citep{Prit77,Walk88}.

The \Halpha\ image contains emission from the \NII\ line as well.
Spectroscopy by \citet{Smi91} indicates that 62\% of the emission in the 
ring is due to \Halpha, and this ratio has been used to correct the fluxes
before conversion to star formation rates.  In the nuclear region
\citet{Smi91} found that only 24\% of the emission is due to \Halpha;
given the large uncertainty in the \Halpha\ flux from this region
anyway, we have not bothered to apply a different correction.  

\subsection{Gas Consumption Timescales\label{sfr_tau}}

The CO profile and the adopted SFR profile are shown in
Figure~\ref{taugas}, at 6\arcsec\ resolution.  The ratio of the two
profiles gives a nominal gas consumption time $\rm \tau_{gas} \equiv
\Sigma_{gas}/\Sigma_{SFR}$.  Due to the neglect of gas recycling by
evolved stars, which may extend these times by factors of $\sim$2
\citep*{KC94}, and uncertainties in the CO-to-H$_2$ and \Halpha-to-SFR
conversions, these times should be considered very approximate.
Nonetheless, it is clear that the gas consumption time within the ring
($\sim$1 Gyr) is much less than a Hubble time, as one would expect in a
starburst.  We neglect the \HI\ contribution to $\Sigma_{\rm gas}$,
which is small ($\lesssim$25\%) in this region.

%%%%%%%%%%%%%%%%%%%%%%%%%%%%%%%%%%%%%%%%%%%%%%%%%%%%%%%%%%%%%
%%%%%%%%%%%%%%%%%%%%%%%%   FIG. 18  %%%%%%%%%%%%%%%%%%%%%%%%%
%%%%%%%%%%%%%%%%%%%%%%%%%%%%%%%%%%%%%%%%%%%%%%%%%%%%%%%%%%%%%

\vskip 0.25truein
\includegraphics[width=3.25in]{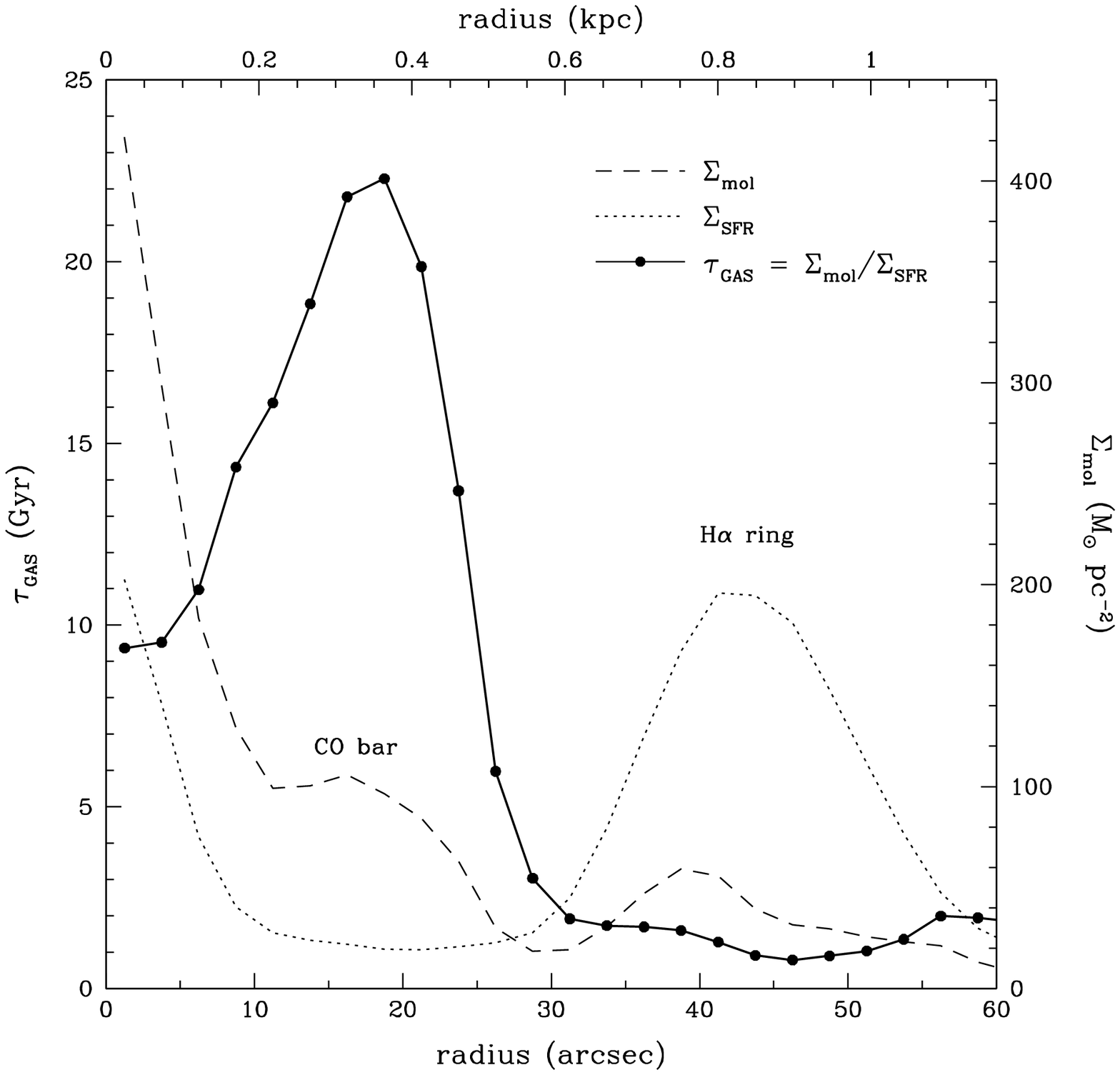}
\figcaption[taugas.eps]{
Radial profiles of CO (robustly weighted cube) and \Halpha, after
conversion into molecular gas surface density and SFR surface density
respectively, and their ratio, the gas consumption time.
\label{taugas}}
\vskip 0.25truein

%%%%%%%%%%%%%%%%%%%%%%%%%%%%%%%%%%%%%%%%%%%%%%%%%%%%%%%%%%%%%

Focusing now on the radial variation in \tgas, the most striking trend
is the very long gas consumption time in the vicinity of the molecular
bar ($r$=10\arcsec--30\arcsec) in comparison with the ring.
Equivalently, the star formation efficiency, as measured by $\rm
\Sigma_{SFR}/\Sigma_{gas} = \tau_{gas}^{-1}$, is much higher in the
ring than in the inner disk.  (We define the SFE here as the star
formation rate per unit gas mass, since the fraction of gas converted
into stars is only well-defined for an individual cloud.) It is possible
that some of this difference is due to dust extinction obscuring star
formation near the nucleus and a standard X-factor overestimating the
amount of molecular gas.  We discuss these possibilities further in
\S\ref{disc_molrad}.

Our interpretation of \tgas\ as the gas consumption time and $\tau_{\rm
gas}^{-1}$ as the star formation efficiency assumes that $\Sigma_{\rm
gas}$ and $\Sigma_{\rm SFR}$ are close to their time-averaged values.
This will probably not be true on a point-by-point basis, due to
evolutionary effects and a bias towards observing massive stars in
regions which have been cleared of dense gas, but it may be
approximately true when averaging azimuthally within rings.  However,
in regions where star formation on large scales is episodic (e.g.,
controlled by a threshold density), unusually high or low star
formation efficiencies will be inferred depending on whether we are
observing the galaxy in an active or quiescent star-forming phase.  

%%%%%%%%%%%%%%%%%%%%%%%%%%%%%%%%%%%%%%%%%%%%%%%%%%%%%%%%%%%%%
%%%%%%%%%%%%%%%%%%%%%%%%   FIG. 19  %%%%%%%%%%%%%%%%%%%%%%%%%
%%%%%%%%%%%%%%%%%%%%%%%%%%%%%%%%%%%%%%%%%%%%%%%%%%%%%%%%%%%%%

\vskip 0.25truein
\includegraphics[width=3.25in]{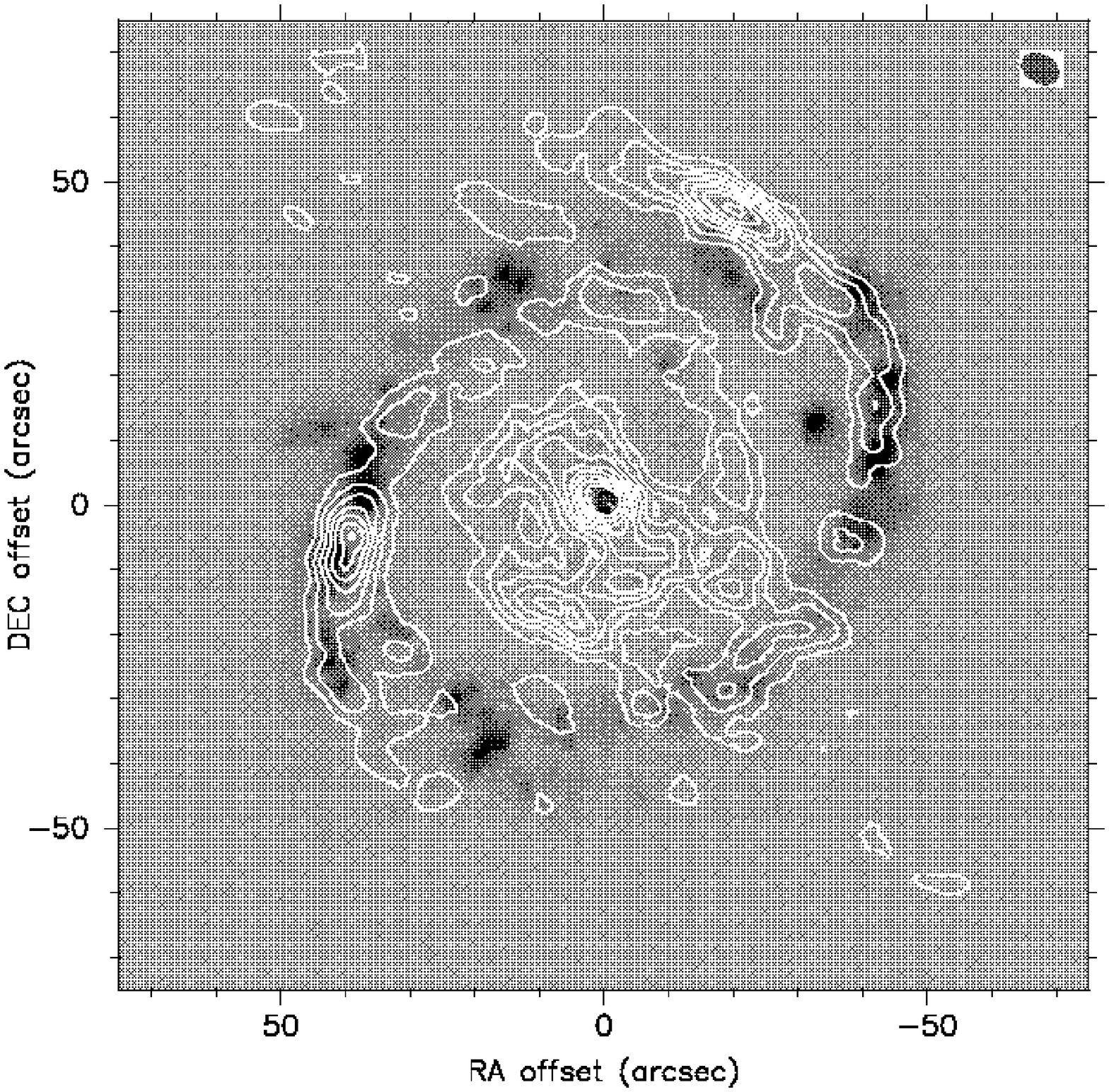}
\figcaption[coha.bit.ps]{
CO contours as in Figure~\ref{bimom0} overlaid on the \Halpha\ image of
\citet{Pog89}.
\label{coha}}
\vskip 0.25truein

%%%%%%%%%%%%%%%%%%%%%%%%%%%%%%%%%%%%%%%%%%%%%%%%%%%%%%%%%%%%%

\subsection{Azimuthal Symmetry\label{sfr_az}}

A feature that is not apparent in the radial profiles is the remarkable
180\arcdeg\ rotational symmetry in the \Halpha\ ring
(Figure~\ref{coha}), in stark contrast to the distribution of CO
emission.  This is truly unexpected if, as we expect, star formation is
associated with molecular clouds.  We compare the symmetry properties
of the \Halpha, and \HI, and CO in Figure~\ref{azprof}, which shows the
peak intensity in the ring as a function of azimuth from an arbitrarily
chosen reference point.  The \Halpha\ and \HI\ show a good twofold
symmetry (dashed and solid lines agree well, with a linear correlation
coefficient of 0.4--0.5) while the CO distribution shows poor twofold
symmetry (linear correlation coefficient of $-0.4$).  Instead, the CO
distribution suggests a threefold symmetry as remarked on in
\S\ref{morph_mom}, although the peak at 30\arcdeg\ azimuth is weaker
than the other two.  A cross-correlation analysis shows that the
\HI\ and \Halpha\ distributions match best if we assume that the
\HI\ is shifted $\sim$45\arcdeg\ upstream of the \Halpha\ (i.e.\ to the
right in Figure~\ref{azprof}).  This corresponds to a time lag of 3.5
Myr using the adopted rotation curve, but the improvement in the
correlation after making this shift is only marginal.

There is no obvious explanation for the differences in these
symmetries, although the symmetries themselves are likely to be
mainfestations of density wave resonances---a clear indication that the
star formation process is controlled in part by the global dynamics of
the galaxy.  The presence of twofold symmetry in both \HI\ and
\Halpha\ distributions is consistent with the earlier suggestion
(\S\ref{morph_mom}) that \HI\ is mainly a photodissociation product,
but Figure~\ref{azprof} also shows that the detailed correspondence
between \Halpha\ and \HI\ is poor.  We examine this issue further in
\S\ref{disc_at}.  The presence of a threefold symmetry in the CO
distribution may result from an interaction of underlying $m$=1 and
$m$=2 modes, as has been suggested by \citet*{Elm92} from studies
of other spiral galaxies.

%%%%%%%%%%%%%%%%%%%%%%%%%%%%%%%%%%%%%%%%%%%%%%%%%%%%%%%%%%%%%
%%%%%%%%%%%%%%%%%%%%%%%%   FIG. 20  %%%%%%%%%%%%%%%%%%%%%%%%%
%%%%%%%%%%%%%%%%%%%%%%%%%%%%%%%%%%%%%%%%%%%%%%%%%%%%%%%%%%%%%

\vskip 0.25truein
\includegraphics[width=3.25in]{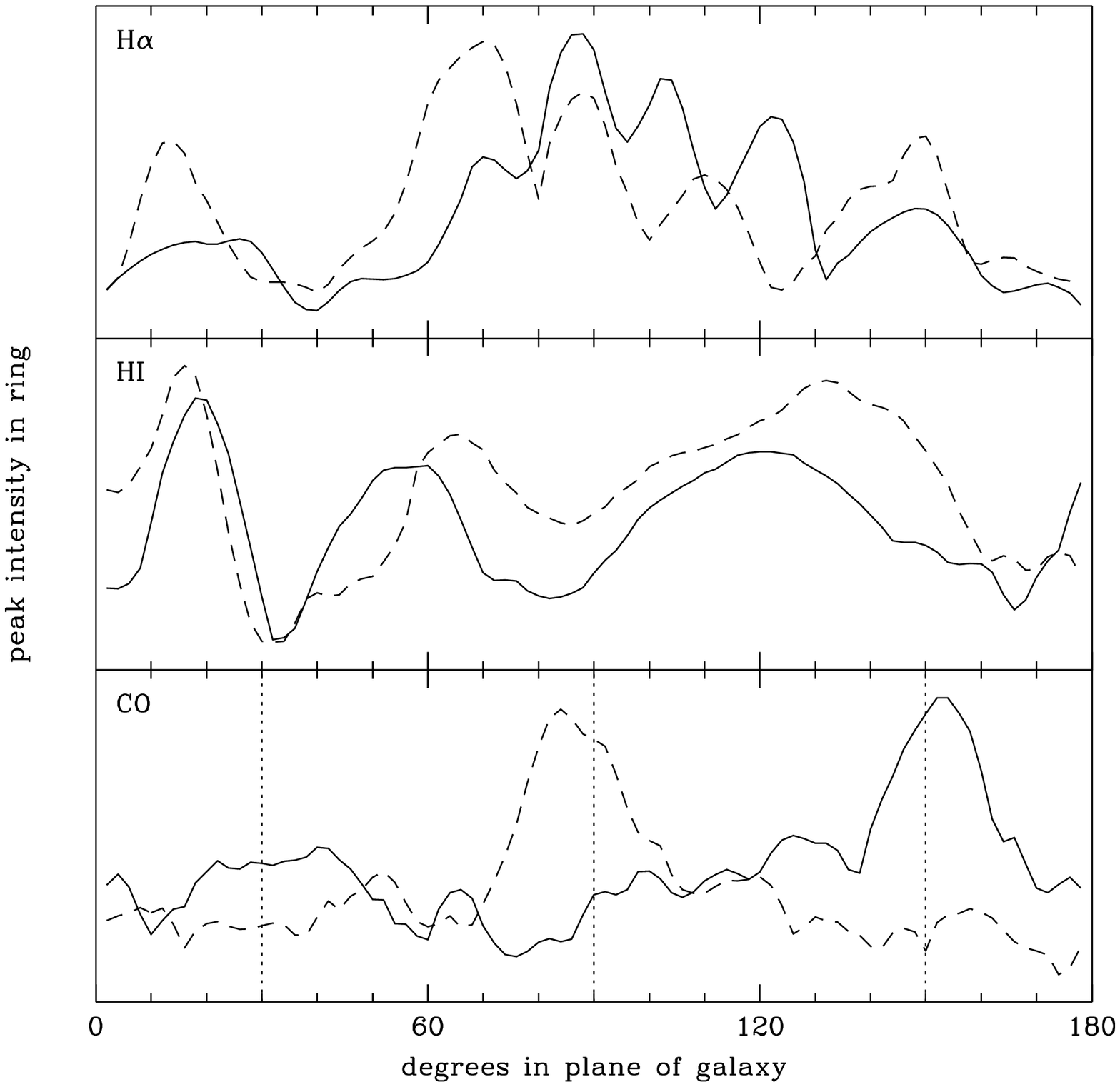}
\figcaption[azprof.eps]{
Azimuthal profiles of \Halpha, \HI, and CO peak intensity in the ring
($r$=30--60\arcsec).  The azimuthal angle increases counterclockwise
from due south (PA=180\arcdeg).  The dashed lines are the continuation
of the profile on the other side of the galaxy.  The dotted lines in
the bottom panel are spaced by 120\arcdeg\ and highlight the threefold
symmetry of the CO.
\label{azprof}}
\vskip 0.25truein

%%%%%%%%%%%%%%%%%%%%%%%%%%%%%%%%%%%%%%%%%%%%%%%%%%%%%%%%%%%%%

\subsection{Toomre Stability Criterion\label{sfr_crit}}

Can the radial variation in star formation efficiency be explained in
terms of a simple gravitational stability analysis?  In
Figure~\ref{qprof} the SFR profile is plotted along with the ratio $Q$
of the critical density for gravitational instability, $\Sigma_{\rm
crit} \equiv \kappa\sigma_v/\pi G$, to the gas surface density
$\Sigma_{\rm gas}$.  Thus, a value of $Q<1$ corresponds to instability
\citep{Saf60,Toom64}.  This formula is strictly appropriate for
axisymmetric (ring-like) perturbations only, and we discuss some of the
caveats in the interpretation of $Q$ in \S\ref{disc_crit}.  For the
velocity dispersion $\sigma_v$ we have taken a value of 7 \kms\ as
determined from the Gaussian fits to the robust CO datacube where the
major axis intersects the ring.  This is compatible with the value of 6
\kms\ assumed by \citet{KC89} as well as observations of \HI\ in
face-on galaxies \citep[e.g.][]{vdK84}.  Although beam smearing will
lead to broadening of the CO line profiles by the galaxy's rotation,
this effect is minimized near the major axis once the rotation curve
has flattened.  For a 6\arcsec\ beam located 45\arcsec\ from the
center, which subtends an angle of $\sim$10\arcdeg\ as measured in the
plane of the galaxy, the azimuthal velocity gradient across the beam is
negligible ($<$1 \kms) and the radial gradient is $\sim$4 \kms, based
on the CO rotation curve.  Correcting for this would only bring
$\sigma_v$ down to 6 \kms, and would be excessive since the edges of
the beam actually carry less weight in the smearing.  We also neglect
the likely rise in velocity dispersion towards the galaxy center, but
note that it would lead to a sharper rise in $Q$ than that shown in
Figure~\ref{qprof}.

%%%%%%%%%%%%%%%%%%%%%%%%%%%%%%%%%%%%%%%%%%%%%%%%%%%%%%%%%%%%%
%%%%%%%%%%%%%%%%%%%%%%%%   FIG. 21  %%%%%%%%%%%%%%%%%%%%%%%%%
%%%%%%%%%%%%%%%%%%%%%%%%%%%%%%%%%%%%%%%%%%%%%%%%%%%%%%%%%%%%%

\vskip 0.25truein
\includegraphics[width=3.25in]{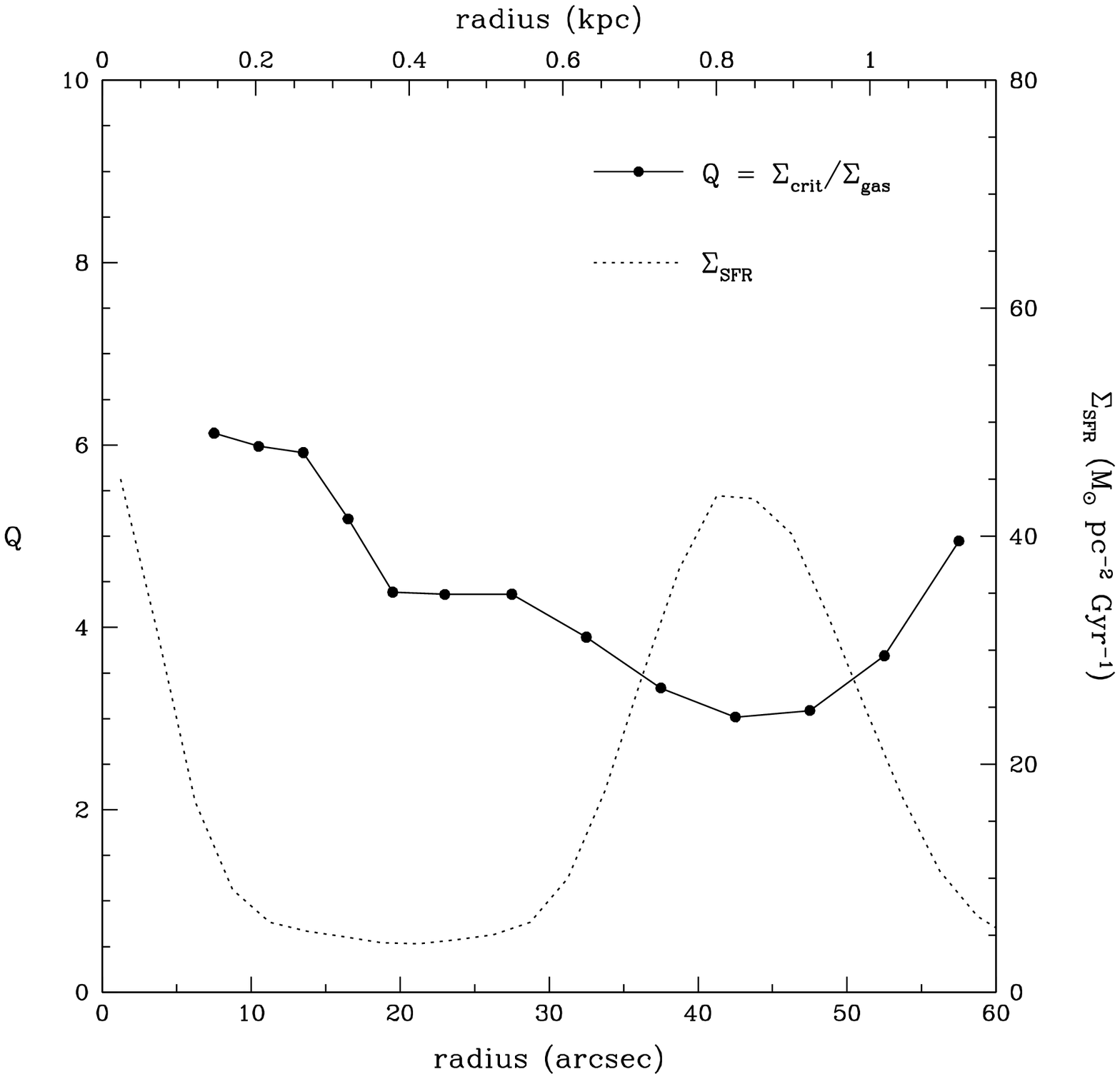}
\figcaption[qprof.eps]{
Radial profile of the Toomre $Q$ parameter (defined in text) superposed
on the SFR profile.
\label{qprof}}
\vskip 0.25truein

%%%%%%%%%%%%%%%%%%%%%%%%%%%%%%%%%%%%%%%%%%%%%%%%%%%%%%%%%%%%%

The resulting $Q$ profile shows a clear minimum in the vicinity of the
star-forming ring, consistent with a general picture in which star
formation in the ring is triggered by gravitational instability.
However, even here the {\it azimuthally averaged} gas density is a
factor of $\sim$3 less than the critical density.  As evident from the
moment-0 map (Figure~\ref{bimom0}), the ring contains a number of
molecular condensations which locally have surface densities (averaged
over a $\sim$100-pc wide beam and corrected to face-on) of up to 300
\Msol\ pc$^{-2}$, somewhat higher than the critical density ($\sim$200
\Msol\ pc$^{-2}$).  Thus it may be possible for star formation to
result from gravitational instability where {\it local} enhancements in
the gas density occur, although a different process (e.g.,
non-axisymmetric instabilities) would be necessary to create these
enhancements.  The large value of $Q$ in the ring is discussed further
in \S\ref{disc_crit}.

\subsection{The SFR--Gas Density Relationship\label{sfr_law}}

As noted in \S\ref{intro}, \citet{KC89,KC98a} has found that the
disk-averaged SFR is much better correlated with the average
\HI\ density than with the average H$_2$ density.  This result is
independent of whether \Halpha, UV continuum, or far-infrared fluxes
are used to derive the SFR, and is especially noticeable in
low-luminosity galaxies, leading \citet{KC98a} to attribute it to the
metallicity dependence of the X-factor.  He also finds that above the
critical density, the relation between the SFR and {\it total} gas
density is well-described by a Schmidt law, $\Sigma_{\rm SFR} \propto
\Sigma_{\rm gas}^N$, with $N \sim$ 1.3--1.5.  To investigate how well
these conclusions apply within the disk of NGC 4736, we compared the
fluxes of the three tracers CO, \Halpha, and \HI\ when averaged in
rings and on a point-by-point basis.

%%%%%%%%%%%%%%%%%%%%%%%%%%%%%%%%%%%%%%%%%%%%%%%%%%%%%%%%%%%%%
%%%%%%%%%%%%%%%%%%%%%%%%   FIG. 22  %%%%%%%%%%%%%%%%%%%%%%%%%
%%%%%%%%%%%%%%%%%%%%%%%%%%%%%%%%%%%%%%%%%%%%%%%%%%%%%%%%%%%%%

\vskip 0.25truein
\includegraphics[width=3.25in]{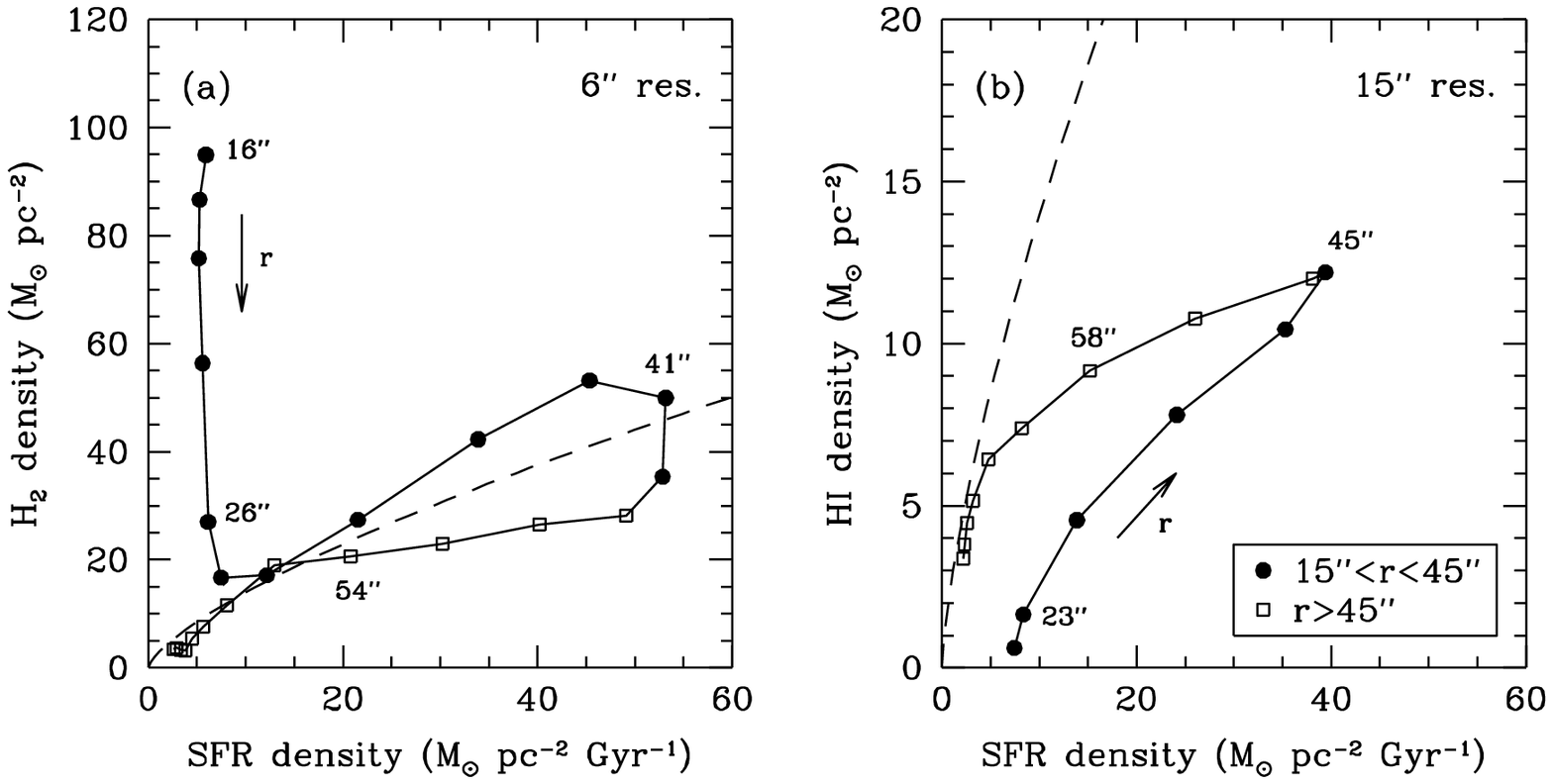}
\figcaption[sfrgas1.eps]{
Comparison of SFR density with gas density, averaged in rings, for (a) CO
data at 6\arcsec\ resolution, (b) \HI\ data at 15\arcsec\ resolution.
The CO data points begin at 16\arcsec\ and are spaced by 2\farcs5; the
\HI\ data points start at 17\farcs5 and are spaced by 5\arcsec.
The dashed line is the composite Schmidt law as given by \citet{KC98a}.
\label{sfrgas1}}
\vskip 0.25truein

%%%%%%%%%%%%%%%%%%%%%%%%%%%%%%%%%%%%%%%%%%%%%%%%%%%%%%%%%%%%%

Figure~\ref{sfrgas1} shows this comparison for rings of constant width
ranging from $r$=15\arcsec--75\arcsec\ for the CO data and
$r$=15\arcsec--90\arcsec\ for the \HI\ data.  (We have excluded the
nuclear region due to uncertainties in the \Halpha\ flux there,
\S\ref{obs_ha} and \S\ref{sfr_cal}.)  
The molecular gas (Figure~\ref{sfrgas1}a) dominates the gas
density over most of this range in radius, and is thus most appropriate
for comparing the SFR and total gas densities.  The overall correlation
between CO and \Halpha\ is poor, since the distribution of points in
the figure is strongly bimodal, with strong CO fluxes in the inner disk
($r<30$\arcsec) where little star formation is occurring, in addition
to the expected correlation in the vicinity of the ring ($r \approx
45$\arcsec).  The dashed line is the composite Schmidt law given by
\citet{KC98a},
\[\rm \Sigma_{SFR} = 0.25 
	\left(\frac{\Sigma_{gas}}{1\,\Msol\,pc^{-2}}\right)^{1.4}
	\;\Msol\,Gyr^{-1}\,pc^{-2}\;.\]
This parametrization is based on averages over the star-forming disks 
of 61 normal galaxies and the $\sim$1 kpc-size inner disks of 36 starbursts,
and has been shown to apply across a wide range of scales and 
galaxy types.  Both the power-law index and the normalization are a
surprisingly good match to the CO data outside of the inner disk, 
with the predicted gas densities lying within a factor of $\sim$1.5 of
their actual values.  Thus, when azimuthally averaged, star formation 
{\it in the ring only} appears to follow a Schmidt law.  Interior to the
ring, star formation falls well below the Schmidt law prediction.
This is in agreement with the low star formation efficiency for the
central region found in \S\ref{sfr_tau}, and is discussed further 
in \S\ref{disc_molrad}.

The azimuthally averaged \HI\ emission, being strongly peaked at the
ring, shows a good correlation with \Halpha, especially at high
densities (Figure~\ref{sfrgas1}b).  However, the SFR density rises
well above the Schmidt law prediction except at large radii
($r>70$\arcsec), which is not surprising given that most of the gas in
the inner region is molecular.  The positive correlation between
\Halpha\ and \HI\ emission within a region where the gas is primarily
molecular lends further support to the hypothesis that \HI\ in the ring
is mostly a dissociation product (see \S\ref{disc_at}).

%%%%%%%%%%%%%%%%%%%%%%%%%%%%%%%%%%%%%%%%%%%%%%%%%%%%%%%%%%%%%
%%%%%%%%%%%%%%%%%%%%%%%%   FIG. 23  %%%%%%%%%%%%%%%%%%%%%%%%%
%%%%%%%%%%%%%%%%%%%%%%%%%%%%%%%%%%%%%%%%%%%%%%%%%%%%%%%%%%%%%

\vskip 0.25truein
\includegraphics[width=3.25in]{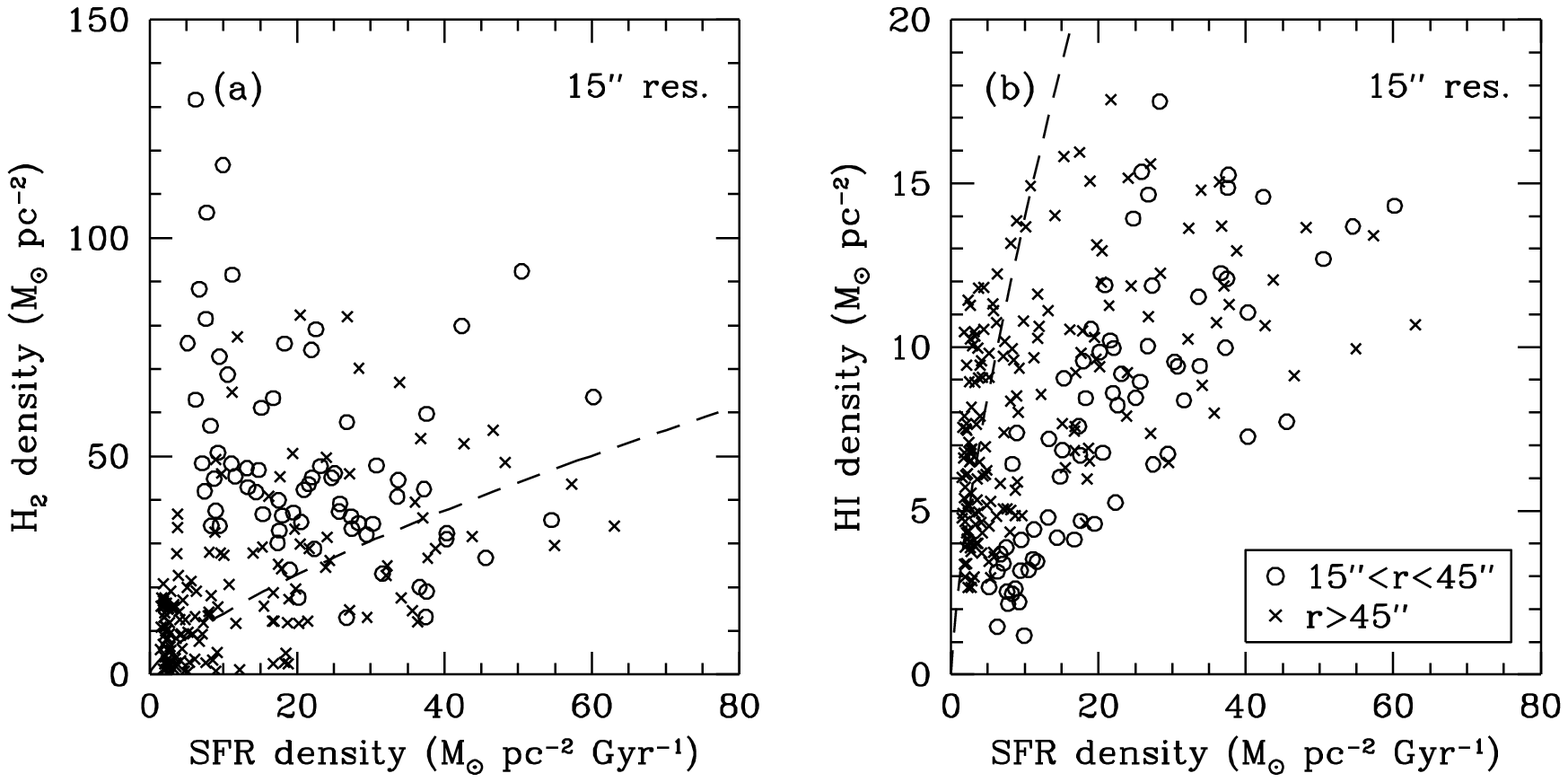}
\figcaption[sfrgas2.eps]{
Comparison of SFR density with gas density, on a pixel-by-pixel basis,
for (a) CO data, (b) \HI\ data, both at 15\arcsec\ resolution.
Data have been output in 8\arcsec\ pixels, so not all points are
independent.  The dashed line is the same as in Figure~\ref{sfrgas1}.
\label{sfrgas2}}
\vskip 0.25truein

%%%%%%%%%%%%%%%%%%%%%%%%%%%%%%%%%%%%%%%%%%%%%%%%%%%%%%%%%%%%%

Since averaging in rings causes pixels in the galaxy center to be
over-represented with respect to outer pixels, we also compared the
images on a pixel-to-pixel basis (Figure~\ref{sfrgas2}).  The overall
trends noted above are confirmed:  $\Sigma_{\rm SFR}$ is better
correlated with $\Sigma_{\rm HI}$ (correlation coefficient 0.53) than
with $\Sigma_{\rm H_2}$ (0.28), and $\Sigma_{\rm SFR}$ generally falls
{\it below} the Schmidt Law prediction when only molecular gas is
considered (particularly in the central regions) and {\it above} the
prediction when only atomic gas in considered.  On the other hand,
there is a much larger scatter in the pixel-by-pixel comparison than in
the azimuthal averages (Figure~\ref{sfrgas1}).  Some amount of scatter
will be introduced by measurement error (near zero flux) or pixellation
effects, but mostly it is indicative of the small-scale
anticorrelation between neutral and ionized gas that is apparent in
Figure~\ref{coha}.  This is largely an evolutionary effect (massive
stars dissociate and ionize their natal clouds) and is greatly reduced
by averaging in rings.  Making sense of the additional information
provided by {\it not} averaging in rings will require a more
sophisticated analysis, to be pursued in a future study (Sheth et al.,
in preparation).

%% DISCUSSION

\section{Discussion}\label{disc}

\subsection{Radial Inflow or Elliptical Streaming?\label{disc_kin}}

Our analysis in \S\ref{kin} of inflow, warp, and elliptical streaming
models indicates that no single model seems to account for all features
of the data.  The simultaneous variation of the kinematic and isophotal
PA with radius, and the implied oval distortion in the stellar disk,
point strongly towards elliptical streaming as the origin of the
large-scale \HI\ velocity residuals.  On the other hand, the
near-orthogonality of the major and minor axes and the azimuthal
dependence of the velocity residuals disagrees with a simple model of
gas flowing in aligned oval orbits and favors a radial inflow model,
although the implied inflow speeds are large ($\sim$40 \kms).
Elliptical streaming does seem to describe the residuals in the inner
CO disk quite well, where the influence of the nuclear bar is strong.
It may be that when applied to the weaker oval distortion, our
elliptical streaming model is a poor description of reality.

In fact, oval gas orbits will not be aligned with the bar at all radii
because of the dissipative nature of the gas \citep{But96}.  The
principal orbits in a barred potential change orientations from
parallel to perpendicular to the bar with each resonance crossing.  The
gas, however, cannot make this transition abruptly without collisions
and dissipation, so it will tend to follow elliptical orbits that
change orientation gradually with radius \citep{San76}.  As shown in
Figure~\ref{ellip}, this will generate a spiral pattern in the gas due
to orbit crowding, similar to what is actually observed in the
\HI\ (Figure~\ref{mom1}).  The velocity residuals also become more
complicated in this situation.  Although there should still be equal
amounts of inflow and outflow across a circular ring, the crowding of
orbits into spiral arms where the radial flow is inward (see
Figure~\ref{ellip}) leads to an observational bias against regions of
outflow, where the gas surface density is low.  This bias may
contribute to the apparent inflow signature in NGC 4736, and since the
direction of radial streaming motions along spiral arms changes across
each resonance, it might also explain the reversals in $v_{\rm exp}$
shown in Figure~\ref{vexp}.  However, it appears doubtful whether
sampling bias alone can account for the strong inflow signature just
outside the ring, since the \HI\ velocity field is well-sampled in
azimuth in the region from $r$=40\arcsec--100\arcsec.  Further work
will therefore be needed to determine whether a model of misaligned
oval orbits can reproduce the kinematics of this region.

%%%%%%%%%%%%%%%%%%%%%%%%%%%%%%%%%%%%%%%%%%%%%%%%%%%%%%%%%%%%%
%%%%%%%%%%%%%%%%%%%%%%%%   FIG. 24  %%%%%%%%%%%%%%%%%%%%%%%%%
%%%%%%%%%%%%%%%%%%%%%%%%%%%%%%%%%%%%%%%%%%%%%%%%%%%%%%%%%%%%%

\vskip 0.25truein
\includegraphics[width=3.25in]{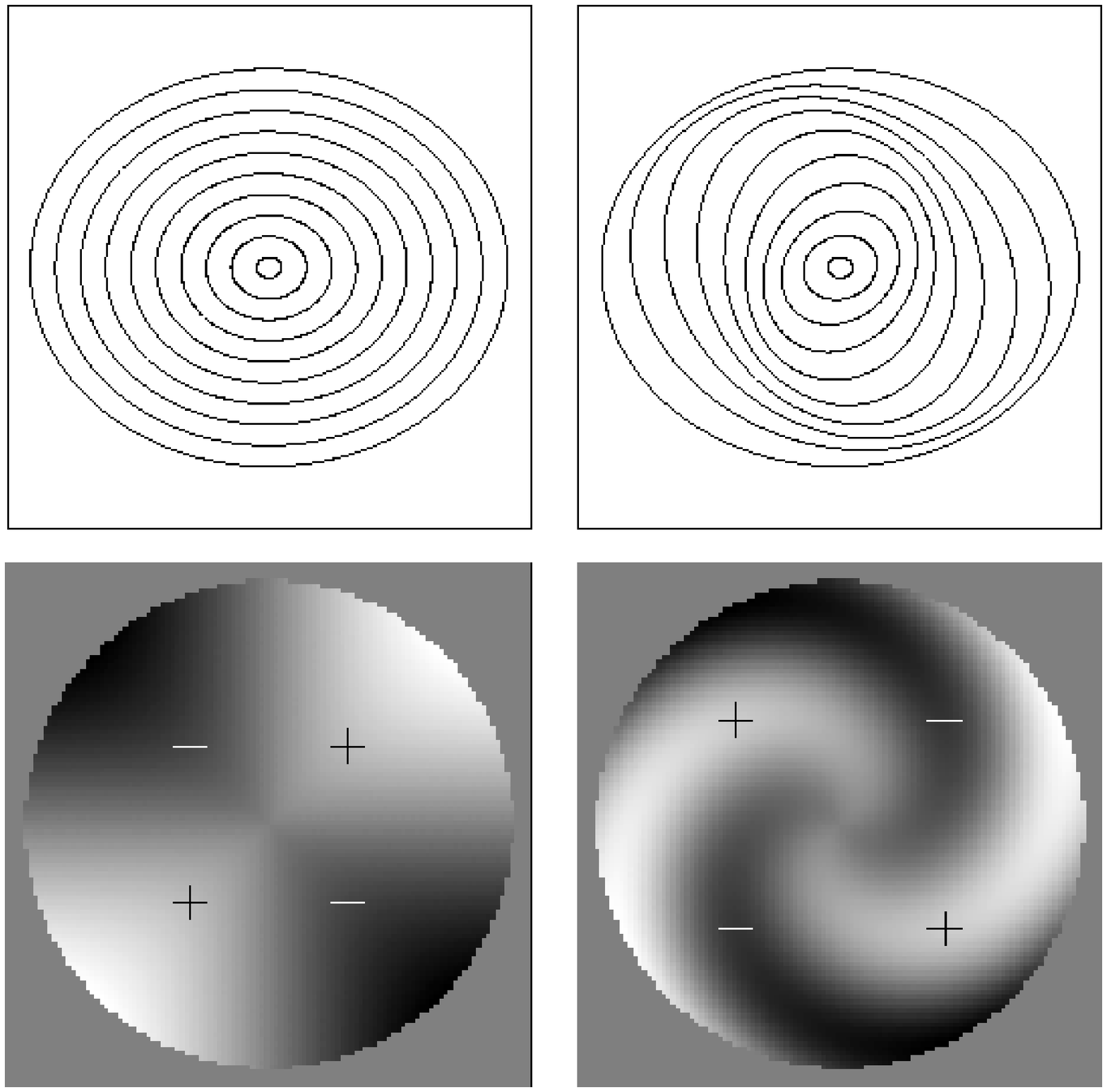}
\figcaption[ellip.bit.ps]{
Orbit orientations ({\it top}) and radial expansion velocity ({\it
bottom}) for two models: ({\it left}) aligned elliptical orbits; ({\it
right}) orbits that have been precessed into a spiral pattern.  The
galaxy plane is viewed face-on, and velocities are measured in the disk
plane, not along the line of sight.  Note that for the aligned orbits,
$v_{\rm exp} \propto \sin 2\theta$, whereas for the precessed orbits
the locations of inflow and outflow form a spiral pattern.  Both models
assume a solid-body rotation curve and clockwise rotation.
\label{ellip}}
\vskip 0.25truein

%%%%%%%%%%%%%%%%%%%%%%%%%%%%%%%%%%%%%%%%%%%%%%%%%%%%%%%%%%%%%

An important point to note is that even if the \HI\ residuals were due
primarily to elliptical streaming in misaligned orbits, a small net
inflow would still result from the torques exerted on the gas by the
oval distortion in the potential \citep{Com88}.  Only orbits that are
aligned parallel or perpendicular to the oval experience no net torque;
gas in misaligned orbits between ILR and CR is forced inward, while
between CR and OLR it is forced outward.  This gradual flow leads to
mass accumulation at the ILR and OLR, but since the flow velocities
required to form and sustain a ring are quite small
(\S\ref{disc_evol}), they may only be observable in locations where
strong shocks occur.

\subsection{A Bar Within Oval System\label{disc_bar}}

The coexistence of a nuclear bar and large-scale oval distortion in NGC
4736 place it among a growing number of galaxies with multiple bars, as
discussed by \citet{Fried93b}.  These authors have conducted detailed
simulations of the interaction and evolution of two bar structures
rotating at different pattern speeds and concluded that significant
radial mass transport can occur.  Whether this scenario is applicable to a
given galaxy, however, depends on the relative orientation of the two
bars.  While parallel or orthogonal nested bars can share the same pattern
speed, for reasons of dynamical stability misaligned bars cannot do so
\citep{Lou88}.  Misaligned bars are therefore a good indication of
different pattern speeds, although the intrinsic orientation of a bar
structure is difficult to determine without prior knowledge of the
inclination and position angle of the galaxy.

In the case of NGC 4736, the nuclear bar is oriented roughly along the
kinematic minor axis of the galaxy, whereas the large-scale oval has a
PA ($\sim$95\arcdeg) that is 20\arcdeg--25\arcdeg\ less than the
kinematic major axis (line of nodes), and will be even further away
when deprojected with our assumed disk parameters (if the oval has an
axis ratio of 0.75 as found by MMG95, its major axis will be
45\arcdeg\ from the kinematic major axis when deprojected.)  Thus the
two bar structures in NGC 4736 appear to be neither aligned nor
orthogonal, and we conclude that they are rotating at different pattern
speeds, in agreement with MMG95.

Observational determination of the actual pattern speeds is difficult, and
most methods have been designed for application to grand-design spiral
galaxies (\citealt{But96} and references therein).  Our choice of
pattern speeds places both the OLR of the central bar and the ILR of the
oval distortion at the ring ($r\approx 45\arcsec$), consistent with 
previous studies \citep[MMG95;][]{Ger91}.  This scenario is supported 
by the following arguments:
\begin{enumerate}
\item Theoretical studies and modeling of observations suggest that
bars end somewhat inside their corotation radius
(\citealt{Cont89}; \citealt{Ath92}; \citealt*{Laine98}).  
In NGC 4736 the nuclear bar extends out
to $r_{bar} \approx 20$\arcsec\ (deprojected), thus a corotation radius of
$r$=26\arcsec\ ($1.3\,r_{bar}$) is in agreement with this general rule.

\item For a flat rotation curve, the ratio of the OLR radius to the
ILR radius is $(\sqrt{2}+1)/(\sqrt{2}-1) = 5.8$, roughly the ratio
of the outer ring radius ($\sim$300\arcsec) to the inner ring radius 
($\sim$50\arcsec).

\item As a result of changes in the orientation of the dominant stable
orbits, the radial direction of spiral arm streaming motions and
gravitational torques changes across each principal resonance
\citep{Com88}.  Corresponding changes in the inflow velocity are
seen in the data (Figure~\ref{vexp}).  Regardless of whether the fitted
$v_{\rm in}$ is due to net inflow or to streaming motions
(see \S\ref{disc_kin}), these changes are suggestive of resonance
crossings.

\item In simulations the region around the CR is depopulated due to
torques, as well as a lack of stable orbits between the CR and OLR
in the case of
strong bars.  In NGC 4736 this is consistent with the gap in CO
brightness around $r$=30\arcsec\ and the gap in optical brightness
from $r$=200\arcsec--300\arcsec\ \citep{Bos77}.
\end{enumerate}
On the other hand, our pattern speed for the nuclear bar conflicts with
evidence from dynamical simulations \citep[e.g.,][]{Tag87} that for
nested bars the ILR of the outer bar corresponds to the CR of the inner
bar.  If this were the case in NGC 4736, it would be harder to
understand the apparent outflow between the bar and the ring, unless
the ring is sandwiched between two ILRs of the oval \citep{Com88},
which is questionable (an inner ILR tends to occur where the rotation
curve flattens, which for this galaxy is well inside the ring) but not
ruled out by the present data.  A stronger constraint could perhaps be
provided by a dynamical (e.g. N-body) simulation of the galaxy, but such
modeling is beyond the scope of this paper.

An intriguing alternative possibility, suggested by the threefold
symmetry evident in the CO distribution, is that the ring corresponds
to both the ILR of the oval and the outer 3:1 resonance of the bar,
where $\Omega_p=\Omega+\kappa/3$.  In this scenario, the arc-like
features at the bar ends would occur at the inner 3:1 resonance, and
the absence of threefold symmetry outside the zone between these
resonances indicates that the $m$=3 waves do not propagate beyond them
\citep{Elm92}.  It is unclear how the direction of radial streaming
motions (discussed in item 3 above) would change in such a model.

As bars are a convenient way to transport angular momentum outward so
material can accrete toward the center \citep{Fried93a}, the existence
of nested bars may provide a solution to the problem of fueling nuclear
activity in galaxies \citep{Shl89}, although observations to date do
not show a significant tendency for active galaxies to contain bars
\citep*{Ho97,Mulch97}.  In such a ``bar-within-bar'' scenario, gas
brought in to the ILR of the outer bar could be driven in further by
a rapidly rotating inner bar.  However, since the inner bar can
only bring in material that is already inside its CR, for our adopted
pattern speed (OLR$_s$=ILR$_p$) it cannot draw in material from the ILR
ring of the outer bar.  Rather, it adds material to the ring by
depleting gas near its CR.  We speculate that this may be a general
property of double-barred galaxies with strong rings, whereas in
systems with pattern speeds such that CR$_s$=ILR$_p$, gas may continue 
on toward the nucleus instead of collecting in a ring. 

\subsection{Molecular Gas and Star Formation\label{disc_mol}}

\subsubsection{Azimuthal differences in the ring\label{disc_molaz}}

The results of \S\ref{sfr} clearly indicate that the link between
molecular gas and massive star formation in NGC 4736 is not as tight as
might na\"{\i}vely be expected.  The contrast in the morphologies of CO
and \Halpha\ along the ring is difficult to account for in terms of
patchy extinction or local variations in the X-factor, as the CO and
\Halpha\ distributions display fundamentally different symmetries.  It
is also difficult to attribute to recent changes in the H$_2$
distribution, since the OB stars responsible for the \Halpha\ emission
are $\lesssim$10 Myr old, which is less than the orbital timescale
(20--30 Myr).  Here we consider the possibility that certain locations
in the ring are more conducive to the formation of massive stars,
irrespective of the total gas content in these regions.

Why would massive stars preferentially form in certain locations in the
ring?  A recent study by \citet*{Croc96} of 32 ringed galaxies revealed
a tendency for \HII\ regions in intrinsically oval rings to be
``bunched up'' along the major axis of the ring, presumably where gas
slows down in its orbits.  Using our assumed disk parameters and the
parameters of the ring given by MMG95, the intrinsic major axis of the
ring has a PA of 320\arcdeg\ in the sky, which appears to be too large
to account for the concentration of \Halpha\ emission at PA
260\arcdeg--305\arcdeg.  A second mechanism for producing strongly
bisymmetric star formation in rings is the crowding together of gas in
orbits associated with the bar and ring respectively to produce a
``twin peaks'' morphology \citep{Ken92}.  Star formation would then be
enhanced in these gas concentrations.  This picture may be more
appropriate for NGC 4736: the bright \HII\ regions west of the nucleus
are just downstream of where the outer CO arm (which may trace orbits
associated with the oval distortion) intersects the ring
(Figure~\ref{coha}).  Either of these mechanisms could be aided by the
tendency for a self-gravitating ring to collapse azimuthally as well as
radially: \citet{Elm94a} has analyzed the growth of azimuthal
perturbations in a magnetic ring and found that the fastest growing
wavelength is roughly 4 times the radial thickness of the ring.

In all of these scenarios, however, peaks in molecular gas are expected
to correspond to peaks in star formation, so the problem of the different
azimuthal symmetries in CO and \Halpha\ remains unresolved.
One solution for NGC 4736 would be for the distribution of
{\it dense} molecular gas, as traced by HCN(1--0) emission, to resemble
the bisymmetric \Halpha\ distribution rather than the CO, since star
formation will presumably be concentrated in the densest gas.  In NGC
6951, for instance, \citet*{Koh99} found that the HCN peaks are shifted
downstream of the CO peaks, and agree better with the positions of
\Halpha\ and radio continuum emission than the CO does.  To investigate
this possibility for NGC 4736, relatively deep observations will be
needed: short BIMA observations at 18\arcsec\ resolution detect no HCN
emission at a 3$\sigma$ upper limit of 0.15 Jy bm$^{-1}$ (0.1 K).

Alternatively, the triggering of massive star formation may be enhanced
by the compression or collision of gas clouds, as has been suggested by
\citet*{Sco86}.  Since star formation is observed to be an inefficient
process \citep{Myers86}, the total gas mass may be less important if
forces besides self-gravity can aid in collapse.  In starburst rings,
convergent gas flow in regions where streamlines intersect may provide
the necessary trigger, resulting in the formation of unusually compact
clusters \citep[e.g.][]{Barth95}.  Whether cloud collisions tend to
lead to the formation of gravitationally bound clouds or to cloud
disruption is a controversial issue \citep{Lar88}, but simulations of
cloud collisions which take into account the clumpy structure of
molecular clouds succeed in producing gravitational instabilities
\citep{Kim96}.  The notion that gas kinematics can affect the SFE is
also in accord with star formation patterns observed in galactic bars
(see \S\ref{disc_molrad} below) and merging galaxies \citep*{Mihos93}.

\subsubsection{Radial differences\label{disc_molrad}}

The lack of star formation interior to the ring, despite the large
quantities of molecular gas in the inner disk, is not predicted by
Schmidt-type laws ($\Sigma_{\rm SFR} \propto \Sigma_{\rm gas}^N$), or
even alternative formulations where the SFR scales with the orbital
frequency \citep{Wyse89} or the stellar surface density \citep{Dop94},
both of which peak toward the center.  As discussed in \S\ref{profile},
radial variations in the X-factor due to a metallicity gradient may
lead us to overestimate the H$_2$ mass in this region, but for a
realistic metallicity gradient this is unlikely to lead to the sharp
rise in \tgas\ observed inside the ring (Figure~\ref{taugas}).  While a
dramatic increase in extinction is possible, in \S\ref{sfr_cal} we
argued that obscured massive star formation was unlikely to be
widespread in the inner disk.  Here we consider three other factors
which could strongly affect either the SFE or the X-factor in the inner
disk: (1) inhibition of star formation due to the kinematic effects of
the bar; (2) a high critical density for gravitational instability; or
(3) a change in both the X-factor and star formation efficiency due to
increased velocity dispersions.

\begin{enumerate}

\item A depressed level of star formation along galactic bars has been
noted for some time \citep[for recent work see][]{Phil96}, with massive
star formation tending to concentrate at the bar ends or in
circumnuclear rings, especially in early-type galaxies.
\citet{Tubbs82} has suggested that star formation is inhibited along
bars because clouds are disrupted when they hit the bar shock.
Alternatively, the short orbital timescales in the vicinity of the
nuclear bar may prevent clouds from collapsing gravitationally before
being driven apart by diverging streamlines as they exit the bar.
Observations of the galaxies NGC 1097 and NGC 6574, both of which
contain abundant molecular gas associated with a nuclear bar but 
relatively little star formation \citep{Kot00}, also point to the 
presence of a bar being a key factor.

\item The deficit of star formation in the inner disk may be related to
the large critical density for gravitational instability ($\Sigma_{\rm
crit}$).  This density typically increases towards a galaxy's center
due to the stabilizing effects of rapid rotation, and can lead to very
large gas densities being required before star formation is triggered.
Once star formation commences, it proceeds at a rapid rate due to the
nonlinearity of the Schmidt law, creating a nuclear starburst
\citep{KC89}.  The large values of $Q$ in the nuclear region of NGC
4736, as well as evidence for a past starburst, are consistent with
this scenario.  Of course, $Q$ may not be a good measure of
gravitational instability in the vicinity of a bar where the potential
is far from axisymmetric.

\item The CO(1--0) line may have low to moderate ($\tau \sim 1$)
optical depth, due to high velocity dispersions (and possibly
excitation temperatures) in the molecular gas.  Such a situation has
been inferred in the Galactic Center by \citet{Dah98} based on large
$^{12}$CO/C$^{18}$O ratios, and a similar conclusion was reached by
\citet{Aalto95} for the central regions of starburst galaxies.  High
velocity dispersions are especially likely in the vicinity of a nuclear
bar.  A low CO optical depth will result in the CO intensity
overestimating the H$_2$ column density when a standard X-factor is
applied.  Furthermore, since high velocity dispersions would raise the
virialization density, there is likely to be a greater fraction of CO
emission that is not associated with virialized clouds and is hence
unrelated to star formation.  Both of these factors would contribute to
the high CO/\Halpha\ ratio in the inner disk of NGC 4736.

\end{enumerate}

All of these explanations seem plausible based on the existing
evidence.  Further studies of nuclear and ring starbursts would be
useful for understanding what factors are most important in triggering
or inhibiting star formation near the nucleus.  The recent work of
\citet{Jogee99}, for instance, suggests that the suppression of star
formation in regions of strong non-circular kinematics may be quite
common.  In addition, much remains to be learned about how star
formation occurs in barred galaxies and in the Galactic Center, and
high-resolution observations of $^{13}$CO in NGC 4736 should be
undertaken to probe the conditions of the molecular gas in both the
inner disk and the ring.

\subsection{Atomic Gas and Star Formation\label{disc_at}}

The similarities in the \HI\ and \Halpha\ radial (\S\ref{sfr_law}) and
azimuthal (\S\ref{sfr_az}) distributions in the ring, where most of the
gas is molecular, suggest that \HI\ there is largely a product of H$_2$
dissociation.  Similar conclusions have been reached for the
grand-design spiral galaxies M51 \citep{Vogel88,Til89} and M83
\citep*{Allen86}, in regions where the ISM is also predominantly
molecular.  On the other hand, the azimuthal profiles of \HI\ and
\Halpha\ do not agree in detail, even though both are bisymmetric.
This may result from the depletion of \HI\ in favor of \HII\ near the
brightest \HII\ regions, or the presence of an older stellar component
in the ring that contributes to photodissociation but not photoionization.
A direct probe of photodissociation regions is
provided by observations of the [\ion{C}{2}] line at 158 $\mu$m, which
\citet{Stac91} have used to argue that photodissociated gas constitutes
a large fraction of the gas mass in star-forming galaxies.  However,
the resolution of their observations (55\arcsec) was too low to allow a
detailed comparison of [\ion{C}{2}] and \HI.  Future high-resolution
observations of [\ion{C}{2}] by the SOFIA airborne observatory will
provide a strong test of this model.

If this interpretation is correct, then the strong correlation between
the \HI\ surface density ($\Sigma_{\rm HI}$) and SFR per unit area 
($\Sigma_{\rm SFR}$) found in this study and in
previous work \citep[][and references therein]{KC98a} is simply a
result of the \HI\ being produced by star formation \citep{Shaya87}.
This may only apply in the inner, H$_2$ dominated regions of massive
spirals: in the outer regions the preferred phase for the neutral gas
will be atomic \citep[e.g.,][]{Elm93}, and one would expect the
formation of dense \HI\ clouds to precede (rather than follow) the
formation of molecular clouds and stars, as is apparently the case in
the Perseus arm of our Galaxy \citep{Heyer98}.  Hence the correlation
between \HI\ and SFR should break down in the outer parts of disks,
which may explain why the correlation between the {\it total}
\HI\ content and SFR in galaxies is comparatively poor \citep{Young89}.

\subsection{Star Formation at Subcritical Gas Densities\label{disc_crit}}

As shown in \S\ref{sfr_crit}, massive star formation in NGC 4736
appears to be occurring at gas densities that, when azimuthally
averaged, are a factor of $\sim$3 below the expected threshold density
for gravitational instability.  A similar conclusion had been
tentatively reached by \citet{Shi98} based on lower resolution data.
Hence star formation is evidently quite vigorous at $Q \sim 3$, whereas
\citet{KC89} found for his sample that star formation was suppressed
when $Q>1.5$.  Previous studies of the low surface brightness galaxies
M33 \citep*{Wils91} and NGC 2403 \citep{Thorn95} have also found star
formation occurring at ``subcritical'' gas densities, but in contrast
NGC 4736 is a bright, early-type galaxy.

One should bear in mind, however, that the derived value of $Q$ is
subject to considerable uncertainty because it is the product of
several measured quantities.  Since $Q \propto
\kappa\sigma_v/\Sigma_{\rm gas}$, we could overestimate it by
overestimating $\kappa$ or $\sigma_v$ or underestimating $\Sigma_{\rm
gas}$.  The errors in $\kappa$ are formally $\sim$30\%, but they are
harder to quantify for $\sigma_v$ from our rough analysis (although
$\sigma_v \ll 7$ \kms\ seems unrealistic) and largely unknown for
$\Sigma_{\rm gas}$, which relies on the X-factor.  Since comparable
or greater uncertainties are likely to exist for the data used by 
\citet{KC89}, a factor of $\sim$2 discrepancy between our
results may not be particularly meaningful.

Even if the basic $Q<1$ criterion does break down in this galaxy, this
need not imply that star formation has occurred in regions that were
gravitationally stable.  Rather, it may just be an indication that a
one-parameter model is an incomplete description of how gravitational
instability ultimately leads to star formation.  As mentioned in
\S\ref{sfr_crit}, there are molecular condensations in the ring where
$Q<1$ locally.  Such condensations probably resulted from processes
other than axisymmetric gravitational instability, but their subsequent
collapse to form stars may still be governed in part by $Q$.  Moreover,
the derivation of $Q$ assumes a thin isothermal one-component disk;
more realistic models show that the interaction of gas and stars
creates greater instability \citep{Jog84,Elm95}.  Finally, we emphasize
that there remains considerable uncertainty about the physical basis
for the $Q$ threshold \citep[see e.g.][]{Lar88}, in particular whether
it is in fact related to an axisymmetric instability ($Q$ threshold
$\sim 1$) or to swing amplification of density waves ($Q$ threshold
1--2).

Considering both the observational and theoretical uncertainties, it is
remarkable how well the location of the dip in $Q$ matches the location
of the ring (Figure~\ref{qprof}).  We conclude that our data support
the general picture of a link between star formation and gravitational
instability in NGC 4736, although with some uncertainty in the exact
value of the $Q$ threshold.  However, firm conclusions on the general
applicability of the star formation threshold can only be drawn once a
large sample of galaxies has been examined in detail.

\subsection{Radial Gas Flows and Galaxy Evolution\label{disc_evol}}

Star formation is occurring in the ring at a rate which can only be
sustained for $\sim$1 Gyr (Figure~\ref{taugas}).  In fact,
Figure~\ref{taugas} already gives some indication that substantial gas
consumption has occurred, in that the positions of the CO and
\Halpha\ rings are not coincident.  Based on the current SFR, roughly
0.2 \Msol\ yr$^{-1}$ of additional gas needs to be added to the ring
($30\arcsec<r<60\arcsec$) to replenish the gas used by star formation
\citep{Smi91}.  (This is probably a lower limit since extinction has
not been taken into account in deriving the SFR.)  This level of
refueling could be accomplished by an outflow velocity of approximately
1 \kms\ at $r$=30\arcsec\ or a comparable inflow velocity at
$r$=60\arcsec, based on the gas surface densities at those radii.  One
can equivalently view this as the inflow velocity required to build up
the ring on a timescale of $\sim$1 Gyr, in the absence of star
formation.  Note that replenishment requires much smaller radial flows
than those predicted by the radial inflow model (Figure~\ref{vexp}),
which would tend to rapidly build up the ring.

Our estimate of the gas consumption time is admittedly simplistic,
since it neglects the effects of stellar recycling, which can increase
gas consumption times by factors of $\sim$2 \citep{KC94}, and is quite
sensitive to uncertainties in the IMF, extinction, X-factor, etc.
However, it does indicate that even small radial flow rates can have a
dramatic impact on extending the lifetimes of star-forming rings.  Thus,
as long as bars continue to drive radial gas flows, star formation in
rings can probably continue indefinitely.  The mass concentration in
the ring inferred from the rotation curve (\S\ref{kin_rot}) indicates
that it is not a very young feature.  On the other hand, it does not
appear old enough to have strongly affected the $K$-band light profile
(\S\ref{profile}).  A detailed photometric study aimed at determining
the approximate age of the ring, as has been performed by \citet{But91}
for NGC 7702, would be quite useful in understanding its evolution.

How did a strong central peak in the H$_2$ surface density arise, if gas
inflow stops at the ring (ILR)?  Continued mass loss from stars in the
bulge may build up significant amounts of gas, which the nuclear bar
then redistributes toward the ring and nucleus.  Star formation might
be active in the ring because of the lower critical density (small
$\kappa$) there, but be sporadic at the nucleus because the
critical density there is much higher.  This is consistent with the
spectroscopic evidence that the nucleus has undergone a recent
starburst event.  Alternatively, gas could have been driven directly
into the central region by the oval early in its evolution, before
matter became centrally concentrated enough to produce an ILR.

%% CONCLUSIONS

\section{Conclusions}\label{conc}

We have presented an analysis of CO, \HI, and \Halpha\ data for 
the ringed galaxy NGC 4736.  Our main results can be summarized 
as follows:

{\it Gas distribution.}---A 
molecular counterpart is seen to the nuclear stellar bar, as has been
previously reported by \citet{Sak99}.  Arm-like extensions from the bar
ends, probably associated with a spiral dust arc seen in color index
maps, connect the inner molecular disk with the star-forming ring,
which is associated with tightly wound CO arms.  Overall, the
distribution of neutral gas appears to be dominated by \HI\ outside the
ring and H$_2$ interior to the ring.  The \HI\ intensity along the ring
shows a strong twofold symmetry which is completely lacking in the
CO.\@ Instead, the CO shows indications of a threefold symmetry, both
in the ring and the inner disk.  The radial distribution of gas does
not follow a simple exponential but shows distinct humps, possibly due
to mass redistribution by radial flows.  The stellar profile derived
from a $K$-band image is steeper than the gas profile within
$r$=60\arcsec.

{\it Gas kinematics.}---The
rotation curve is consistent with a scenario in which the ring
corresponds to the OLR of the nuclear bar and the ILR of the
large-scale oval distortion.  Such a coupling of nested bars may
contribute to the accumulation of gas in the ring.  The directions of
radial gas flow suggested by the velocity residuals are also in good agreement
with this picture.  The influence of the nuclear bar on the kinematics
of the inner disk is readily apparent in the CO velocity residuals,
which show the expected $\cos 3\theta$ signature for streaming in
aligned elliptical orbits.  The \HI\ velocity residuals outside the
ring do not show this signature, resembling instead a model of uniform
radial inflow.  However, the very high ($\sim$40 \kms) inflow
velocities inferred raise difficulties for this model as well.  A more
realistic model in which elliptical orbits are precessed (i.e.,
change orientation with radius) may warrant further study.

{\it Star formation.}---The
radial distributions of CO and \Halpha\ are radically different and
are consistent with a simple Schmidt law only when the inner CO disk is
excluded.  Some possible explanations for this discrepancy are
inhibition of star formation in the vicinity of the nuclear bar, a high
threshold density for gravitational instability in the central regions,
or a change in the X-factor related to a change in CO(1--0) opacity.
The correlation between \HI\ and \Halpha\ emission and their similar
azimuthal profiles suggest that \HI\ may be largely a product of H$_2$
dissociation, at least in regions that are dominated by molecular gas.
We find that the Toomre $Q$ parameter has a local minimum of $\sim$3 at
the location of the star-forming ring, when derived from azimuthally
averaged gas densities.  This minimum is in general agreement with the
theory in which star formation is initiated by global gravitational
instabilities, although $Q$ is expected to be a factor of $\sim$2
smaller than observed.  The azimuthal \Halpha\ profile, unlike the CO profile,
displays a strong twofold symmetry, indicating that massive star
formation is enhanced at certain locations in the ring, independent
of the molecular gas mass.  Such locations may be associated with
dense gas not traced by CO or with convergent gas flows.  We conclude
that the star formation rate is not determined solely by the available
gas mass, but rather that large-scale dynamics play a significant role
in organizing and possibly triggering star formation.

%% ACKNOWLEDGMENTS

\acknowledgments

We thank the referee, Jeff Kenney, for insightful comments and
our collaborators on the BIMA SONG team for helpful
discussions and for use of SONG data prior to publication.  Special
thanks go to Tamara Helfer for her assistance in the early phases of
data taking and analysis.  We thank Axel Wei{\ss} for fruitful
discussions on combining single-dish and interferometer data.  We also
thank Peter Teuben for his assistance with the kinematical modeling,
and the many researchers who contributed data for this project (Robert
Braun, Rosa Gonz\'{a}lez Delgado, Neb Duric, Maryvonne Gerin, Rob
Kennicutt, Claus M\"{o}llenhoff, and Rick Pogge).  This research was
supported by NSF grant AST 9613998 to the UC Berkeley Radio Astronomy
Laboratory.  T. W. acknowledges support from a Phi Beta Kappa graduate
fellowship.  We have made extensive use of NASA's Astrophysics Data
System Abstract Service (ADS), as well as the the NASA/IPAC
Extragalactic Database (NED) which is operated by the Jet Propulsion
Laboratory, California Institute of Technology, under contract with the
National Aeronautics and Space Administration.  This publication also
makes use of data products from the Two Micron All Sky Survey, which is
a joint project of the University of Massachusetts and the Infrared
Processing and Analysis Center, funded by the National Aeronautics and
Space Administration and the National Science Foundation.

%% BIBLIOGRAPHY

\bibliographystyle{apj}
\bibliography{tony}

\begin{thebibliography}{123}
\expandafter\ifx\csname natexlab\endcsname\relax\def\natexlab#1{#1}\fi

\bibitem[{Aalto {et~al.}(1995)Aalto, Booth, Black, \& Johansson}]{Aalto95}
Aalto, S., Booth, R.~S., Black, J.~H., \& Johansson, L. E.~B. 1995, A\&A, 300,
  369

\bibitem[{Allen {et~al.}(1986)Allen, Atherton, \& Tilanus}]{Allen86}
Allen, R.~J., Atherton, P.~D., \& Tilanus, R. P.~J. 1986, \nat, 319, 296

\bibitem[{Athanassoula(1992)}]{Ath92}
Athanassoula, E. 1992, MNRAS, 259, 345

\bibitem[{Barth {et~al.}(1995)Barth, Ho, Filippenko, \& Sargent}]{Barth95}
Barth, A.~J., Ho, L.~C., Filippenko, A.~V., \& Sargent, W. L.~W. 1995, AJ, 110,
  1009

\bibitem[{Becker {et~al.}(1995)Becker, White, \& Helfand}]{Bec95}
Becker, R.~H., White, R.~L., \& Helfand, D.~J. 1995, ApJ, 450, 559

\bibitem[{Begeman(1989)}]{Beg89}
Begeman, K.~G. 1989, A\&A, 223, 47

\bibitem[{Benedict {et~al.}(1996)Benedict, Smith, \& Kenney}]{Ben96}
Benedict, G.~F., Smith, B.~J., \& Kenney, J. D.~P. 1996, AJ, 111, 1861

\bibitem[{Block {et~al.}(1994{\natexlab{a}})Block, Bertin, Stockton,
  Grosb{\o}l, Moorwood, \& Peletier}]{Block94a}
Block, D.~L., Bertin, G., Stockton, A., Grosb{\o}l, P., Moorwood, A. F.~M., \&
  Peletier, R.~F. 1994{\natexlab{a}}, A\&A, 288, 365

\bibitem[{Block {et~al.}(1994{\natexlab{b}})Block, Witt, Grosb{\o}l, Stockton,
  \& Moneti}]{Block94b}
Block, D.~L., Witt, A.~N., Grosb{\o}l, P., Stockton, A., \& Moneti, A.
  1994{\natexlab{b}}, A\&A, 288, 383

\bibitem[{{Bloemen} {et~al.}(1986){Bloemen}, {Strong}, {Mayer-Hasselwander},
  {Blitz}, {Cohen}, {Dame}, {Grabelsky}, {Thaddeus}, {Hermsen}, \&
  {Lebrun}}]{Bloem86}
{Bloemen}, J. B. G.~M., {Strong}, A.~W., {Mayer-Hasselwander}, H.~A., {Blitz},
  L., {Cohen}, R.~S., {Dame}, T.~M., {Grabelsky}, D.~A., {Thaddeus}, P.,
  {Hermsen}, W., \& {Lebrun}, F. 1986, A\&A, 154, 25

\bibitem[{Bohlin {et~al.}(1978)Bohlin, Savage, \& Drake}]{Bohl78}
Bohlin, R.~C., Savage, B.~D., \& Drake, J.~F. 1978, ApJ, 224, 132

\bibitem[{Bosma(1978)}]{Bos78}
Bosma, A. 1978, PhD thesis, Univ.\ of Groningen

\bibitem[{Bosma {et~al.}(1977)Bosma, {van der Hulst}, \& Sullivan}]{Bos77}
Bosma, A., {van der Hulst}, J.~M., \& Sullivan, W.~T. 1977, A\&A, 57, 373

\bibitem[{Braun(1995)}]{Brn95}
Braun, R. 1995, A\&AS, 114, 409

\bibitem[{Braun(1997)}]{Brn97}
---. 1997, ApJ, 484, 637

\bibitem[{Briggs(1995)}]{Briggs95}
Briggs, D. 1995, PhD thesis, New Mexico Inst.\ of Mining and Tech.

\bibitem[{Burton \& Liszt(1978)}]{Burt78}
Burton, W.~B. \& Liszt, H.~S. 1978, ApJ, 225, 815

\bibitem[{Buta(1988)}]{But88}
Buta, R. 1988, ApJS, 66, 233

\bibitem[{Buta(1991)}]{But91}
---. 1991, ApJ, 370, 130

\bibitem[{Buta \& Combes(1996)}]{But96}
Buta, R. \& Combes, F. 1996, Fund.~Cosmic Phys., 17, 95

\bibitem[{{Byrd} {et~al.}(1994){Byrd}, {Rautiainen}, {Salo}, {Buta}, \&
  {Crocker}}]{Byrd94}
{Byrd}, G., {Rautiainen}, P., {Salo}, H., {Buta}, R., \& {Crocker}, D.~A. 1994,
  AJ, 108, 476

\bibitem[{Combes(1988)}]{Com88}
Combes, F. 1988, in Galactic and Extragalactic Star Formation, ed. R.~E.
  Pudritz \& M.~Fich (Dordrecht: Kluwer), 475

\bibitem[{Combes \& Gerin(1985)}]{Com85}
Combes, F. \& Gerin, M. 1985, A\&A, 150, 327

\bibitem[{Contopoulos {et~al.}(1989)Contopoulos, Gottesman, Hunter, \&
  England}]{Cont89}
Contopoulos, G., Gottesman, S.~T., Hunter, J.~H., \& England, M.~N. 1989, ApJ,
  343, 608

\bibitem[{Crocker {et~al.}(1996)Crocker, Baugus, \& Buta}]{Croc96}
Crocker, D.~A., Baugus, P.~D., \& Buta, R. 1996, ApJS, 105, 353

\bibitem[{Cui {et~al.}(1997)Cui, Feldkhun, \& Braun}]{Cui97}
Cui, W., Feldkhun, D., \& Braun, R. 1997, ApJ, 477, 693

\bibitem[{Dahmen {et~al.}(1998)Dahmen, H\"{u}ttemeister, Wilson, \&
  Mauersberger}]{Dah98}
Dahmen, G., H\"{u}ttemeister, S., Wilson, T.~L., \& Mauersberger, R. 1998,
  A\&A, 331, 959

\bibitem[{{de Vaucouleurs} \& Buta(1980)}]{dV80}
{de Vaucouleurs}, G. \& Buta, R. 1980, ApJS, 44, 451

\bibitem[{{de Vaucouleurs} {et~al.}(1991){de Vaucouleurs}, {de Vaucouleurs},
  Corwin, Buta, Paturel, \& Fouqu\'{e}}]{RC3}
{de Vaucouleurs}, G., {de Vaucouleurs}, A., Corwin, H.~G., Buta, R.~J.,
  Paturel, G., \& Fouqu\'{e}, P. 1991, Third Reference Catalogue of Bright
  Galaxies (New York: Springer-Verlag)

\bibitem[{Dopita(1985)}]{Dop85}
Dopita, M.~A. 1985, ApJ, 295, L5

\bibitem[{Dopita \& Ryder(1994)}]{Dop94}
Dopita, M.~A. \& Ryder, S.~D. 1994, ApJ, 430, 163

\bibitem[{Duric \& Dittmar(1988)}]{Dur88}
Duric, N. \& Dittmar, M.~R. 1988, ApJ, 332, L67

\bibitem[{Elmegreen(1993)}]{Elm93}
Elmegreen, B.~G. 1993, ApJ, 411, 170

\bibitem[{Elmegreen(1994)}]{Elm94a}
---. 1994, ApJ, 425, L73

\bibitem[{Elmegreen(1995)}]{Elm95}
---. 1995, MNRAS, 275, 944

\bibitem[{Elmegreen {et~al.}(1992)Elmegreen, Elmegreen, \& Montenegro}]{Elm92}
Elmegreen, B.~G., Elmegreen, D.~M., \& Montenegro, L. 1992, ApJS, 79, 37

\bibitem[{Friedli \& Benz(1993)}]{Fried93a}
Friedli, D. \& Benz, W. 1993, A\&A, 268, 65

\bibitem[{Friedli \& Martinet(1993)}]{Fried93b}
Friedli, D. \& Martinet, L. 1993, A\&A, 277, 27

\bibitem[{Gerin {et~al.}(1991)Gerin, Casoli, \& Combes}]{Ger91}
Gerin, M., Casoli, F., \& Combes, F. 1991, A\&A, 251, 32

\bibitem[{{Gonz\'{a}lez Delgado} {et~al.}(1997)}]{Gon97}
{Gonz\'{a}lez Delgado}, R.~M. {et~al.} 1997, ApJS, 108, 155

\bibitem[{Heckman(1980)}]{Hek80}
Heckman, T.~M. 1980, A\&A, 87, 152

\bibitem[{Heyer \& Terebey(1998)}]{Heyer98}
Heyer, M.~H. \& Terebey, S. 1998, ApJ, 502, 265

\bibitem[{Ho {et~al.}(1997)Ho, Filippenko, \& Sargent}]{Ho97}
Ho, L.~C., Filippenko, A.~V., \& Sargent, W. L.~W. 1997, ApJ, 487, 591

\bibitem[{Huchtmeier \& Seiradakis(1985)}]{Hucht85}
Huchtmeier, W.~K. \& Seiradakis, J.~H. 1985, A\&A, 143, 216

\bibitem[{Jog \& Solomon(1984)}]{Jog84}
Jog, C.~J. \& Solomon, P.~M. 1984, ApJ, 276, 114

\bibitem[{Jogee(1999)}]{Jogee99}
Jogee, S. 1999, PhD thesis, Yale University

\bibitem[{Kenney(1994)}]{Ken94}
Kenney, J. D.~P. 1994, in Mass-Transfer Induced Activity in Galaxies, ed.
  I.~Shlosman (Cambridge: Cambridge U. Press), 78

\bibitem[{Kenney(1997)}]{Ken97}
Kenney, J. D.~P. 1997, in The Interstellar Medium in Galaxies, ed. J.~M.
  van~der Hulst (Dordrecht: Kluwer), 33

\bibitem[{Kenney {et~al.}(1993)Kenney, Carlstrom, \& Young}]{Ken93}
Kenney, J. D.~P., Carlstrom, J.~E., \& Young, J.~S. 1993, ApJ, 418, 687

\bibitem[{Kenney {et~al.}(1991)Kenney, Scoville, \& Wilson}]{Ken91a}
Kenney, J. D.~P., Scoville, N.~Z., \& Wilson, C.~D. 1991, ApJ, 366, 432

\bibitem[{Kenney {et~al.}(1992)Kenney, Wilson, Scoville, Devereux, \&
  Young}]{Ken92}
Kenney, J. D.~P., Wilson, C.~D., Scoville, N.~Z., Devereux, N.~A., \& Young,
  J.~S. 1992, ApJ, 395, L79

\bibitem[{Kennicutt(1989)}]{KC89}
Kennicutt, R.~C. 1989, ApJ, 344, 685

\bibitem[{Kennicutt(1998{\natexlab{a}})}]{KC98a}
---. 1998{\natexlab{a}}, ApJ, 498, 541

\bibitem[{Kennicutt(1998{\natexlab{b}})}]{KC98b}
---. 1998{\natexlab{b}}, ARA\&A, 36, 189

\bibitem[{Kennicutt {et~al.}(1994)Kennicutt, Tamblyn, \& Congdon}]{KC94}
Kennicutt, R.~C., Tamblyn, P., \& Congdon, C.~W. 1994, ApJ, 435, 22

\bibitem[{Kimura \& Tosa(1996)}]{Kim96}
Kimura, T. \& Tosa, M. 1996, A\&A, 308, 979

\bibitem[{Kohno {et~al.}(1999)Kohno, Kawabe, \& Vila-Vilar\'{o}}]{Koh99}
Kohno, K., Kawabe, R., \& Vila-Vilar\'{o}, B. 1999, ApJ, 511, 157

\bibitem[{Kormendy(1982)}]{Korm82}
Kormendy, J. 1982, in Morphology and Dynamics of Galaxies, ed. L.~Martinet \&
  M.~Mayor (Sauverny: Geneva Obs.), 113

\bibitem[{Kotilainen {et~al.}(2000)Kotilainen, Reunanen, Laine, \&
  Ryder}]{Kot00}
Kotilainen, J.~K., Reunanen, J., Laine, S., \& Ryder, S.~D. 2000, A\&A, 353,
  834

\bibitem[{Kutner \& Ulich(1981)}]{Kut81}
Kutner, M.~L. \& Ulich, B.~L. 1981, ApJ, 250, 341

\bibitem[{Laine {et~al.}(1998)Laine, Shlosman, \& Heller}]{Laine98}
Laine, S., Shlosman, I., \& Heller, C.~H. 1998, MNRAS, 297, 1052

\bibitem[{Larson(1988)}]{Lar88}
Larson, R. 1988, in Galactic and Extragalactic Star Formation, ed. R.~E.
  Pudritz \& M.~Fich (Dordrecht: Kluwer), 459

\bibitem[{Liszt \& Burton(1980)}]{Liszt80}
Liszt, H.~S. \& Burton, W.~B. 1980, ApJ, 236, 779

\bibitem[{Louis \& Gerhard(1988)}]{Lou88}
Louis, P.~D. \& Gerhard, O.~E. 1988, MNRAS, 233, 337

\bibitem[{Maoz {et~al.}(1995)Maoz, Filippenko, Ho, Rix, Bahcall, Schneider, \&
  Macchetto}]{Maoz95}
Maoz, D., Filippenko, A.~V., Ho, L.~C., Rix, H.-W., Bahcall, J.~N., Schneider,
  D.~P., \& Macchetto, F.~D. 1995, ApJ, 440, 91

\bibitem[{Mihos {et~al.}(1993)Mihos, Bothun, \& Richstone}]{Mihos93}
Mihos, J.~C., Bothun, G.~D., \& Richstone, D.~O. 1993, ApJ, 418, 82

\bibitem[{M\"{o}llenhoff {et~al.}(1995)M\"{o}llenhoff, Matthias, \&
  Gerhard}]{Mol95}
M\"{o}llenhoff, C., Matthias, M., \& Gerhard, O.~E. 1995, A\&A, 301, 359 (MMG95)

\bibitem[{Mulchaey \& Regan(1997)}]{Mulch97}
Mulchaey, J.~S. \& Regan, M.~W. 1997, ApJ, 482, L135

\bibitem[{Mulder(1995)}]{Mul95}
Mulder, P.~S. 1995, A\&A, 303, 57

\bibitem[{Mulder \& {van Driel}(1993)}]{Mul93}
Mulder, P.~S. \& {van Driel}, W. 1993, A\&A, 272, 63

\bibitem[{Myers {et~al.}(1986)Myers, Dame, Thaddeus, Cohen, Silverberg, Dwek,
  \& Hauser}]{Myers86}
Myers, P.~C., Dame, T.~M., Thaddeus, P., Cohen, R.~S., Silverberg, R.~F., Dwek,
  E., \& Hauser, M.~G. 1986, ApJ, 301, 398

\bibitem[{Oey \& Kennicutt(1993)}]{Oey93}
Oey, M.~S. \& Kennicutt, R.~C. 1993, ApJ, 411, 137

\bibitem[{Phillips(1996)}]{Phil96}
Phillips, A.~C. 1996, in Barred Galaxies, ASP Conference Series Vol.\ 91, ed.
  R.~Buta, D.~A. Crocker, \& B.~G. Elmegreen (San Francisco: ASP), 44

\bibitem[{Pogge(1989)}]{Pog89}
Pogge, R.~W. 1989, ApJS, 71, 433

\bibitem[{Pritchet(1977)}]{Prit77}
Pritchet, C. 1977, ApJS, 35, 397

\bibitem[{Quillen {et~al.}(1995)Quillen, Frogel, Kenney, Pogge, \&
  DePoy}]{Qui95}
Quillen, A.~C., Frogel, J.~A., Kenney, J. D.~P., Pogge, R.~W., \& DePoy, D.~L.
  1995, ApJ, 441, 549

\bibitem[{Quirk(1972)}]{Quirk72}
Quirk, W.~J. 1972, ApJ, 176, 9

\bibitem[{Regan {et~al.}(1997)Regan, Vogel, \& Teuben}]{Reg97}
Regan, M.~W., Vogel, S.~N., \& Teuben, P.~J. 1997, ApJ, 482, L143

\bibitem[{Rieke \& Lebofsky(1985)}]{Rieke85}
Rieke, G.~H. \& Lebofsky, M.~J. 1985, ApJ, 288, 618

\bibitem[{Roberts(1969)}]{Rob69}
Roberts, W.~W. 1969, ApJ, 158, 123

\bibitem[{Roche \& Aitken(1985)}]{Roche85}
Roche, P.~F. \& Aitken, D.~K. 1985, MNRAS, 213, 789

\bibitem[{Safranov(1960)}]{Saf60}
Safranov, V.~S. 1960, Ann.\ d'Ap., 23, 979

\bibitem[{Sakamoto {et~al.}(1995)Sakamoto, Okumura, Minezaki, Kobayashi, \&
  Wada}]{Sak95}
Sakamoto, K., Okumura, S., Minezaki, T., Kobayashi, Y., \& Wada, K. 1995, AJ,
  110, 2075

\bibitem[{Sakamoto {et~al.}(1999)Sakamoto, Okumura, Ishizuki, \&
  Scoville}]{Sak99}
Sakamoto, K., Okumura, S.~K., Ishizuki, S., \& Scoville, N.~Z. 1999, ApJS, 124,
  403

\bibitem[{Sandage(1961)}]{San61}
Sandage, A. 1961, The Hubble Atlas of Galaxies (Washington, D.C.: Carnegie
  Inst. of Washington)

\bibitem[{Sanders \& Huntley(1976)}]{San76}
Sanders, R.~H. \& Huntley, J.~M. 1976, ApJ, 209, 53

\bibitem[{Sanders \& Tubbs(1980)}]{San80}
Sanders, R.~H. \& Tubbs, A.~D. 1980, ApJ, 235, 803

\bibitem[{Sault {et~al.}(1995)Sault, Teuben, \& Wright}]{Sau95}
Sault, R.~J., Teuben, P.~J., \& Wright, M. C.~H. 1995, in Astronomical Data
  Analysis Software and Systems IV, ASP Conference Series Vol.\ 77, ed. R.~A.
  Shaw, H.~E. Payne, \& J.~J.~E. Hayes (San Francisco: ASP), 433

\bibitem[{Schmidt(1959)}]{Schmidt59}
Schmidt, M. 1959, ApJ, 129, 243

\bibitem[{Schwarz(1981)}]{Sch81}
Schwarz, M.~P. 1981, ApJ, 247, 77

\bibitem[{Scoville {et~al.}(1986)Scoville, Sanders, \& Clemens}]{Sco86}
Scoville, N.~Z., Sanders, D.~B., \& Clemens, D.~P. 1986, ApJ, 310, L77

\bibitem[{Shaya \& Federman(1987)}]{Shaya87}
Shaya, E.~J. \& Federman, S.~R. 1987, ApJ, 319, 76

\bibitem[{Shioya {et~al.}(1998)Shioya, Tosaki, Ohyama, Murayama, Yamada,
  Ishizuki, \& Taniguchi}]{Shi98}
Shioya, Y., Tosaki, T., Ohyama, Y., Murayama, T., Yamada, T., Ishizuki, S., \&
  Taniguchi, Y. 1998, PASJ, 50, 317

\bibitem[{Shlosman {et~al.}(1989)Shlosman, Frank, \& Begelman}]{Shl89}
Shlosman, I., Frank, J., \& Begelman, M.~C. 1989, \nat, 338, 45

\bibitem[{Shu {et~al.}(1973)Shu, Milione, \& Roberts}]{Shu73}
Shu, F.~H., Milione, V., \& Roberts, W.~W. 1973, ApJ, 183, 819

\bibitem[{Simkin {et~al.}(1980)Simkin, Su, \& Schwarz}]{Sim80}
Simkin, S.~M., Su, H.~J., \& Schwarz, M.~P. 1980, ApJ, 237, 404

\bibitem[{Smith {et~al.}(1991)Smith, Lester, Harvey, \& Pogge}]{Smi91}
Smith, B.~J., Lester, D.~F., Harvey, P.~M., \& Pogge, R.~W. 1991, ApJ, 373, 66

\bibitem[{Sodroski {et~al.}(1995)}]{Sod95}
Sodroski, T.~J. {et~al.} 1995, ApJ, 452, 262

\bibitem[{Stacey {et~al.}(1991)Stacey, Geis, Genzel, Lugten, Poglitsch,
  Sternberg, \& Townes}]{Stac91}
Stacey, G.~J., Geis, N., Genzel, R., Lugten, J.~B., Poglitsch, A., Sternberg,
  A., \& Townes, C.~H. 1991, ApJ, 373, 423

\bibitem[{Stanimirovi\'{c} {et~al.}(1999)Stanimirovi\'{c}, Staveley-Smith,
  Dickey, Sault, \& Snowden}]{Stan99}
Stanimirovi\'{c}, S., Staveley-Smith, L., Dickey, J.~M., Sault, R.~J., \&
  Snowden, S.~L. 1999, MNRAS, 302, 417

\bibitem[{Steer {et~al.}(1984)Steer, Dewdney, \& Ito}]{Steer84}
Steer, D.~G., Dewdney, P.~E., \& Ito, M.~R. 1984, A\&A, 137, 159

\bibitem[{Strong \& Mattox(1996)}]{Str96}
Strong, A.~W. \& Mattox, J.~R. 1996, A\&A, 308, L21

\bibitem[{Tagger {et~al.}(1987)Tagger, Sygnet, Athanassoula, \& Pellat}]{Tag87}
Tagger, M., Sygnet, J.~F., Athanassoula, E., \& Pellat, R. 1987, ApJ, 318, L43

\bibitem[{Teuben(1991)}]{Teu91}
Teuben, P. 1991, in Warped Disks and Inclined Rings Around Galaxies, ed.
  S.~Casertano, P.~D. Sackett, \& F.~H. Briggs (Cambridge: Cambridge U. Press),
  40

\bibitem[{Teuben(1995)}]{Teu95}
Teuben, P.~J. 1995, in Astronomical Data Analysis Software and Systems IV, ASP
  Conference Series Vol.\ 77, ed. R.~A. Shaw, H.~E. Payne, \& J.~J.~E. Hayes
  (San Francisco: ASP), 398

\bibitem[{Thornley \& Wilson(1995)}]{Thorn95}
Thornley, M.~D. \& Wilson, C.~D. 1995, ApJ, 447, 616

\bibitem[{Tilanus \& Allen(1989)}]{Til89}
Tilanus, R. P.~J. \& Allen, R.~J. 1989, ApJ, 339, L57

\bibitem[{Toomre(1964)}]{Toom64}
Toomre, A. 1964, ApJ, 139, 1217

\bibitem[{Tubbs(1982)}]{Tubbs82}
Tubbs, A.~D. 1982, ApJ, 255, 458

\bibitem[{Turner \& Ho(1994)}]{Tur94}
Turner, J.~L. \& Ho, P. T.~P. 1994, ApJ, 421, 122

\bibitem[{{van der Hulst} {et~al.}(1992){van der Hulst}, Terlouw, Begeman,
  Zwitser, \& Roelfsema}]{vdH92}
{van der Hulst}, J.~M., Terlouw, J.~P., Begeman, K., Zwitser, W., \& Roelfsema,
  P.~R. 1992, in Astronomical Data Analysis Software and Systems I, ASP
  Conference Series Vol.\ 25, ed. D.~M. Worall, C.~Biemesderfer, \& J.~Barnes
  (San Francisco: ASP), 131

\bibitem[{{van der Kruit}(1976)}]{vdK76}
{van der Kruit}, P.~C. 1976, A\&A, 52, 85

\bibitem[{{van der Kruit} \& Allen(1978)}]{vdK78}
{van der Kruit}, P.~C. \& Allen, R.~J. 1978, ARA\&A, 16, 103

\bibitem[{{van der Kruit} \& Shostak(1984)}]{vdK84}
{van der Kruit}, P.~C. \& Shostak, G.~S. 1984, A\&A, 134, 258

\bibitem[{Vogel {et~al.}(1988)Vogel, Kulkarni, \& Scoville}]{Vogel88}
Vogel, S.~N., Kulkarni, S.~R., \& Scoville, N.~Z. 1988, \nat, 334, 402

\bibitem[{Walker {et~al.}(1988)Walker, Lebofsky, \& Rieke}]{Walk88}
Walker, C.~E., Lebofsky, M.~J., \& Rieke, G.~H. 1988, ApJ, 325, 687

\bibitem[{Walterbos \& Braun(1996)}]{Walt96}
Walterbos, R. A.~M. \& Braun, R. 1996, in The Minnesota Lectures on
  Extragalactic Neutral Hydrogen, ASP Conference Series Vol.\ 106, ed. E.~D.
  Skillman (San Francisco: ASP), 1

\bibitem[{Wilson(1995)}]{Wils95}
Wilson, C.~D. 1995, ApJ, 448, L97

\bibitem[{Wilson {et~al.}(1991)Wilson, Scoville, \& Rice}]{Wils91}
Wilson, C.~D., Scoville, N.~Z., \& Rice, W.~R. 1991, AJ, 101, 1293

\bibitem[{Wong(2000)}]{Mythesis}
Wong, T. 2000, PhD thesis, Univ.\ of California at Berkeley

\bibitem[{Wong \& Blitz(1999)}]{Wong99}
Wong, T. \& Blitz, L. 1999, \apss, in press

\bibitem[{Wyse \& Silk(1989)}]{Wyse89}
Wyse, R. F.~G. \& Silk, J. 1989, ApJ, 339, 700

\bibitem[{Young {et~al.}(1989)Young, Xie, Kenney, \& Rice}]{Young89}
Young, J.~S., Xie, S., Kenney, J. D.~P., \& Rice, W.~L. 1989, ApJS, 70, 699

\end{thebibliography}

\end{document}